\documentclass[showpacs,amsmath,amssymb,aps,rmp]{revtex4-1}
\usepackage{graphicx}
\usepackage{subfigure}

\usepackage{dcolumn}
\usepackage{bm}

\begin{document}

\title{A Materials Perspective on Casimir and van der Waals Interactions}

\author{$^{1}$L. M. Woods}
\author{$^2$D. A. R. Dalvit}
\author{$^3$A. Tkatchenko}
\author{$^{4}$P. Rodriguez-Lopez}
\author{$^5$A. W. Rodriguez}
\author{$^{6,7}$R. Podgornik}

\affiliation{$^{1}$Department of Physics, University of South Florida, Tampa FL, 33620, USA}
\affiliation{$^2$Theoretical Division, MS B123, Los Alamos National Laboratory, Los Alamos NM, 87545, USA}
\affiliation{$^3$Fritz-Haber-Institut der Max-Planck-Gesellschaft, D-14195 Berlin, Germany}
\affiliation{$^4$Laboratoire de Physique Th\'eorique et Mod\`eles Statistiques, CNRS UMR 8626, B\^at.~100, Universit\'e Paris-Sud, 91405 Orsay cedex, France}
\affiliation{$^5$Princeton Univ, Dept Elect Engn, Princeton, NJ 08540 USA}
\affiliation{$^6$Jozef Stefan Inst, Dept Theoret Phys, SI-1000 Ljubljana, Slovenia}
\affiliation{$^7$Univ Massachusetts, Dept Phys, Amherst, MA 01003 USA}

\begin{abstract}
Interactions induced by electromagnetic fluctuations, such as van der Waals and Casimir forces, are of universal nature present at any length scale between any types of systems with finite dimensions. Such interactions are important not only for the fundamental science of materials behavior, but also for the design and improvement of micro- and nano-structured devices. In the past decade, many new materials have become available, which has stimulated the need of understanding their dispersive interactions. The field of van der Waals and Casimir forces has experienced an impetus in terms of developing novel theoretical and computational methods to provide new insights in related phenomena. The understanding of such forces has far reaching consequences as it bridges concepts in materials, atomic and molecular physics, condensed matter physics, high energy physics, chemistry and biology.
In this review, we summarize major breakthroughs and emphasize the common origin of van der Waals and Casimir interactions. We examine progress related to novel {\it ab initio} modeling approaches and their application in various systems, interactions in materials with Dirac-like spectra, force manipulations through nontrivial boundary conditions, and applications of van der Waals forces in organic and biological matter. The outlook of the review is to give the scientific community a materials perspective of van der Waals and Casimir phenomena and stimulate the development of experimental techniques and applications.

\end{abstract}

\maketitle

\section{Introduction}

\begin{figure}[h]
\centerline{\includegraphics[width=11.0cm]{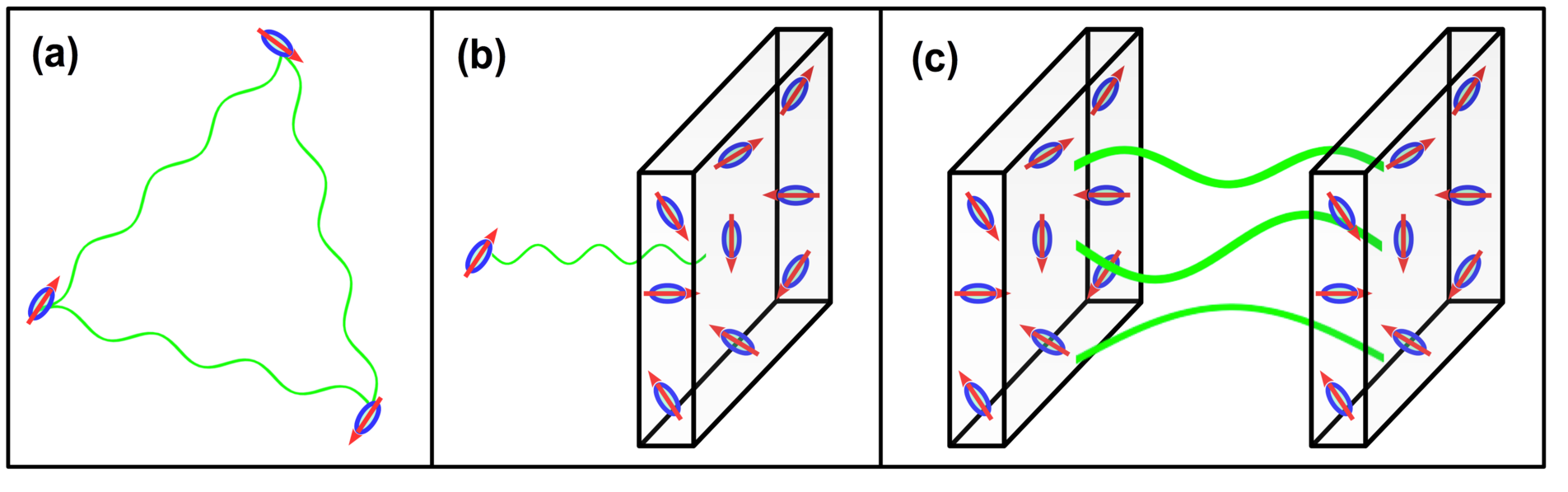}}
\caption{(Color Online) Schematic representation of dispersive interactions induced by electromagnetic fluctuations for: (a) the dipolar vdW force between atoms and molecules; (b) the Casimir-Polder force between atoms and large objects; and (c) the Casimir force between large objects. For small enough separations one can neglect retardation effects due to the finite speed of light $c$, which corresponds to the vdW regime. For large enough separations, retardation effects become important, which is charactersitic for the Casimir regime.  }
\label{fig:Fig_Intro1}
\end{figure}

Phenomena originating from electromagnetic fluctuations play an important role in many parts of science and technology. The Casimir effect, first predicted as an attractive force between neutral perfect metals \cite{Casimir1948}, has made an especially large impact. This non-classical electromagnetic force is typically associated with the coupling between objects with macroscopic dimensions. The same type of interaction known as a Casimir-Polder force concerns atom/surface configurations \cite{Polder1948}. The conceptual realization of the Casimir and Casimir-Polder effects, however, is much more general. The connection of such interactions with broader definitions of "dispersion forces" establishes a close relationship with the van der Waals (vdW) force  \cite{Mahanty1976,Parsegian,Barton1999}. The common origin of vdW and Casimir interactions is directly related to their fluctuations nature, since at thermodynamic equilibrium the electromagnetic energy of dipoles (associated with the vdW force) can also be associated with the energy stored in the electromagnetic fields (the Casimir regime), as illustrated schematically in Fig. \ref{fig:Fig_Intro1}. The point that these constitute the same phenomenon was realized by several authors, including \cite{Barash1984} who write, "The fluctuation nature of the van der Waals forces for macroscopic objects is largely the same as for individual atoms and molecules. The macroscopic and microscopic aspects of the theory of the van der Waals forces are therefore intimately related."

This ubiquitous force, present  between any types of objects, has tremendous consequences in our understanding of interactions and stability of materials of different kinds, as well as in the operation of devices at the micro- and nano-scales. The Casimir force becomes appreciable for experimental detection at sub-micron separations. This is especially relevant for nano- and micro-mechanical devices, such as most electronic gadgets we use everyday, where stiction and adhesion appear as parasitic effects \cite{Buks2001}. The Casimir force, on the other hand,  can be used to actuate components of small devices without contact \cite{Chan2001}. 

vdW interactions are recognized to play a dominant role in the stability and functionality of materials with chemically inert components, especially at reduced dimensions. The most interesting recent example has been graphene and its related nanostructures \cite{Geim2004}. 
The graphene Dirac-like spectrum together with the reduced dimensionality are responsible for novel behaviors in their Casimir/vdW forces. The graphene “explosion” in science and technology has stimulated discoveries of other surface materials, including 2D dichalcogenides, 2D oxides or other honeycomb layers, such as silicene, germanene, or stanene, where dispersive forces are of primary importance. Engineering heterostructures with stacking different types of layers is an emerging field with technological applications via vdW assembly \cite{Geim-2013}. Other materials with Dirac spectra are also being investigated. For example, topological insulators, Chern insulators, and Weyl semimetals are very interesting for the Casimir/vdW field as the surface of such materials has a distinct nature from the bulk. 

The importance of the vdW interaction extends to organic and biological matter. Perhaps the adhesion of the {\it Gecko}, a type of lizzard from the {\it Gekkota} infraorder, has become a pop-cultural poster child for such interactions after Autumn and coworkers \cite{Autumn-2002} in a series of experiments showed that complex hierarchical nano-morphology of the gecko's toe pads \cite{MLee-book} allows them to adhere to hydrophobic substrates \cite{Autumn-2002a}. Dispersion forces play an important role in the organization of other bio-systems, such as cellulose, lignin, and proteins. The stability of biological matter via an array of lipid membranes coupled through the vdW force is a fundamental problem of much current interest in soft matter physics. 

The stability of many hard materials, including composites and heterostructures, is also closely related to their Casimir and vdW interactions. 
The electromagnetic nature makes this phenomenon inherently long-ranged  as it depends in a complicated manner upon the electromagnetic boundary conditions and response properties of the materials. Metallic and dielectric structures of nontrivial shapes lend themselves as a platform where this aspect can be investigated in order to tailor this force in terms of its magnitude and sign. The payoff is highly beneficial in the context of being able to reduce the unwanted stiction and adhesion in nano electro-mechanical  and micro electro-mechanical devices and improve their performance. Structured materials, including metamaterials, photonic crystals, and plasmonic nanostructures, on the other hand, allow the engineering of the optical density of states and magnetic response, which is also useful for manipulating the Casimir force.

Research published in the past decade has shown that the role of materials can hardly be overestimated when it comes to the description and understanding of dispersive interactions. In addition to recent books discussing basic concepts \cite{Parsegian,Bordag:book,Dalvit:book,Simpson:book,BuhmannI:book,BuhmannII:book},  
there are several existing topical reviews on the Casimir effect with different emphasis.  Aspects such as the quantum  field theory nature \cite{Bordag2001,Milton2004},  the quantum electrodynamics (QED) method \cite{Buhmann07}, experimental progress  \cite{Lamoreaux2005}, the Lifshitz theory and proximity force approximation (PFA) with related experiments \cite{Mostepanenko2009}, and non-trivial boundary conditions \cite{Rodriguez11:review,Dalvit11:review,BuhmannI:book,ReidRo12:review,Rodriguez14:review} have been summarized.

 Nevertheless, the materials perspective of vdW/Casimir interactions has not been considered so far. With recent advances in materials science, especially in novel low-dimensional materials, composites, and biosystems, this field has become a platform for bridging not only distance scales, but also concepts from condensed matter, high energy, and computational physics. There is an apparent need for discussing progress beyond the existing topical reviews via a materials perspective and give a broader visibility of this field. The purpose of this article is to summarize advances in the development and application of theoretical and computational techniques for the description of Casimir and vdW interactions guided and motivated by progress in materials discoveries. Each section of this review describes a separate direction defined by the type of systems, length scales, and applications of vdW/Casimir phenomena. An integral part is a succinct presentation of first principles and coarse grained computational methods highlighting  how the distance scale is interconnected with adequate micro and macroscopic description of the materials themselves.  Our intention is to stay within the equilibrium conditions and not include  thermal non-equilibrium vdW/Casimir effects, nor critical Casimir interactions.
 The complexity of these omitted aspects of the vdW/Casimir science and amount of published work warrant a separate  review.

We begin the discussion with the vdW regime (Sec. II). The most significant advances in the past decade  have been in the development of novel first principles methods for vdW calculations. 
Much of this progress has been motivated by the need for an accurate description of vdW interactions in materials as well as relevant experimental measurements. In the next  section (Sec. III), we move on to larger separation scales and focus on emerging materials with Dirac-like spectra, such as graphene and systems with non-trivial topological phases. By summarizing results obtained via the Lifshitz theory, QED approach, and perturbative Coulomb interaction calculations we discuss how the Dirac spectra affect various characteristics of the vdW/Casimir force. The following two sections (Sec. IV and Sec. V) are devoted to Casimir interactions in structured materials. We discuss how the force can be manipulated via response properties engineering and non-trivial boundary conditions. For this purpose, we summarize not only important work in metamaterials, photonic crystals, and plasmonic nanostructures, but we also describe the progress in relevant computational tools. Biological materials are included in Sec. VI by highlighting results obtained via the Lifshitz and Hamaker theory calculations. Fluctuation phenomena for bio-systems are also discussed in light of other, Casimir-like phenomena. Much of this review is focused on the rapid expansion of theoretical and computational advances applied to the vdW/Casimir interactions. 
Although we concentrate on theoretical and computational work, key experiments giving us unprecedented insight into vdW and Casimir interactions are reviewed throughout the paper as well as in Sec VII.  In the last section, we give our outlook for the future by discussing open problems in this field.

\section{{\bf Ab Initio} Methods for van der Waals Forces}

Non-covalent interactions originating from correlated electron fluctuations between materials at separations on the $\AA$ to a few $nm$ scale 
play a key role in understanding their stability and organization. In recent years, important advances have been made towards computational methods for calculating vdW interactions with sufficient accuracy. These state-of-the-art methods are firmly based on a microscopic description of vdW interactions. Based on the treatment of the electron degrees of freedom of the atomistic system, we distinguish between two types of approaches: exact and approximate formulations of the many-body correlation energy. We discuss the essentials in terms of the adiabatic connection fluctuation-dissipation theorem (ACFDT), as both approaches rely on it. Based on the substantial evidence accumulated
over the last few years, we argue that ubiquitous many-body effects in the vdW energy are crucial for accurate
modeling of realistic materials. The inclusion of these effects in first-principles calculations and comparative performance evaluation for a wide range of materials, including finite and periodic molecular systems, (hard) insulating and 
semiconducting solids, and interfaces between organic and inorganic systems are also discussed.
 
\subsection{Exact non-relativistic treatment of microscopic vdW interactions}

The exact energy of a microscopic system obtained via the solution of its Schr\"{o}dinger equation seamlessly includes the vdW contribution. Explicitly solving the Schr\"{o}dinger equation for more than
a few electrons, however, is still a prohibitive task due to the complexity of the many-body problem. Therefore, first-principles 
modeling of realistic materials often starts with more tractable mean-field models, such as the Hartree-Fock approximation (HFA), or density-functional approximations (DFAs), which utilize the three-dimensional electron charge density, $n(\mathbf{r})$, in lieu of the more 
complicated many-electron wavefunction. Unfortunately, these commonly utilized approximations are unable to describe the long-range electronic 
correlation energy and therefore fail to treat vdW interactions. 

The vdW energy is directly related to the electron correlation energy, $E_c$, which can be constructed exactly by invoking the ACFDT~\cite{ACFD1,ACFD2}
\begin{equation}
\label{eqACFD}
E_c = -\frac{\hbar}{2\pi} \int_{0}^{\infty}d\omega \, \int_0^{1}d\lambda \, {\bf Tr}[(\chi_{\lambda}({\bf r},{\bf r'}, i\omega)-\chi_0({\bf r},{\bf r'}, i\omega))v({\bf r},{\bf r'})] ,
\end{equation}
where $\chi_{\lambda}({\bf r},{\bf r'},  i\omega)$ and $\chi_0({\bf r},{\bf r'},  i\omega)$ are respectively the interacting and bare (non-interacting) response functions at Coulomb coupling strength $\lambda$. Here, $\omega$ is the frequency of the electric field,  $v({\bf r},{\bf r'})=|{\bf r}-{\bf r'}|^{-1}$ is the Coulomb potential, and $\bf{Tr}$ denotes the spatial trace operator (six-dimensional integral) over the  spatial electronic coordinates ${\bf r}$ and ${\bf r'}$. The essential idea is that Eq.~\ref{eqACFD} is an adiabatic connection between a reference non-interacting mean-field system with $\lambda=0$ and the fully interacting many-body system with $\lambda=1$~\cite{ACFD1,ACFD2}. The vdW contribution can be found from the so-obtained $E_c$ in a tractable manner provided that a set of single-particle
orbitals computed with DFAs or HFA can be used to construct $\chi_{0}({\bf r},{\bf r'}, i\omega)$. This is still a formidable computational task for systems with thousands of electrons. In addition, approximations are needed to obtain $\chi_{\lambda}({\bf r},{\bf r'}, i\omega)$ for $0 < \lambda \leq 1$. 
  
The power and significance of the ACFDT approach is that essentially all existing vdW modeling methods  can be derived from 
approximations to Eq.~\ref{eqACFD}. For example, the widely employed pairwise approximation is obtained by truncating the
ACFDT expression to second order in the perturbative expansion of the Coulomb interaction. The simple addition 
of inter-atomic vdW potentials that is used to compute the vdW energy in classical force fields and DFA calculations can
be recovered from Eq.~\ref{eqACFD} by further approximating the response function as a sum of independent dipole oscillators located at 
every nucleus in a given material~\cite{TAD-JCP-2013}. The vdW-DF approach originated by Langreth, Lundqvist, and 
collaborators~\cite{LL-PRL,LL-PRB,LL-review} that has become widely used to correct semi-local DFAs can also be 
derived from Eq.~\ref{eqACFD} by making a local approximation to the response function in terms of the electron density and then 
employing second-order perturbation theory. However, the main shortcoming of all these rather efficient approximations is that they 
are unable to capture the non-trivial many-body effects contained in the interacting response function $\chi_{\lambda}({\bf r},{\bf r'},i\omega)$ 
as well as the infinite-order nature of the ACFDT expression in Eq.~\ref{eqACFD}. 

\subsection{Response functions and polarization waves}

The interacting response function is defined self-consistently \textit{via} the Dyson-like equation
\begin{equation}
\chi_{\lambda}=\chi_{0}+\chi_{0}(\lambda v + f_{\lambda}^{xc}) \chi_{\lambda},
\label{eqDyson}
\end{equation}
which contains the exchange correlation kernel $f_{\lambda}^{xc}({\bf r},{\bf r'},i\omega)$, an unknown quantity which must be approximated in practice. 
 Neglecting the explicit dependence 
of $f_{\lambda}^{xc}$ on the coupling constant allows for an analytic integration over $\lambda$ 
in Eq.~(\ref{eqACFD}), and forms the basis for the most widely employed approximation, namely the random-phase 
approximation (RPA)~\cite{Bohm-Pines,GellMann-Brueckner}.

\begin{figure}[ht]
\includegraphics[scale=0.4]{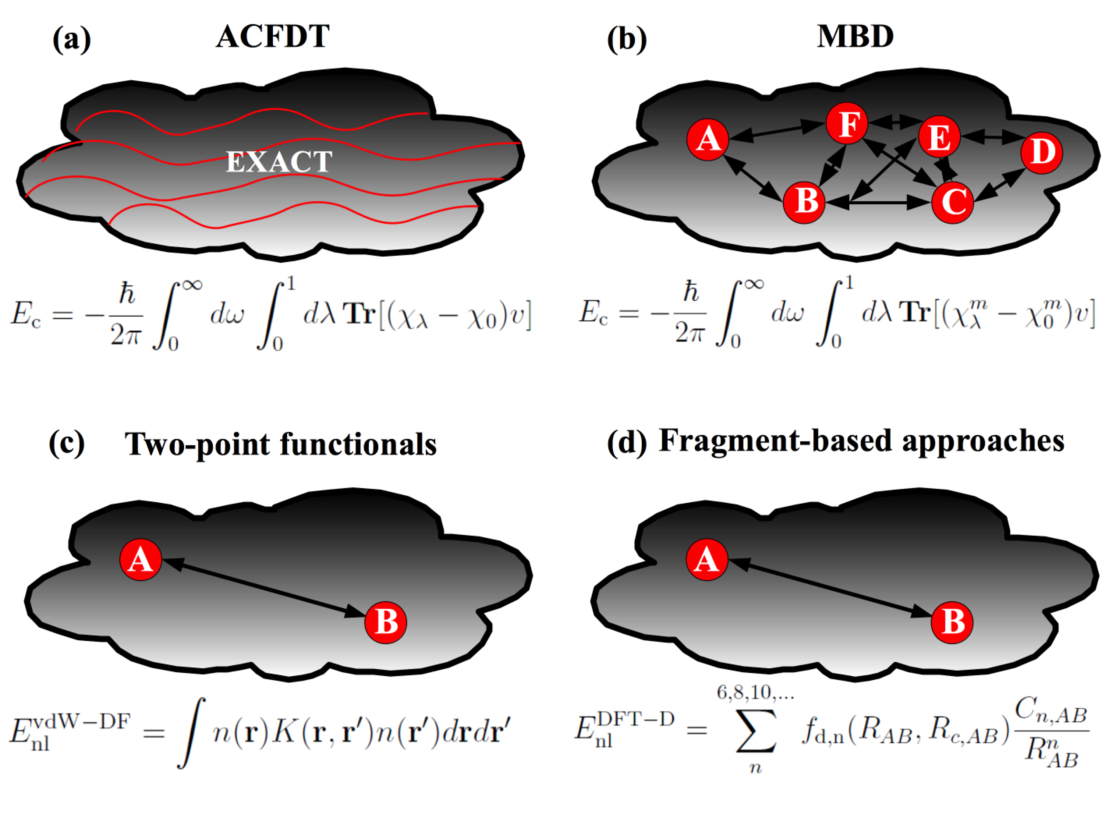}
\caption{(Color online) Schematic representation of first principles methods for the (a) exact formulation of the electronic correlation energy $E_c$ from the adiabatic connection fluctuation-dissipation theorem; (b) formulation based on coupled dipolar fluctuations, such as the many-body dispersion (MBD) methods \cite{MBD,MBD-JCP}; (c) $E_c$ using two-point functionals obtained by approximating the non-homogeneous system with a homogeneous-like response; and (d) fragment-based correlation energy obtained from multipolar expansions.}
\label{Fig1_AT}
\end{figure}

The non-interacting response function can be obtained using the  Adler-Wiser formalism~\cite{Adler:1962,Wiser:1963}, given a set of occupied and unoccupied electronic orbitals $\left\{\phi_i\right\}$ with 
corresponding energies $\left\{\epsilon_i\right\}$ and occupation numbers $\left\{f_i\right\}$ determined from semi-local DFT, Hartree-Fock, 
or hybrid self-consistent field calculations, \textit{i.e.},
\begin{equation}
\label{eqchi0}
\chi_0({\bf r},{\bf r'},i\omega)=\sum_{ij}(f_i-f_j) \frac{\phi_i^*({\bf r})\phi_i({\bf r}')\phi_j^*({\bf r}')\phi_j({\bf r})}{\epsilon_i-\epsilon_j+i\omega} .
\end{equation}
This mean-field $\chi_0$ can exhibit relatively long-range fluctuations (polarization waves), the extent of which is determined by the overlap between occupied and rather delocalized unoccupied electronic states used in  Eq.~\ref{eqchi0}. In this framework, the fluctuations in $\chi_1$ may be shorter-ranged
than in $\chi_0$, especially in 3D solids where the Coulomb interaction leads to significant
screening effects. The situation is generally very different in anisotropic nanostructured materials, where
the Coulomb interaction might lead to so-called \textit{anti-screening} effects, i.e. significantly farsighted 
polarization waves (see Fig. \ref{Fig1_AT}(a) for illustration). 

So far, the general understanding of polarization waves comes from coarse-grained approximations to the density-density response function. For example,  Dobson \textit{et al.} ~\cite{Dobson-PRL} found that the asymptotic vdW interaction between two low-dimensional
metallic objects differs qualitatively from the 
commonly employed sum-over-pairs expressions. Another example is the vdW graphene-graphene interaction energy decays as $d^{-3}$, instead of the conventionally expected $d^{-4}$ 
power law. Here, however,  many-body renormalization of the Dirac graphene carriers 
beyond the RPA might lead to vdW interaction power law between $d^{-3}$ and $d^{-4}$~\cite{Dobson-PRX} (also discussed in Sec. IIIC). This is a matter of ongoing debate.

For some time it was assumed that complete delocalization of fluctuations is required to identify interesting deviations
from the otherwise pairwise-additive behavior. However, Misquitta \textit{et al.} ~\cite{Misquitta-2010, Misquitta-2014} demonstrated that semiconducting wires
also exhibit unusual asymptotics, which becomes more pronounced with the decrease of
the band gap~\cite{Misquitta-2010,Misquitta-2014}. In this case, the vdW interaction exhibited a power law of $d^{-2}$
at large but finite distances, converging to the standard $d^{-5}$ behavior for large inter-wire separations. Ambrosetti \textit{et al.} ~\cite{AmbrosettiF-2014} have analyzed the spatial extent of dipole polarization waves in a wide range
of systems and demonstrated a continuous variation of the power law for finite distances between 1D wires and 2D layers with visibly enhanced non-local responses due to the  collective many-body 
effects. Such relative \textit{farsightedness} of vdW interactions provides an avenue for appropriately tuning the interactions 
between complex polarizable nanostructures.

Another way to understand polarization waves in materials consists in studying the renormalization (non-additivity) of polarizability and vdW
coefficients for different systems as a function of their size and topology. Ruzsinzsky \textit{et al.} ~\cite{Ruzsinsky-2012} modeled the polarizability
of fullerenes employing a hollow shell model with a finite thickness. They demonstrated that the polarizability scales superlinearly 
as a function of fullerene size. This leads to a super-quadratic increase in the vdW $C_6$ coefficients, clearly demonstrating
the importance of long-range fluctuations. Tao and co-workers extended these findings to a wide range of nanoclusters~\cite{Tao-2014}.
Recently, Gobre and Tkatchenko ~\cite{Gobre-2013} studied the dependence of carbon-carbon vdW coefficients for a variety of carbon nanomaterials
and they found that vdW $C_6$ 
coefficients could change from 20 hartree$\cdot$bohr$^6$ to 150 hartree$\cdot$bohr$^6$ depending on the dimensionality,
topology, and size of the carbon nanostructure. This clearly demonstrates the extreme non-additivity of vdW interactions in low-dimensional materials, and highlights the need to include collective effects in vdW interactions when modeling the self-assembly of such nanostructures.

\subsection{Approximate microscopic methods for van der Waals interactions}

 Since the vdW energy is  a tiny part of the total
energy of a many-electron system, vdW methods have to be coupled to an underlying
electronic structure method that provides an adequate treatment of hybridization,
charge transfer, electrostatics, and induced polarization, among other electronic
structure effects. Density-functional theory (DFT) with approximate 
exchange-correlation functionals provides an optimal approach in this regard. DFT is able to correctly describe short-range quantum-mechanical interactions,
and also treats classical electrostatic and polarization effects rather accurately. The total energy $E_t$ of a many-electron system is 
\begin{equation}
E_t = E_{\rm{kin}} + E_{\rm{es}} + E_{\rm{x}} + E_{\rm{c}} ,
\label{eqDFTEnergy}
\end{equation}
where $E_{\rm{kin}}$ is the electronic kinetic energy (corresponding to mean-field kinetic energy in the Kohn-Sham framework),
$E_{\rm{es}}$ is the electrostatic energy (including nuclear repulsion, electron-nucleus attraction, and Hartree electron-electron
repulsion), and $E_{\rm{x}}$ and $E_{\rm{c}}$ are the non-classical exchange and correlation terms, respectively. Most DFT methods utilize semi-local approximations by using information about the electron density (local density approximation, LDA) and its gradients (generalized gradient approximation, GGA). Other approaches are based on  the Laplacian of the electron density in the so-called meta-GGA functionals~\cite{Perdew-2014,Truhlar-Minesotta}. It may also be advantageous to include a certain amount of exact Hartree-Fock exchange in DFA, leading to so-called hybrid functionals.

We note that from ACFDT $E_{\rm{x}}$ and $E_{\rm{c}}$ are non-local (Eq.~\ref{eqACFD}). The correlation energy $E_c$, which is of relevance to the vdW interaction, can be written as $E_{\rm{c}} = E_{\rm{sl}} + E_{\rm{nl}}$, where  $E_{\rm{sl}}$  is the  semi-local correlation energy and $E_{\rm{nl}}$ is the non-local part. The fact that such partition is not unique has led  to a flurry of heuristic approaches that aim to construct a reliable approximation to the full electronic correlation energy. The different classes of methods for the non-local correlation energy are schematically shown in Fig. \ref{Fig1_AT} and are summarized in what follows.

\subsubsection{Two-point density functionals for vdW interactions}

Obtaining an exact expression for  $\chi_{\lambda}({\bf r},{\bf r'},\omega)$ in general is not possible. However, for a 3D homogeneous electron gas the correlation energy can be written \textit{exactly} in terms of the electron density $n(\mathbf{r})$. Approximating the polarization of a non-homogeneous system assuming homogeneous-like response is possible in certain situations~\cite{Rapcewicz-Ashcroft,Dobson-Dinte}. These ideas have led to the derivation of the vdW-DF approach. In addition to the approximation of the interacting polarizability as a local quantity, one also takes a second-order approximation in Eq.~\ref{eqACFD} assuming $\chi_{\lambda} = \chi_1$. Thus the non-local correlation energy is obtained as
\begin{equation}
E^{\rm{vdW-DF}}_{\rm{nl}} = \int n(\mathbf{r}) K(\mathbf{r},\mathbf{r}') n(\mathbf{r}') d\mathbf{r} d\mathbf{r}' \quad ,
\label{eqLL}
\end{equation} 
where $K(\mathbf{r},\mathbf{r}')$ is a ``vdW propagator'' (Fig. \ref{Fig1_AT}(c)). Note that Eq.~\ref{eqLL}  constitutes a great simplification over the exact Eq.~\ref{eqACFD}  since only $n(\mathbf{r})$ and its gradient (utilized in $K$) are required.

The original implementation of this additive non-local correlation energy to the total DFT energy was proposed to couple $E^{\rm{vdW-DF}}_{\rm{nl}}$ to a revised Perdew-Burke-Ernzerhof functional~\cite{revPBE}, the  rationale being that this functional yields repulsive binding-energy for prototypical vdW-bound systems, such as rare-gas dimers~\cite{LL-PRL}. A follow-up implementation using the DFT functional PW86~\cite{LL-PRB} has generated a revised, vdW-DF2 functional. While the vdW-DF2 approach was shown to perform much better for intermolecular interactions, its behavior at
vdW distances is significantly deteriorated when compared to vdW-DF~\cite{Vydrov-PRA-2010}. Specifically, while the 
vdW $C_6$ coefficients in the vdW-DF method are accurate to 19\%, the error increases to 60\% when using vdW-DF2. These approaches illustrate  the challenging problem of balancing between semi-local and non-local interactions in a meaningful
manner.
 
Following the success of the vdW-DF approach, Vydrov and Van Voorhis (VV) provided a significantly simplified vdW functional derivation
and revised the definition of local polarizability, by employing a semiconductor-like dielectric function along with the Clausius--Mossotti relation between polarizability and dielectric 
function~\cite{Vydrov-2009,Vydrov-JCTC}. The VV approach requires one parameter for the  local polarizability and a second one  for the coupling between the non-local vdW energy with the parent DFA approach. The VV functional was assessed
with a wide range of semi-local and hybrid functionals, and by tuning these two parameters, it yielded remarkable performance
for intermolecular interactions compared to benchmark data from high-level quantum-chemical calculations~\cite{Vydrov-JCTC}. Other approaches, such as the C09x functional of Cooper~\cite{Cooper-2010} and the ``opt'' family  by Klime{\v{s}} and Michaelides~\cite{Klimes-2010,Klimes-2011}, rely on the same definition in Eq.~\ref{eqLL}, however the coupling with the DFA is revised by adjusting one or more parameters in the semi-local functional. The "opt" functionals parameters in particular  were adjusted to a benchmark database of intermolecular interaction energies showing  a good performance for cohesive properties of solids~\cite{Klimes-2011}. There are indications, however, that the "opt" functionals overestimate the binding in larger and more complex molecular systems~\cite{Angelos-review}. 

These recent developments have led to many novel insights into the nature of vdW interactions. However, the drastic approximations in $E^{\rm{vdW-DF}}_{\rm{nl}}$  in terms of the additive polarizability and the dependence on the electron density on two points only must be assessed carefully for realistic materials. The neglected non-additive effects can play a very important role in many systems~\cite{Dobson-PRL,Misquitta-2010,Ruzsinsky-2012,Gobre-2013,AmbrosettiF-2014,Tkatchenko-AFM,Tao-2014,Misquitta-2014}. Also, the neglected  three-body Axilrod-Teller and higher-order terms may be quite prominent as well 
~\cite{Donchev,Woods-JPCL,MBD,MBD-PNAS2012,Marom-Angew,Ambrosetti-JPCL,MBD-ACR}.  At this point it is unclear how 
to incorporate higher-order terms in existing non-local vdW functionals without substantially increasing their cost. One possibility is going towards RPA-like approaches, but this would mean departing from a pure density functional picture. 
Another possibility entails further coarse graining of the 
system to a fragment-based description, the progress of which is summarized below.

\subsubsection{Fragment-based methods for vdW interactions}

Fragment-based methods can be traced back to the original work of London~\cite{London}, in which case utilizing second-order perturbation theory for the Coulomb interaction the dispersion energy between two spherical atoms $A$ and $B$ can be obtained. The London expression, often using just the dipolar term $C_{6,AB} / R^6_{AB}$, is the basis for calculating
vdW dispersion energies in a wide range of atomistic methods, including Hartree-Fock calculations~\cite{Scoles1975,Scoles1977},  DFA 
calculations~\cite{Grimme-D2,Grimme-D3,Becke-Johnson,TS-PRL,Corminboeuf-2011} and  quantum chemistry methods~\cite{MP2-DeltavdW}. $E_{\rm{vdW}}^{(2)}$ is valid only at large separations for which the overlap between orbitals of atoms $A$ and $B$ can be neglected. At shorter sepations the overlap naturally reduces the interaction~\cite{Koide}, which can be conviniently included by a  \emph{damping} function \cite{Tang-Toennies,Grimme-D2,Grimme-D3,Becke-Johnson,TS-PRL}
\begin{equation}
E_{\rm{vdW}}^{(2)} = \sum_{n}^{6,8,10,...} f_{\rm{d},n}(R_{AB},R_{c,AB}) \frac{C_{n,AB}}{R^n_{AB}} \quad ,
\label{eqDDisp}
\end{equation}
where $f_{\rm{d},n}(R_{AB},R_{c,AB})$ is the damping function that depends on a cutoff radius $R_{c,AB}$ (Fig. \ref{Fig1_AT}(d)). This type of approach can be quite useful in DFA-GGA functionals, such as  PBE~\cite{PBE}, which perform very well for chemical bonds. In this case, the dipolar approximation to Eq.~\ref{eqDDisp} is sufficient~\cite{ScolesJCP2001,ElstnerHobzaJCP2001,WuYang,Parrinello2004,Grimme-D1}. These "DFT-D" approaches have experienced tremendous developments.  In particular, Grimme~\cite{Grimme-D2} published a set of empirical
parameters for a range of elements and demonstrated
that the addition of dispersion energy to a wide range of functionals yields remarkably accurate
results for intermolecular interactions. In the latest DFT-D3 method, Grimme has extended his empirical set of parameters to cover elements from H to Pu~\cite{Grimme-D3}.

Considerable efforts have been dedicated towards determining vdW parameters directly from electronic structure calculations also. Becke and Johnson~\cite{Becke-Johnson} 
developed an approach based on the exchange-hole dipole moment (XDM) to determine vdW coefficients, which can be computed by  using Hartree-Fock orbitals. Later, Steinmann and Corminboeuf presented an alternative derivation based on electron density and its first and second derivatives~\cite{Corminboeuf-2011}. The alternative derivation of the fragment-based vdW-DF functional by Sato and Nakai has demonstrated an interesting connection
(and potential equivalence) between fragment-based methods and explicit non-local functionals~\cite{Sato-2009,Sato-2010}. Tkatchenko and Scheffler (TS) developed a DFA based approach to determine both $C_{6,AB}$ coefficients and $R_{c,AB}$ radii as functionals of the electron density~\cite{TS-PRL}, which implies that the  vdW parameters respond to changes due to  hybridization, static charge transfer, and other electron redistribution processes.  The TS approach demonstrated that by utilizing the electron density 
of a molecule and high-level reference data for the free atoms, it is possible to obtain asymptotic vdW coefficients with 
accuracy of 5.5\%, improving by a factor of 4-5 on other existing approaches at the time. 
Bu\v{c}ko and co-workers have pointed out that an iterative Hirshfeld partitioning scheme for the
electron density can significantly extend the applicability of the TS method to ionic materials~\cite{Bucko-2013,Bucko-2014}.

This field is still developing at a rather quick pace, therefore revised and completely new fragment-based approaches are still being introduced.

\subsubsection{Efficiently beyond pairwise additivity: Explicit many-body vdW methods}

In more complex and heterogeneous systems, it is necessary to go beyond the simple additive models and 
further efforts of atomistic vdW modeling must be directed towards the inclusion of many-body effects. In principle, RPA using DFA orbitals provides a good model, however the dependence of $\chi_0$ on the exchange-correlation functional and the high computational cost in the $\chi_0$ computations may be limiting factors. The main challenge is to construct reliable approximations for the long-ranged vdW correlations, since the short-ranged correlation effects are well accounted for in DFA. Therefore, the full $\chi_0({\bf r},{\bf r'},i\omega)$ is often unnecessary as is the case in nonmetallic or weakly metallic systems. In such situations, it is possible to describe $\chi_0$ by a set of localized atomic response functions (ARFs), which can be constructed to accurately capture the electronic response beyond a certain cutoff distance.

Although the ARF concept has been employed in \textit{model systems} starting 50 years ago~\cite{Bade,Donchev,Cole-CFDM,Woods-JPCL,Dobson-chains}, only recently has this idea been extended to non-local vdW interactions in \textit{realistic materials}~\cite{MBD-JPCM14}. For this purpose, spatially-extended ARFs that increase the applicability 
of the model to include close contact have been used within the so-called many-body dispersion (MBD) method, schematically illustrated in  Fig. \ref{Fig1_AT}(b) ~\cite{MBD,MBD-JCP}. Within this approach, each $p$-th atom in the material is represented by a single dipole oscillator with a frequency-dependent polarizability $\alpha_p(i\omega) = \frac{\alpha_{p,0}\omega_{p,0}^2}{\omega_{p,0}^2+\omega^2}$, where $\alpha_{p,0}$ is the static polarizability and $\omega_{p,0}$ is an effective excitation (or resonant) frequency. The bare ARF response then is written as 
\begin{equation}
\label{eqChiPol}
\chi_{0,p}({\bf r},{\bf r'},i\omega) = - \alpha_p(i\omega) \nabla_{{\bf r}}\delta^3({\bf r} - {\bf R}_p) \otimes \nabla_{{\bf r'}}\delta^3({\bf r'} - {\bf R}_p) , 
\end{equation}
where ${\bf R}_p$ is the location of the $p$-th atom and $\otimes$ signifies a tensor product. 
The bare response function for a collection of atoms follows simply as the direct sum over the individual ARFs, 
$\chi_0({\bf r},{\bf r'},i\omega)=\chi_{0,p}({\bf r},{\bf r'},i\omega) \oplus \chi_{0,q}({\bf r},{\bf r'},i\omega) \oplus \cdots$. 
The ARF response contains the infinite-order correlations from the start and it can be used in Eq.~\ref{eqACFD} to calculate the interaction energy. It has been demonstrated that the RPA correlation energy is equivalent to the exact diagonalization of the Hamiltonian corresponding to ARFs coupled 
by a long-range dipole potential~\cite{TAD-JCP-2013}. Using second-order perturbation theory  one also recovers the well-known pairwise-additive formula for the vdW energy~\cite{TAD-JCP-2013}.

We note that  the solution of Eq.~\ref{eqACFD} for a model system of ARFs yields an expression for
the long-range correlation energy beyond what would simply be called ``vdW dispersion energy'' in the traditional London picture~\cite{London}. Even for two atoms described by dipole-coupled ARFs, the correlation energy contains an infinite numbers of terms $C_{n,AB} / R_{AB}^{n}$. The polarizability of the combined $AB$ system in general  is  not
equal to the sum of polarizabilities of isolated atoms $A$ and $B$, and higher-order correlation terms account precisely for this fact.  It has been found that the convergence of the perturbative 
series expansion in Eq.~\ref{eqDDisp} can be extremely slow, especially for systems which have either high polarizability density or
low dimensionality. This is clearly illustrated by the binding energy in supramolecular complexes or double-walled nanotubes, 
for which even 8-body terms make a non-negligible contribution to the correlation energy on the order of 2-3\%~\cite{Ambrosetti-JPCL}.

\subsection{Applications of atomistic vdW methods to materials}
\begin{figure}[ht]
\includegraphics[scale=0.55]{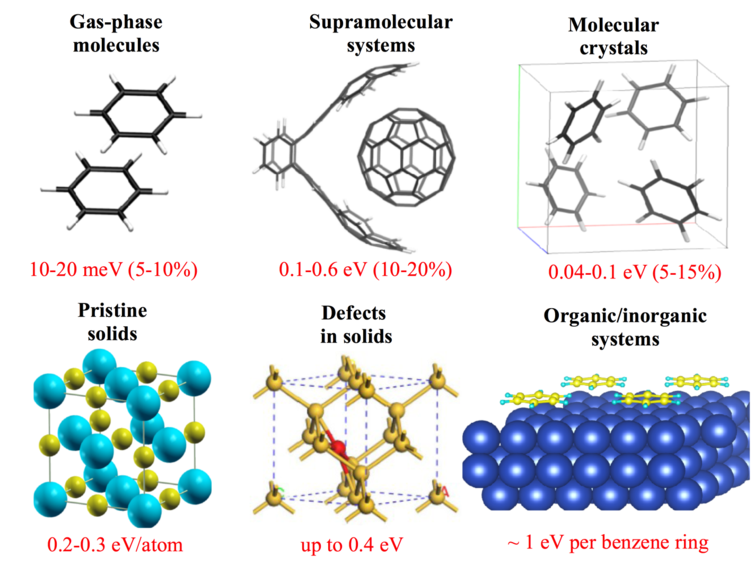}
\caption{(Color online) Various types of materials with calculated cohesion energies and errors as compared to available reference data. The top row of values shows typical errors of atomistic vdW methods compared to benchmark bending energies, while the bottom row of values shows contributions of the vdW energy to the binding energy of the corresponding materials.}
\label{Fig2_AT}
\end{figure}

Our discussions above  show that the developments of novel many-body methods and understanding of many-body effects in the vdW energy is an area of significant current
interest~\cite{MBD,MBD-JCP,Silvestrelli-MBD,otero2013many,modrzejewski2014range}. This is highly motivated from an experimental point of view as well. Being able to describe vdW interactions in different systems is highly desirable in order to explain existing and predict new experimental findings. Fig.\ref{Fig2_AT} summarizes typical results from $ab$  $initio$ calculations, as described below, for cohesion energies and error ranges as compared to available reference data. 

\subsubsection{Finite and periodic molecular systems}

For smaller molecules, high-level  quantum-chemical benchmarks using coupled-cluster calculations are now 
widespread \cite{S22,S22-Sherrill,S66}. 
In particular, the coupled-cluster method with single, double, and perturbative triple
excitations [CCSD(T)] is currently considered as the ``gold standard'' of quantum chemistry.
For yet larger molecules (up to 200 light atoms), it is possible to carry out Diffusion 
Quantum Monte Carlo (DQMC) calculations~\cite{Ambrosetti-JPCL,OAvL-DMC} using massively-parallel computer
architectures. DQMC calculations in principle yield the exact solution (within statistical sampling
accuracy) for the Schr\"{o}dinger equation within the fixed-node approximation~\cite{DQMC-review}.
Examples of small molecules benchmark databases are those for the S22~\cite{S22,S22-Sherrill} and S66~\cite{S66} dimers, containing 22 and 66
dimers, respectively.
For supramolecular systems, the S12L database has been recently introduced
by Grimme~\cite{Grimme-S12L} and benchmark binding energies for 6 of these 12 complexes have been calculated 
using DQMC~\cite{Ambrosetti-JPCL}. For extended periodic molecular crystals, one can rely on experimental
lattice enthalpies, extrapolated to 0 K and with zero-point energy subtracted. Two databases, C21~\cite{Johnson-C21} and X23~\cite{Reilly-X23}, 
have been recently introduced for molecular crystals.
Both of these databases include molecular crystals bound primarily by either hydrogen bonds or vdW dispersion,
including a few crystals with mixed bonding nature. The X23 database extended the C21 one and improved the 
calculation of vibrational contributions required to convert between experimental sublimation enthalpies and
lattice energies.

Initially, the development of atomistic methods for vdW interactions has been largely driven
by their performance for small molecules in the S22 and S66 databases. Currently, the best methods
are able to achieve accuracies of 10-20 meV (better than 10\%) in the binding energies compared to reference CCSD(T) values (Fig. \ref{Fig2_AT}). The errors are due to inaccuracy in the asymptotic vdW coefficients, empirical parameters in damping
functions, and errors in the exchange-correlation functional. 

Due to such rather uniform performance of different methods for small molecules, the focus has 
 shifted to assessing the performance for larger systems. 
Here, in fact, the differences are more prominent, because the vdW energy makes a much larger relative
contribution to cohesion. For example, for polarizable supramolecular systems, such as the
``buckyball catcher'' complex, pairwise dispersion corrections overestimate the binding
energy by 0.4--0.6 eV~\cite{catcher-JCTC} compared to reference DQMC values. Only upon accurately including
many-body dispersion effects one obtains results within 0.1 eV from the best available benchmark~\cite{Ambrosetti-JPCL}.
So far, vdW-DF functionals have not been applied to study binding energetics in the S12L database.

For periodic molecular crystals, some pairwise and many-body fragment-based methods are able 
to achieve remarkable accuracy of 40--50 meV per molecule (5\% mean absolute relative error), compared to experimental results~\cite{Reilly-JPCL,Reilly-X23}. 
Since the difference in lattice energies between various available experiments is on the same order of magnitude, this highlights the mature state 
of vdW dispersion corrections to DFA. The vdW-DF2 approach yields a somewhat larger error of $\approx$ 70 meV (7.5\%) on the C21 database~\cite{Johnson-C21}.
Understanding the performance of different vdW-inclusive methods for large molecular systems is still a subject of ongoing
research. Some of the pairwise correction approaches have been specifically fitted to periodic systems, trying to mimic
many-body screening effects by changing the short-range damping function. This procedure seems to work well for 
certain molecular crystals with high symmetry, but this is obviously not a transferable approach.

Many-body vdW correlations become even more relevant for the \textit{relative} energetics of molecular
systems, which are essential to predict the correct polymorphic behavior of molecular crystals. Marom \emph{et al.} have
demonstrated that only upon including many-body effects one is able to correctly reproduce the structures and relative
stabilities of glycine, oxalic acid and tetrolic acid~\cite{Marom-Angew}. 
Another interesting example is the aspirin crystal, for which a long-standing controversy 
has been about the relative stability of polymorphs form I and form II~\cite{Price-aspirin}. 
Reilly and Tkatchenko have recently demonstrated that the stability of the most abudant form I arises 
from an unexpected coupling between collective vibrational and electronic degrees of freedom
(dynamic plasmon--phonon coupling)~\cite{Reilly-PRL}. In this case, many-body vdW correlations renormalize phonon frequencies 
 leading to low-frequency phonon modes that increase the entropy  and ultimately determine the
stability of this ubiquitous form of aspirin in comparison to the metastable form II. Furthermore, the bulk and
shear moduli of both forms are substantially modified and become in better agreement with experiments when calculated with DFA+MBD. 
The aspirin example illustrates how the inclusion of many-body vdW effects may lead to novel 
qualitative predictions for the polymorphism and elastic response of molecular materials.  

\subsubsection{Condensed materials}
For hard solids (ionic solids, semiconductors, and metals), the role of vdW interactions 
was considered to be negligible for a long time, as judged for example by classical condensed-matter textbooks~\cite{Ashcroft-Mermin,Kittel}. 
The rather strong cohesion in hard solids stems from covalent and metallic bonds, or from classical Coulomb interaction between localized charges (Fig. \ref{Fig2_AT}).  
Early estimates of vdW interactions in hard solids varied substantially from being negligible to being very 
important~\cite{Mayer-1933,RZK,Ashcroft-Mermin,Richardson-Mahanty,Perdew-vdW-2010}. 
Recently, this issue has been systematically revisited by employing DFA with vdW interactions. Zhang et al.~\cite{Zhang-PRL} demonstrated
that long-range vdW interactions account for $\approx$ 0.2 eV/atom in the cohesive energy for Si, Ge, GaAs, NaCl, and MgO, and 9--16 GPa in the bulk modulus. This amounts to a contribution of 10--15\% in the cohesive energy and bulk modulus -- far from being negligible if one aims at an accurate description of these properties. Klime{\v{s}} and co-workers applied their ``opt'' functionals based on the vdW-DF approach to a large database of solids~\cite{Klimes-2011} finding  that vdW interactions play an important role for an accurate description of cohesive properties. Overall, their conclusion
is that  vdW interactions allow improving the performance of many different xc functionals, achieving good performance
for both molecules and hard solids. 

Because vdW interactions are important for absolute cohesive properties of solids, any property that depends on energy \textit{differences} is 
also likely to be influenced by vdW effects. 
Therefore, vdW interactions will often play an important role in the relative stabilities of different 
solid phases, phase transition pressures, and phase diagrams as demonstrated for polymorphs of 
TiO$_2$~\cite{Grimme-TiO2}, ice~\cite{Santra-2011}, different reconstructed phases of the oxidized Cu(110) surface~\cite{Stich-2013}, and alkali borohydrites~\cite{Huan-2013}. 

The properties of many solids are substantially affected by the presence of simple and complex defects,
such as neutral and charged interstitials 
and vacancies~\cite{Neugebauer-RMP}. The formation of defects entails a modification
of polarization around defect sites and this can have a substantial effect on the contribution of vdW energy
to the stability and mobility of defects. Gao and Tkatchenko have demonstrated that the inclusion of many-body vdW interactions
in DFA improves the description of defect formation energies, significantly changes the barrier geometries for defect
diffusion, and brings migration barrier heights into close agreement with experimental values~\cite{Gao-PRL}. 
In the case of Si, the vdW energy substantially decreases the migration barriers of interstitials and impurities
by up to 0.4 eV, qualitatively changing the diffusion mechanism~\cite{Gao-PRL}. 
Recently, the proposed mechanism has been confirmed by explicit RPA calculations~\cite{Kresse-PRB}.

\subsubsection{Interfaces between molecules and solids}
The predictive modeling and understanding of hybrid systems formed between molecules and solids are an essential prerequisite for tuning 
their electronic properties and functions. The vdW interactions often make a substantial contribution to the
stability of molecules on solids~\cite{MRSBull}. Indeed, until recent developments for efficiently 
incorporating the long-range vdW energy within state-of-the-art DFAs, it was not possible to study the structure 
and stability of realistic interfaces~\cite{WLiu-ACR}.
Exposed surfaces of solid materials are characterized by formation of collective electronic states, thus the long-range screening effects should be treated at least in an effective way, as done for example in the
DFT+vdW$^{\rm{surf}}$ method that accounts for the collective electronic response effects 
by a combination between an interatomic dispersion expression and the Lifshitz-Zaremba-Kohn theory~\cite{Ruiz-PRL}. 
This method was demonstrated to be reliable for the structure and stability of a broad class of organic molecules adsorbed on metal surfaces,
including benzene, naphthalene, anthracene, diindenoperylene, C$_{60}$, and sulfur/oxygen-containing molecules 
(thiophene, NTCDA, and PTCDA) on close-packed and stepped metal surfaces, leading to an overall accuracy of 0.1~{\AA} 
in adsorption heights and 0.1--0.2 eV in binding energies with respect to state-of-the-art experiments~\cite{WLiu-ACR}.

A particularly remarkable finding is that vdW interactions can contribute more to the binding of strongly bound molecules on 
transition-metal surfaces than they do for molecules physisorbed on coinage metals~\cite{WLiu-PRB,Michaelides-NatureMat}. 
The accurate inclusion of vdW interactions also significantly improves molecular tilting angles and adsorption heights, 
and can qualitatively change the potential-energy surface for adsorbed molecules with flexible functional groups. 
Activation barriers for molecular switches~\cite{WLiu-NatureComm} and reaction precursors~\cite{Siler-JACS} are modified as well. Ongoing work concentrates on understanding the interplay between many-body effects within the solid material and
collective effects within the adsorbed molecular layers.

\section{Dirac Materials beyond atomic scale separations}

Without a doubt, novel $ab$ $initio$ methods have advanced our understanding of vdW interactions between systems at atomic scale separations. When the objects are taken further apart, however, other approaches become more appropriate. Dispersion interactions involving objects with macroscopic dimensions at distances for which the electronic distribution overlapping effects are not important are typically described by the Lifshitz formalism \cite{Lifshitz1956,Dzyalosh}, which has been the mainstream theory for conventional metals and dielectrics with planar extensions for over several decades. New materials with Dirac spectra are emerging, however, and the Lifshitz approach is an excellent tool to capture the signatures of the Dirac carriers in the vdW/Casimir interactions.    

\subsection{Lifshitz formalism}

A generalized Lifshitz formula can be obtained from the ACFDT expression in Eq. \ref{eqACFD} for distance separations larger than several nm-s, where the overlap of the electronic distribution residing on each object can be neglected. In this case, the response properties are independent of each other, thus they are described by the individual response functions $\chi^{(1,2)}_0$   (1,2 denote the two objects) and the mutual Coulomb potential can be taken as a perturbation. Although  $\chi^{(1,2)}_0$ do not include electronic correlations from the overlap, they contain the electronic correlations within each object. When $\chi^{(1,2)}_0$ are calculated via the RPA approach and the mutual Coulomb interaction is described by the RPA ring diagrams \cite{Lifshitz1956, Dzyalosh,Fetter1971}, the ACFDT expression (Eq.\ref{eqACFD}) is transformed to the  non-retarded Lifshitz formula given in Fourier basis for planar homogeneous objects \cite{Dobson-review, Despoja2007}
\begin{equation}
E^{(L)}=- \hbar \int \frac{d\textbf{k}_{\parallel}}{(2\pi)^2} \int_0^{\infty} \frac{d\omega}{2\pi}\ln [1-\chi_0^{(1)}({\textbf{k}_{\parallel}}, i\omega)V_{12}({\textbf{k}_{\parallel}}, i\omega)\chi_0^{(2)}({\textbf{k}_{\parallel}}, i\omega)V_{21}({\textbf{k}_{\parallel}}, i\omega)] ,
\label{lifshitz_nonret}
\end{equation}
where $\textbf{k}_{\parallel}$ is the 2D wave vector and $V_{12}({\textbf{k}_{\parallel}}, i\omega)$ is the Coulomb interaction between the two objects. 

Distance separations on the order of sub-$\mu$m and $\mu$ m scales are characteristic for the Casimir regime, where retardation becomes prominent. For such separations one has to include all photon interactions being exchanged with  the finite speed of light $c$. The Casimir energy can be derived using scattering methods by solving the boundary conditions arising from the electromagnetic Maxwell equations. The interaction energy can written in the form
~\cite{Rahi09:PRD,Lambrecht06,Lambrecht09}
\begin{equation}
E^{(C)}=- \hbar \int \frac{d\omega}{2\pi}\int \frac{d\textbf{k}_{\parallel}}{2\pi^{2}}\ln \mathrm{Det} [{\mathbb{I} - \mathbb{R}_{1}(i\omega)\mathbb{R}_{2}(i\omega)e^{-2d\sqrt{ \frac{\omega^{2}}{c^2} + \textbf{k}_{\parallel}^{2} }}}] ,
\label{Lifshitz_ret}
\end{equation}  
where $\mathbb{R}_{1,2}$ are the reflection matrices of the individual objects evaluated at imaginary frequencies. The reflection matrices describe appropriate boundary conditions and they are expressed in terms of the macroscopic response properties of the objects. Eq. \ref{Lifshitz_ret} can be obtained equivalently via QED techniques relying on the evaluation of the Maxwell stress tensor whose components represent the vacuum expectation of the electromagnetic field and they are given in terms of the dyadic Green's function (more details on this approach are found in Sec. V) \cite{Dzyalosh,Buhmann07}. Utilizing the fluctuation-dissipation theorem and standard complex contour integration techniques, Matsubara frequencies $i\omega_n=in2\pi k_BT/\hbar$ are introduced in the description. As a result, the Casimir interaction energy $E^{(C)}$ can be cast into a temperature-dependent form using the relation $\hbar \int_0^\infty\frac{d\omega}{2\pi} \rightarrow k_BT{\sum_{n=0}^{\infty}}'$ (the prime in the sum means that the $n=0$ term is multiplied by $1/2$). We further note that setting $c=\infty$ in Eq. \ref{Lifshitz_ret} the non-retarded Lifshitz expression (Eq. \ref{lifshitz_nonret}) is recovered.

Being able to utilize independently calculated response properties with different models (including RPA or the Kubo formalism) or even use experimental data in Eqs. \ref{lifshitz_nonret},\ref{Lifshitz_ret} has been especially useful for the versatility of the Lifshitz macroscopic approach. In addition, the Matsubara frequencies give the means to take into account temperature in the interaction unlike the $ab$ $initio$ methods (Sec. II), which calculate the vdW energy at zero temperature. Much of the progress in theoretical and experimental work concerning typical metals and dielectrics interactions, captured by the macroscopic vdW/Casimir approach, has been summarized in several books and recent reviews \cite{Parsegian,Lamoreaux2005,Mostepanenko2009,Bordag:book,Dalvit:book,Buhmann07}.

The materials library is expanding, however. A new subset of systems, characterized by Dirac fermions in their low-energy spectra, has emerged recently \cite{Balatsky-2014}. This distinct class of materials has properties markedly different from the ones of conventional metals and semiconductors whose fermions obey the Schr\"{o}dinger's equation. Recent discoveries have shown that there are many types of systems with Dirac nodes in the band structure, including 2D graphene, topological insulators (TIs) and Weyl semimetals. Research efforts on vdW/Casimir interactions involving graphene and  related systems have shown that fluctuation-induced phenomena are strongly influenced by the Dirac nature of the carriers.  As discussed previously, $ab$ $initio$ calculations have been indispensable for the demonstration of atomic registry-dependent effects, unusual scaling laws, farsightedness and many-body nature of their vdW interaction  in graphitic systems \cite{Bucko-2013,Bucko-2014,Bucko:PRB,Gobre-2013,Woods-JPCL,Lebegue-2010}.  
Nevertheless, it is very important from a fundamental point of view to consider regimes where the dispersion interactions are determined primarily by the low-energy Dirac carriers. Unlike $ab$ $initio$ methods which take into account the entire band-structure of the interaction materials, the Lifshitz/Casimir formalism relying on response functions calculated via low-energy models  gives us an excellent opportunity to study the emergent physics of the Dirac carriers in vdW/Casimir forces.

\subsection{Basic properties of graphene nanostructures}

After the discovery of graphene \cite{Geim2004}, significant progress has been made towards understanding its properties. For example, basic science in terms of the 2D Dirac-like nature \cite{CastroNeto2009}, electronic transport  \cite{DasSarma2011}, collective effects due to electron-electron interactions \cite{Kotov2012}, and spectroscopy \cite{Basov2014} have been studied. Quasi-1D allotropes, such as carbon nanotubes (CNTs) and graphene nanoribbons (GNRs) are also available  \cite{Saito1998,Ma2013},  and key scientific breakthroughs have been summarized \cite{Charlier2007,Yaziev2010}.  In addition to the internal properties, understanding how chemically inert carbon nanostructures interact at larger length scale separations (more than several $\AA$-s) is of primary importance. Much progress in the past several years has been achieved towards learning how such dispersive forces are influenced by the graphitic internal properties and external factors, such as temperature, doping, and applied fields. This knowledge is relevant for a variety of phenomena including the formation and stability of materials and composites, adsorption, manipulation of atoms, and operation of devices, among others. Since the description of the interaction via the macroscopic Lifshitz/Casimir formalism depends upon the low-energy electronic structure and optical properties,  here we provide an overview of the relevant characteristics of graphene, CNTs, and GNRs. 

\begin{figure}[h]
\includegraphics[scale=0.15]{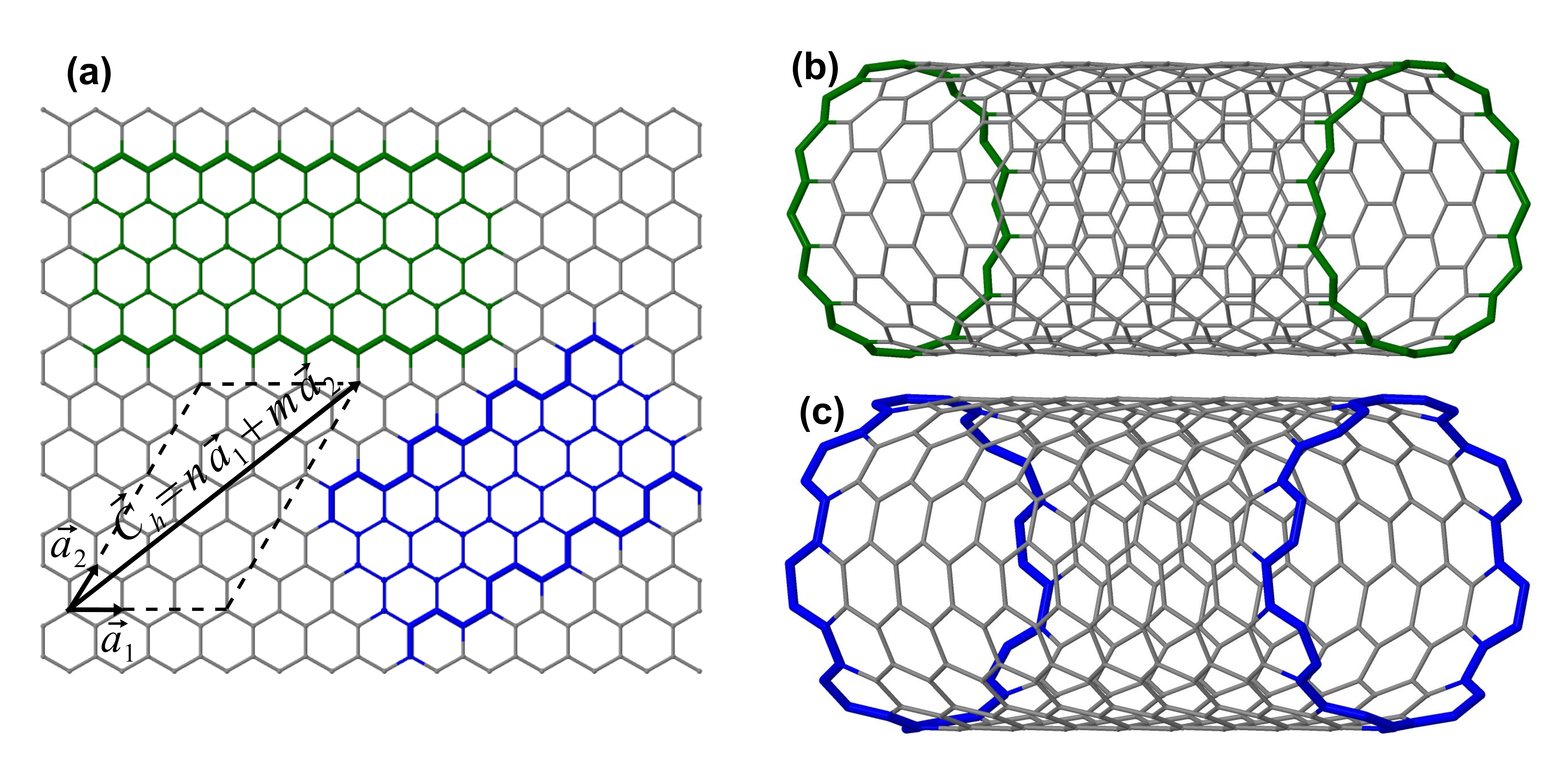}
  \caption{(Color Online) (a) A graphene layer is an atomically thin sheet of honeycomb carbon atoms. 
Zigzag (green) and armchair (blue) graphene nanoribbons can be realized by cutting along the specified edges. 
Carbon nanotubes can be obtained by folding along the chirality vector $\vec{C}_h$, determined   by the indices $n$ and $m$ and the lattice unit vectors $\vec{a}_{1,2}$. The chirality index $(n,m)$ uniquely specifies each nanotube with two achiral examples shown: (b) zigzag $(m,0)$ and (c) armchair $(n,n)$. 
Alternatively, folding a zigzag nanoribbon along the axial direction results in an armchaired nanotube, while folding an armchair nanoribbon gives a zigzag nanotube. }
   \label{fig:Fig_Gr_NStr}
\end{figure}

Graphene is a 2D atomic layer composed of hexagonally oriented rings (Fig. \ref{fig:Fig_Gr_NStr}). Many of its properties can be captured by  a nearest-neighbor tight-binding model within the first Brillouin zone with two inequivalent $K$ points at ${\bf K}=(\pm \frac{4\pi}{\sqrt{3}a},0)$ ($a$ is the graphene lattice constant), which describes  a $\pi$ valence bonding band with one electron and an empty $\pi^{*}$ anti-bonding band with one hole.  The linearization of the energy spectrum around the  ${\bf K}$ centered valleys yields the low-energy massless chiral Dirac-like Hamiltonian in 2D
\begin{equation}
H_{gr}=\hbar v_F {\bf \sigma} \cdot {\bf q} -\mu, 
\label{gra-Hamiltonian}
\end{equation}
where $\mu$ is the chemical potential and ${\bf  \sigma}$ is the 2D spinor. The nonzero spinor components $\sigma_{x},  \sigma_{y}$ are the Pauli matrices, which refer to the graphene pseudospin rather than the real spin. The energy spectrum $E_{gr}$  is linear with respect to the wave vector ${\bf q}={\bf k}-{\bf K}$ according to $E_{gr}=\hbar v_F q$ with the  electronic group velocity being $v_F=\sqrt{3}t_0 a/2\hbar\sim 10^6$ $m/s$ ($t_0$ is the nearest-neighbor tight-binding hopping integral).  It is interesting to note that although graphene was synthesized not long ago, theoretical insight in terms of the low-energy massless Dirac-like $H_{gr}$ was discussed much earlier by \cite{Semenoff1984}, who expanded upon the tight-binding description introduced by \cite{Wallace1947}.

The tight-binding model can be extended to CNTs as well. Imposing periodic boundary conditions around the cylindrical circumference, the corresponding  wave functions are zone-folded, meaning that the wave vector in the azimuthal direction takes a set of discrete values. The nomenclature of CNTs is described via a chirality vector ${\bf C}_h = n {\bf a}_1+m {\bf a}_2$, where $n,m$ are integers. As a result, single-walled CNTs are denoted via a chirality index $(n,m)$ with the achiral nanotubes  labeled as armchaired $(n,n)$ or zigzag $(n,0)$, as shown in Fig. \ref{fig:Fig_Gr_NStr}. The CNT energy bands can also be found \cite{Mintmire1992, Tasaki1998} with zone-folding boundary conditions leading to a chirality dependent energy spectrum. 
The tight-binding energy band structure for GNRs, on the other hand, is obtained by requiring the wave function be periodic along the GNR axis and vanish at the edges, which introduces edge dependent (zigzag or armchair) phase factors in the energy spectra \cite{Brey2006, Akhmerov2008, Sasaki2011}.

The electronic structure of graphene systems determines their optical response properties - key components for the vdW/Casimir calculations. The Dirac-like carriers have profound effects on how electromagnetic excitations are handled by graphene. The chiral symmetry for the graphene quasiparticles in a given K-valley is either parallel or antiparallel to the direction of motion of the electrons and holes. An immediate consequence is that in the $k_BT \ll \hbar\omega$ limit  the optical conductivity for undoped graphene is independent of any materials properties, $\sigma_0=e^2/(4\hbar)$ \cite{Kuzmenko2008}. Nevertheless, even at room temperature, it is found experimentally that  the optical absorption is very small $\sim 2.3\%$ and it depends only on the fine structure constant, $\alpha=1/137$ \cite{Nair2008}. Doping and gating influence the optical properties significantly leading to Pauli blocking for photons with energy less than $2E_F$ ($E_F$ is the Fermi level) and achieving carrier concentrations which can modify the transmission in the visible spectrum \cite{Li2008, Liu2011}. Graphene plasmonics is also quite interesting since graphene plasmons are tunable by gating and doping and they are temperature dependent \cite{Gangadharaiah2008}. In addition to longitudinal plasmons, graphene can support a transverse plasmon mode. It is also interesting to note that the longitudinal modes are gapless, however the transverse ones exist in the window $1.7<\hbar\omega/E_F<2$ and can be tunable from radio to infrared frequency by doping and electric fields \cite{Mikhailov2007}.

Plasmons in GNRs can exist in the near-infrared to far-infrared range and further tunability via a gate voltage can be achieved \cite{Freitag2013}. The optical properties of CNTs are also quite unique. The CNT optical activity, such as electron-energy-loss spectroscopy (EELS) spectra and circular dichroism, are chirality dependent \cite{Wang2005, Dresselhaus2007}. Competing effects due to Coulomb interactions and an attractive e-h coupling are strong in CNTs, which has led to the realization that strong excitonic effects need to be taken into account (especially in small diameter nanotubes) to achieve agreement with experimental optical data \cite{Spataru2001, Spataru2004, Spataru2005}.

The optical response of graphene can be described by considering its 2D conductivity tensor calculated within the Kubo formalism \cite{Falkovsky2007}. Evaluating this general expression for the lowest conduction and highest valence energy bands in the ${\bf q}\rightarrow 0$ approximation leads to the intraband (intra) and interband (inter) contributions 
\begin{eqnarray}
\label{gra-response-intra}
\sigma_{intra}(i\omega)&=& \frac{e^2\ln 2}{\hbar^2\pi\beta\omega}+\frac{e^2}{\hbar^2\pi\beta\omega}\ln(\cosh(\beta\Delta)+\cosh(\mu\beta))-\frac{e^2\Delta^2}{\pi\hbar^2\omega}\int_{\Delta}^\infty\frac{dE}{E^2}\frac{\sinh(\beta E)}{\cosh(\mu\beta)+\cosh(\beta E)} \\
\sigma_{inter}(i\omega)&=&\frac{e^2\omega}{\pi}\int_\Delta^\infty dE\frac{\sinh(\beta E)}{\cosh(\mu\beta)+\cosh(\beta E)}\frac{1}{(\hbar\omega)^2+4E^2}+\frac{e^2\omega\Delta^2}{\pi}\int_\Delta^\infty\frac{dE}{E^2}\frac{\sinh(\beta E)}{\cosh(\mu\beta)+\cosh(\beta E)}\frac{1}{(\hbar\omega)^2+4E^2} ,
 \label{gra-response-inter}
\end{eqnarray}
where $\Delta$ is an energy gap in the graphene spectrum. When $\Delta=\mu=0$ and $k_B T \ll \hbar\omega$, $\sigma$ acquires the universal value $\sigma_0=e^2/4\hbar$, also confirmed experimentally \cite{Nair2008, Li2008}. The graphene conductivity is isotropic when spatial dispersion is not taken into account and the difference between $\sigma_{xx}$ and $\sigma_{yy}$ (graphene is in $xy$-plane) is mostly pronounced for larger $q$ \cite{Falkovsky2007, Drosdoff2012}.

The optical response properties can also be characterized by considering the longitudinal polarization function $\chi_l(q,i\omega)$, which corresponds to the longitudinal component of the conductivity -  $\sigma(q,i\omega)=\frac{ie^2\omega}{q^2}\chi_l(q,i\omega)$ for $q\rightarrow 0$. 
Alternatively, the transverse electric (TE) and transverse magnetic (TM) excitations can be captured via the polarization tensor ${\bold \Pi}$ calculated by a (2+1) Dirac model \cite{Bordag2009, Fialkovsky2011, Klimchitskaya2014a, Sernelius2015}.  It is found that the longitudinal polarization function is related to the $\Pi_{00}$ component,  $\chi_l=-\frac{1}{4\pi e^2\hbar}\Pi_{00}$, while the transverse polarization function is $\chi_{tr}=-\frac{c^2}{4\pi e^2 \hbar\omega^2}(k^2\Pi_{tr}-q_{l}^2\Pi_{00})$.

The optical response of the quasi-1D structures, such as GNRs and CNTs, follows from the Kubo formalism for graphene. Taking into account the zone-folded wave functions and chirality dependent energies leads to the intra and interband optical conductivity spectra of CNTs \cite{Tasaki1998}. Similarly, incorporating the edge dependent wave functions with the appropriate TB energies results in the intra and interband conductivities of zigzag and armchair GNRs \cite{Brey2006,Sasaki2011}.

\subsection{Casimir interactions and graphene nanostructures}

\subsubsection{Graphene}

The vdW/Casimir interaction  involving 2D graphene can be calculated using the fully retarded expression in Eq. \ref{Lifshitz_ret} with response properties (Eqs. \ref{gra-response-intra}, \ref{gra-response-inter}) corresponding to its low-energy Dirac spectrum. The boundary conditions are contained in the matrices $\mathbb{R}_{1,2}$, whose non-zero diagonal components for a graphene/semi-infinite medium system reflecting the TE  ($ss$) and TM ($pp$)  modes are 
\begin{eqnarray}
R_{1}^{(ss)} &=&-\frac{2\pi\omega\sigma \bar{q}c^2}{1+2\pi\omega\sigma \bar{q}c^2}; R_{1}^{(pp)} =\frac{2\pi\sigma \bar{q}/\omega}{1+2\pi\sigma \bar{q}/\omega}  \\
R_{2}^{(ss)} &=&\frac{\mu(i\omega)\bar{q}-\bar{k}}{\mu(i\omega)\bar{q}+\bar{k}}; R_{2}^{(pp)}=\frac{\epsilon(i\omega)q- \bar{k}}{\epsilon(i\omega)\bar{q}+\bar{k}} ,
\end{eqnarray}
where  $\bar{q}=\sqrt{k^2_{\parallel}+(\omega/c)^2}$ and $\bar{k}=\sqrt{k^2_{\parallel}+\mu(i\omega)\epsilon(i\omega)\omega^2/c^2}$. The dielectric and magnetic response functions for the semi-infinite medium are $\epsilon(i\omega)$ and $\mu(i\omega)$, respectively. The reflection coefficients here are expressed in terms of the graphene conductivity $\sigma$, however, these can be given equivalently via other response characteristics using the relations discussed above.  For a graphene/graphene system, the components of the $\mathbb{R}_{2}$ matrix are replaced by the components of the $\mathbb{R}_{1}$ matrix.  

One of the first studies of Casimir interactions for graphene was reported in \cite{Bordag2006}, where the authors considered graphene/perfect metallic semi-infinite medium and atom/graphene systems. The graphene is modeled as a plasma sheet leading to results strongly dependent on the plasma frequency. A more suitable representation of the graphene sheet was later considered by a series of authors by taking into account the Dirac-like nature of the carriers explicitly. It is obtained that  the Casimir force is quite weak compared to the one for perfect metals and that it is strongly dependent upon the Dirac mass parameter \cite{Bordag2009}. Describing the graphene response via the 2D universal graphene conductivity $\sigma_0$ valid in the $k_BT \ll \hbar\omega$ limit \cite{Falkovsky2007,Nair2008},  other authors \cite{Drosdoff2009} find a unique form of the graphene/graphene Casimir force per unit area $A$, $\frac{F_0}{A}=-\frac{3\hbar \sigma_0}{8\pi d^4}= -\frac{3e^2}{32\pi d^4}$. 
This result shows that the distance dependence is the same as the one for perfect metals whose Casimir force is $\frac{F_m}{A}=-\frac{\hbar c\pi^2}{240d^4}$, however the magnitude is much reduced, $F_0/F_m\sim 0.00538$. It is  interesting to note that retardation does not affect the interaction (no speed of light $c$) and $\hbar$ is canceled after taking into account that $\sigma_0=e^2/4\hbar$. 

The graphene interaction has also been investigated via the non-retarded Lifshitz formalism in Eq. \ref{lifshitz_nonret} \cite{Dobson-PRL,Gomez-Santos2009,Sernelius2015,Sarabadani2011}, where the polarization and Coulomb interaction are calculated with the RPA approach. The RPA is a useful tool to study long-ranged dispersive interactions as it gives a natural way to take into account the electron correlation effects of each object and spatial dispersion \cite{Fetter1971,Dobson2011}, as discussed earlier. It has been found that for separations $d>50$ $nm$ the non-retarded Lifshitz approach results in a graphene/graphene force of the form $\frac{F_0}{A}= -\frac{B}{d^4}$, where the magnitude of the constant $B$ agrees with the results from the retarded Casimir calculations  \cite{Drosdoff2009,Drosdoff2012}. 
It is thus concluded that the graphene/graphene interaction is determined by the non-retarded TM mode contribution (captured in the longitudinal polarization) even at distances corresponding to the Casimir regime. These results are truly remarkable since the vdW/Casimir interaction appears to be independent of all of the graphene properties in the low $T$ and/or $d>50$ $nm$ regime. A further interpretation can be given by noting that  the electromagnetic fluctuations exchange occurs at speed $v_F$ (Eq. \ref{gra-Hamiltonian}) rather than the speed of light. This means that the typical thermal wavelength $\lambda_T=\hbar c/k_BT$, which sets the scale where quantum mechanical ($d<\lambda_T$) or thermal ($d>\lambda_T$) fluctuations dominate the interaction, becomes  ${\tilde \lambda}_T=\hbar v_F/k_BT$. The quantum mechanical contributions determine the graphene interaction at separations $d<{\tilde \lambda}_T\sim 50$ $nm$ as opposed to $d<\lambda_T\sim 7$ $\mu m$ for typical metals and dielectrics at $T\sim 300$ K. The thermal fluctuations for graphene become relevant at much reduced distances, and for $d>{\tilde \lambda}_T$ the interaction is $\frac{F_T}{A}=-\frac{\zeta(3)}{8\pi}\frac{k_BT}{d^3}$ \cite{Gomez-Santos2009} 
where $\zeta(n)$ is the Riemann zeta-function. 
Essentially, $v_F$ takes the role of the speed of light enhancing the importance of the zero Matsubara frequency at much lower $T$ and smaller $d$ as compared to conventional metals and dielectrics.

For closer separations ($d<50$ $nm$), a more complete model for the graphene properties is needed. 
Deviations from the asymptotic behavior at low $T$ are found \cite{Drosdoff2009} by using the graphene optical conductivity taken into account by a Drude-Lorentz model that corresponds to higher frequency range $\pi\rightarrow\pi^*$ and $\sigma\rightarrow\sigma^*$ transitions. Recently, it has been shown that the Casimir interaction in a stack of identical graphene layers exhibits a fractional distance dependence in the energy ($E\sim d^{-5/2}$) as a result of the Lorentz oscillators \cite{Nail2015}. Other researchers \cite{Lebegue-2010,Gould2009,Gould2013,Gould2008,Dobson2013} have utilized the RPA approach combined with first-principles calculations for the electronic structure to investigate the non-retarded interaction at very short separations ($d<10$ $nm$) for infinite number of parallel graphene layers.  Interestingly, the interaction energy is found to be $E\sim d^{-4}$. This insulator-like behavior is attributed to the full energy band structure (beyond the two-band model in Eq. \ref{gra-Hamiltonian}) and the associated higher transitions in the response properties. Other authors \cite{Sarabadani2011} have also considered the vdW interaction in a multi-layered graphene configuration within the RPA, however, the reported unusual asymptotic distance dependences may be an artifact of the considered finite graphene thickness.

\begin{figure}[h]
\includegraphics[scale=0.55]{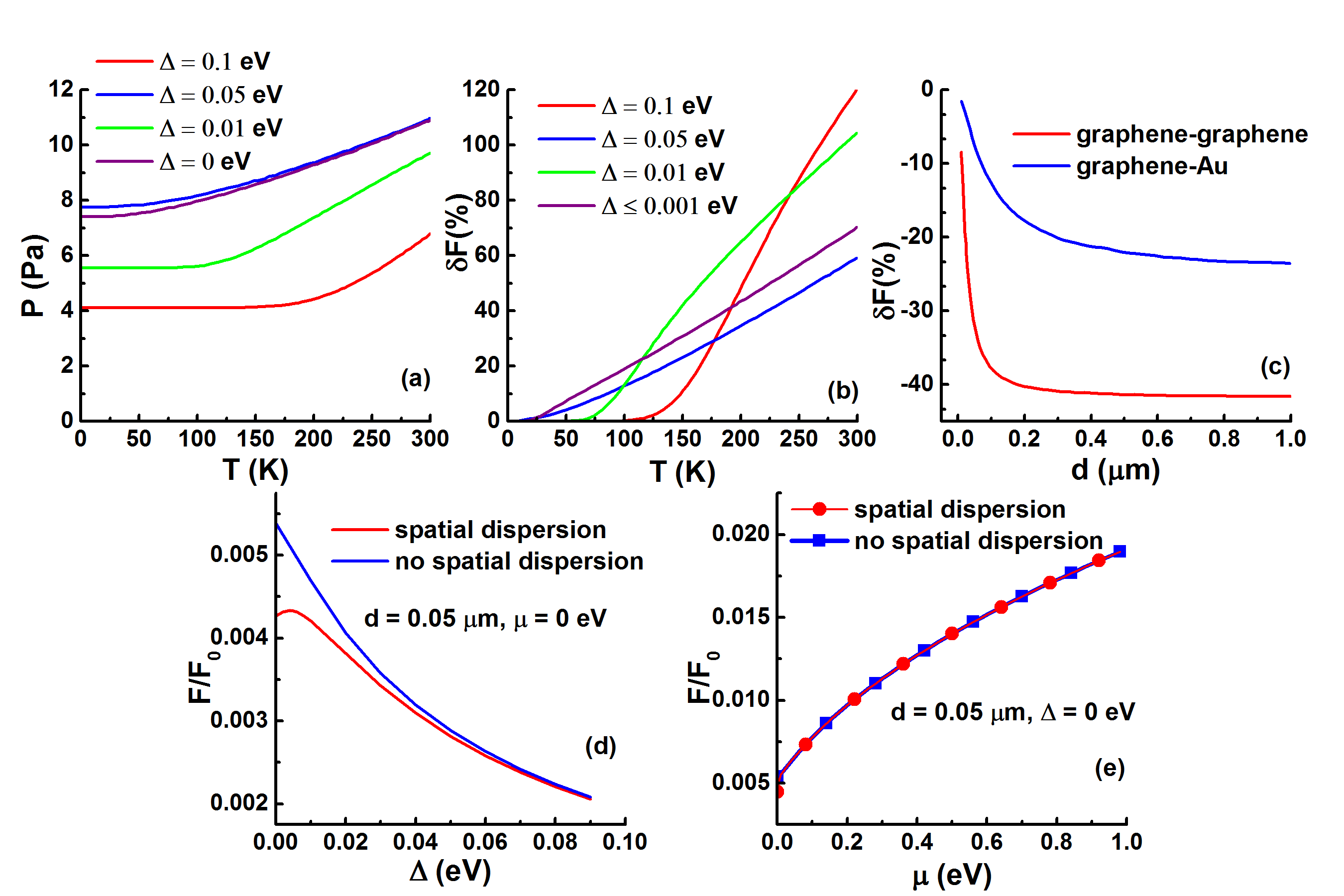}
  \caption{(Color Online) (a) Casimir pressure between two graphene sheets at separation $d=30$ $nm$ as a function of temperature for different values of the gap $\Delta$ (Figure adapted from \cite{Klimchitskaya2013}); (b) Relative thermal correction $\delta F(\%)=(F(T)-F(T=0))/F(T=0)$ for the graphene-Si plate interaction at separation $d=100$ $nm$ (Figure adapted from \cite{Bordag2012}). (c) Relative deviation $\delta F(\%)=(F_{dd}(T)-F_{pt}(T))/F_{pt}(T)$ for graphene-graphene (red) and graphene-Au plate (blue) interactions, where $F_{dd}(T)$ is the Casimir force calculated via the density-density correlation function and $F_{pt}(T)$ is the Casimir force calculated via the polarization tensor (Figure adapted from \cite{Klimchitskaya2014a}). Casimir graphene-graphene force normalized to $F_0=-3e^2/(32\pi d^4)$ with and without spatial dispersion in the graphene conductivity at $T=0$ $K$ as a function of: (d) the gap $\Delta$ and (e) the chemical potential $\mu$ (Figures from \cite{Drosdoff2012}).
  }
   \label{fig:gra-results}
\end{figure}

It has also been shown that temperature together with other factors, such as doping or external fields, affect the graphene thermal and quantum mechanical regimes in an intricate way. In particular, the classical Casimir/vdW interaction determined by the thermal fluctuations  has been examined by several authors in different situations. \cite{Fialkovsky2011} have used the polarization tensor and corresponding reflection coefficients to express the dominating thermal fluctuations regime in terms of the fine structure constant $\alpha$ as $\frac{k_BTd}{\hbar c}\gg \frac{\alpha \ln\alpha^{-1}}{2\zeta(3)}$.  \cite{Sernelius2011} has utilized the longitudinal graphene response in Eq. \ref{lifshitz_nonret} to show that doping plays an important role in the interaction at larger separations, as the force can be increased by an order of magnitude for large degrees of doping. \cite{Bordag2012,Klimchitskaya2013}  have used the fully relativistic Dirac model with the $T$-dependent polarization tensor to investigate how a finite mass gap $\Delta$ in the Dirac model affects these regimes in graphene/graphene and graphene/dielectrics. It is found that for $k_BT \ll \Delta$, thermal fluctuations are not important, while for $\Delta\leq k_BT$ the thermal effects become significant, as shown on Fig. \ref{fig:gra-results}(a, b). The thermal and quantum mechanical regimes have also been studied by \cite{Drosdoff2012} via the longitudinal thermal conductivity, which includes spatial dispersion, an energy gap and chemical potential in the Dirac model.
These authors have shown that tuning $\Delta$ and $\mu$ can be effective ways to modulate the interaction, however, the spatial dispersion does not play a significant role except for the case of small $\Delta$ and low $T$, as shown in 
Fig. \ref{fig:gra-results}(d, e). 

Recent studies \cite{Klimchitskaya2014a, Klimchitskaya2014b, Klimchitskaya2014c, Sernelius2015} provide a thorough analysis of the balance between the thermal and quantum mechanical effects in the graphene Casimir interaction. It is shown that equivalent representations within the temperature dependent longitudinal and transverse polarization and the temperature dependent polarization tensor are possible. The comparison between results from the temperature dependent polarization tensor and density-density correlation function show that at low $T$, both approaches give practically the same results proving  that retardation and TE polarization are unimportant. For $T\neq 0$, deviations are found as shown in Fig. \ref{fig:gra-results}(c) for graphene/graphene and graphene/metal configurations. 

A paper by \cite{Dobson-PRX} reveals that the collective excitations beyond the RPA approximation may be quite important, qualitatively and quantitatively, for the graphene non-retarded vdW interaction. In general, it is assumed that  higher vertex corrections may change the magnitude of the force somewhat, but not the asymptotic distance dependence. However, this may not be the case for graphene as the type of renormalization yields very different results. The renormalization-group method \cite{Kotov2012,Sodemann2012} results in a weak correction to the interaction energy as opposed to the two-loop level in the large-$N$ limit approach \cite{DasSarma2007}, where the characteristic distance dependence has a different power law \cite{Dobson-PRX}. These findings indicate that graphene may be the first type of material for which RPA is not enough to capture the vdW force in the quantum limit. Along the same lines, \cite{Sharma2014} have shown that for strained graphenes, where electron-electron correlations beyond RPA are much more pronounced, corrections to the vdW interaction, consistent with the renormalization-group model, are found.

Besides the fundamental questions regarding basic properties of graphene Casimir/vdW interactions, other and more exotic applications of this phenomenon have been proposed. For example, \cite{Phan2012} have proposed that a graphene flake suspended in a fluid, such as Teflon or bromobenzene, can serve as a prototype system for measuring thermal effects in Casimir interactions. The balance of gravity, buoyancy, and the Casimir force on the flake creates a harmonic-like potential, which causes the flake to be trapped. By measuring changes in the temperature dependent frequency of oscillations, one can potentially relate these changes to the Casimir interaction. Alternative ways to tailor the graphene Casimir interaction have also been recognized. For example, \cite{Svetovoy2011} have shown that the thermal effects can be enhanced or inhibited if one considers the force between graphene and different substrates. \cite{Sernelius2012} finds that retardation due to the finite speed of light can also be made prominent depending on the type of substrates graphene interacts with.  \cite{Drosdoff2011} have proposed that metamaterials with magnetically active components can result in a repulsive Casimir force.  \cite{Phan2013} have shown a regime where repulsion can be achieved with a lipid membrane. Also, Dirac carriers with constant optical conductivity result in unusual Casimir effects behavior in nonplanar objects. For example, the interaction on a spherical shell with $\sigma=$const has markedly different asymptotic behavior and sign when compared to the one for a plasma shell or for planar sheets with $\sigma=$const \cite{Nail2014, Nail2008}.

The Casimir-Polder force involving atoms and graphene sheets has also been of interest. Such studies are relevant not only fundamentally, but also for other phenomena, including trapping or coherently manipulating ultra-cold atoms by laser light \cite{Ito1996, Lukin2009, Kimble2012}. The theoretical description follows from Eq. \ref{Lifshitz_ret} by considering one of the substrates as a rarefied dielectric \cite{Lifshitz1980,Dzyalosh,milonni}.  In addition to atom/graphene \cite{Judd2011}, configurations containing additional substrates have been studied \cite{Mostepanenko2012}.  
Authors have suggested that it may be possible to observe quantum reflection of He and Na atoms via Casimir-Polder interaction as means to discriminate between the Dirac and hydrodynamic model description for graphene \cite{Churkin2012}. Casimir-Polder shifts of anisotropic atoms near multi-layered graphene sheets in the presence of a Huttner-Barnett dielectric (a linearly polarizable medium, which is modeled by microscopic harmonic fields),  
have also been calculated \cite{Eberlein2012}. Thermal fluctuation effects in atom/graphene configurations can also be much stronger  due to the reduced thermal wavelength ${\tilde \lambda}_T$. Thermal Casimir-Polder effects become apparent for $d>50$ $nm$ at room temperature as the interaction is essentially due to the zero Matsubara frequency giving rise to $F_{T}=-\frac{3k_BT\alpha(0)}{4d^4}$ \cite{Drosdoff2012,Mostepanenko2012,Klimchitskaya2014c,Kaur2014}. Interesting possibilities for temporal changes in the atomic spectrum affecting the graphene sheet by creating ripples have also been suggested \cite{Ribeiro2013}. The Casimir-Polder potential has further been explored possibilities for shielding vacuum fluctuations in the framework of the Dirac model \cite{Ribeiro2013a}.  

\subsubsection{Quasi-1D graphene nanostructures}

Investigating atom/CNT interactions is of utmost importance for applications, such as trapping cold atoms near surfaces \cite{Goodsell2010,Petrov2009}, manipulating atoms near surfaces for quantum information processing \cite{Schmiedmayer2002}, and hydrogen storage \cite{Dillon1997}. CNT/CNT interactions are relevant for the stability and growth processes of nanotube composites \cite{Charlier2007}. To calculate the interaction, one must take into account the cylindrical boundary conditions.  Researchers have utilized scattering techniques to study the distance dependence involving metallic wires with Dirichlet, Neumann, and perfect metal boundary conditions \cite{emig06, Noruzifar2011}. Inclined metallic wires have also been considered \cite{Noruzifar2012, Dobson2009}. Calculations for CNT interactions, however, are challenging as one  has to take into account simultaneously the chirality dependent response properties and the cylindrical boundary conditions for the electromagnetic fields. 

The Lifshitz approach has been applied to CNTs via the proximity force approximation, which is typically appropriate at sufficiently close separations \cite{Blocki1977}. The cylindrical surface is represented by an infinite number of plane strips of infinitesimal width, which are then summed up to recover the CNT surface. This method has been applied to atom/single-walled nanotube and atom/multiwall nanotube treated as a cylindrical shell of finite thickness \cite{Blagov2005,Bordag2006,Blagov2007,Klimchitskaya2008,Churkin2011}.
In these studies, the dielectric response of the nanotubes is not chirality dependent. \cite{Blagov2005} uses extrapolated dielectric function for graphite, \cite{Bordag2006} uses the response to be due to a surface density of the $\pi$-electrons smeared over the surface,  \cite{Blagov2007} utilizes a free-electron gas representation for the cylindrical CNT surfaces, while  \cite{Churkin2011} takes the graphene Dirac and hydrodynamic models. In these works, the interaction energy is always of the form $E=-C_3(d)/d^3$, where the coefficient $C_3(d)$ is also dependent on the cylindrical curvature and atomic polarizability.

\begin{figure}[h]
\centerline{\includegraphics[width=8.5cm]{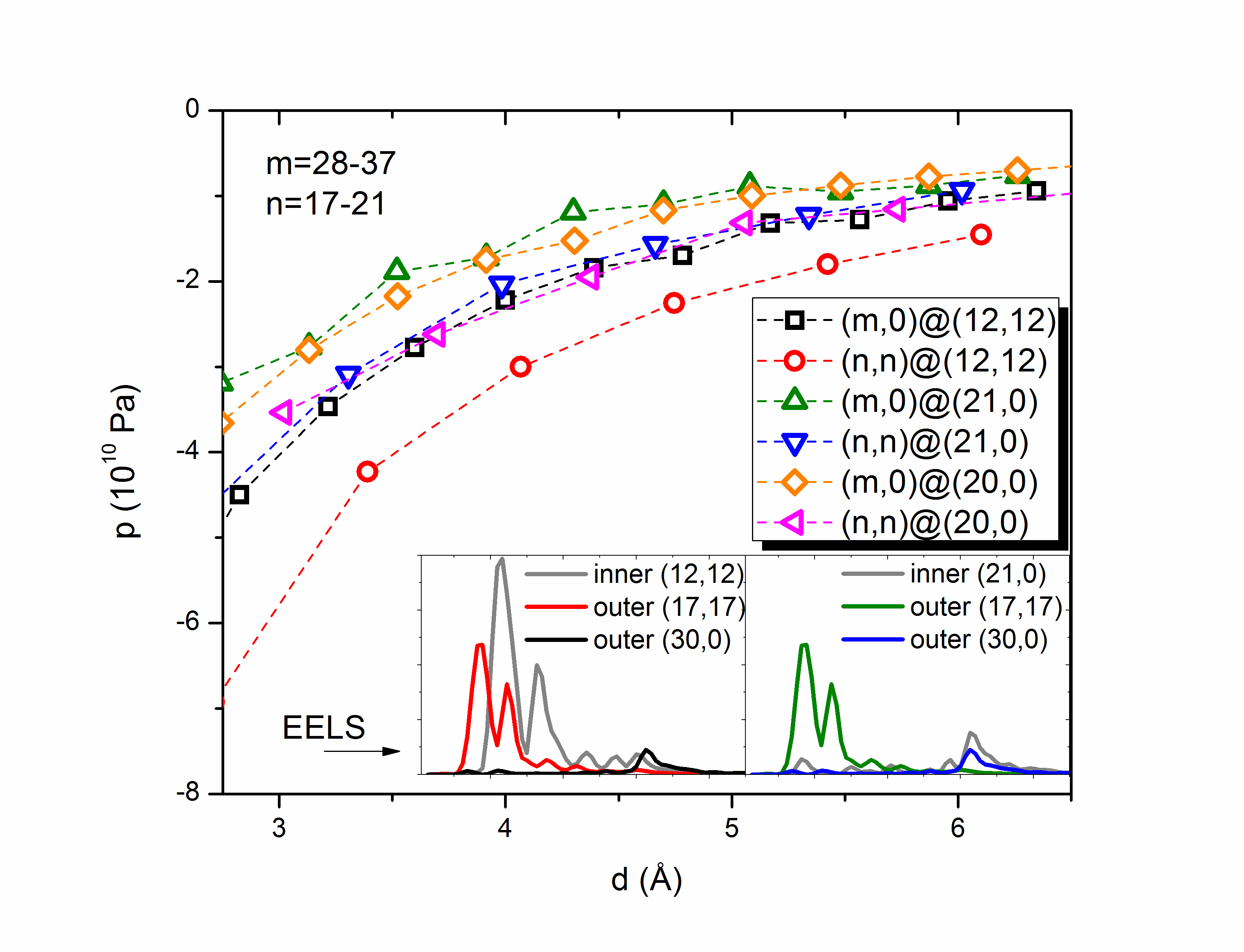}}
\caption{(Color Online) Electromagnetic pressure on each nanotube in a double-wall CNT system as a function of separation. The inset shows calculated EELS for several armchair $(n,n)$ and zigzag $(n,0)$ nanotubes as a function of frequency in eV. The attraction is strongest between two concentric armchair nanotubes due to the presence of strong overlapping low frequency peaks in the spectra. The notation $(m,0)@(n,n)$ corresponds to $(m,0)$ as the inner tube and $(n,n)$ as the outer tube. }
\label{fig:QED-CNT}
\end{figure}

Interactions between nanotubes in a double-wall configuration have been calculated using the QED approach suitable for dispersive and absorbing media as well  \cite{Buhmann07}. Within this formalism  the boundary conditions are taken into account by solving the Fourier domain operator Maxwell equations using a dyadic Green's function, which also allows the inclusion of the chirality dependent response properties of the individual nanotubes. The calculations utilize the fluctuation-dissipation theorem and the force per unit area is the electromagnetic pressure on each surface expressed in terms of the Maxwell stress tensor  \cite{Tomas02}. For planar systems, the QED and the Casimir/Lifshitz theory lead to the same expression (Eq. \ref{Lifshitz_ret}), which has also been shown for systems involving graphene \cite{Hanson2008,Drosdoff2009}. The QED method, applied to the interaction in various double-walled CNTs, revealed  that the chirality dependent low-energy surface-plasmon excitations play a decisive role in the interaction \cite{Popescu2011,Woods2011}. The attractive force is actually dominated by low-energy inter-band plasmon excitations of both nanotubes. The key feature for the strongest attraction is for the CNTs to have overlapping strong plasmon peaks in the electron energy localization spectra (EELS). This is true for concentric $(n,n)$ armchair CNTs, which exhibit the strongest interaction as compared to tubes with comparable radii, but having other chiralities, as shown in 
Fig. \ref{fig:QED-CNT}. 
The results are consistent with electron diffraction measurements showing that the most probable double-walled CNT is the one in which both tubes are of armchair type \cite{Harihara2006}. 
This indicates that the mutual Casimir interaction influenced by the collective excitations may be a potential reason for this preferential formation.

The Casimir-Polder interaction involving CNTs has also been considered via the QED formalism.  For this purpose, one utilizes a generalized atomic polarizability tensor containing dipolar and multipolar contributions and a scattering Green's function tensor expressed in cylindrical wave functions \cite{Buhmann2004,Tai1994,Li2000}. It has been shown that the chirality dependent CNT dielectric function plays an important role determining the strength of the atom/nanotube coupling \cite{Fernari2007}. The QED approach has also been used initially by \cite{Bondarev04,Bondarev05}, where the nonretarded interaction potential is equivalently given in terms of photonic density of states. These studies also show that the interaction is sensitive to the CNT chirality \cite{Rajter1,Rajter2}. It was found that the stronger optical absorption by the metallic CNTs suppresses the vdW atomic attraction, which can be of importance to tailor atomic spontaneous decay near CNT surfaces.

For GNRs the situation is even more technically difficult as compared to nanotubes since analytical results for the boundary conditions for strip-like systems are not available. Nevertheless, a perturbation series expansion of the Lifshitz formula in the dilute limit separates the geometrical and dielectric response properties contributions into convenient factor terms, which can be quite useful to study the finite extensions of the nanoribbons in their vdW interaction \cite{Stedman2014}.   A recent study has also shown that a non-retarded Lifshitz-like formula for the vdW interaction between parallel quasi-1D systems having width $W \ll d$ can also be derived  \cite{Drosdoff2014}. The force per unit length is written in terms of a TM-like "reflection coefficient" containing the GNR response properties, which makes the expression reminiscent of the Lifshitz vdW expression for planar objects in Eq. \ref{Lifshitz_ret}. This is quite appealing as it presents a general way to calculate vdW interactions in any type of 1D parallel systems. 

Applying this theory to GNRs described by their specific response properties \cite{Brey2006} shows that the chemical potential is crucial in the interplay between quantum mechanical and thermal effects in the interaction. A $\mu$-dependent transition between these two regimes is reported correlating with the onset of intraband transitions \cite{Drosdoff2014}. While GNRs with $\mu=0$ behave like typical dielectric materials with a vdW force $F\sim -1/d^6$, when $\mu\sim E_g$ ($E_g$ is the energy gap in the GNR band structure), the interaction becomes completely thermal with a characteristic behavior $F=-\frac{\pi k_BT}{64d (\ln(d/W))^2}$. For semiconductors, such as GaAs wires however, the thermal fluctuations dominate the interaction completely. This is at complete odds with the dispersive interaction involving 2D or 3D materials, where thermal effects are typically very small \cite{Mostepanenko2009}. It turns out that for GaAs quantum wires the plasma frequency is much reduced as compared to 3D GaAs \cite{DasSarma1996} resulting in the force being dominated by the $n=0$ Matsubara term.

\subsection{Materials with topologically nontrivial phases}

In addition to graphene, there are other materials with Dirac spectra and topological insulators (TIs) have a special place in this class of systems. TIs are a new phase of matter with nontrivial topological invariants in the bulk electronic wave function space. The topological invariants are quantities that do not change under continuous deformation, and they lead to a bulk insulating behavior and gapless surface Dirac states in the band structure \cite{Review-TI1,Review_TI2,Review_TI3,Review_TI4,Top_FT_of_TR_invariant_insulators}. The modern history of TIs had started with the realization that a strong spin-orbit coupling can result in  a TI phase with several materials being proposed as possible candidates, including $\mathrm{Bi_{x}Sb_{1-x}}$, $\mathrm{Bi_{2}Se_{3}}$, $\mathrm{Bi_{2}Te_{3}}$, $\mathrm{TlBiSe_{2}}$ among others~\cite{Experimental_TI2,Review_TI4}. In the low momentum limit the 2D states, which are topologically protected by symmetry invariance, are described by a helical version of the massless Dirac Hamiltonian \cite{First_principles_TIs,Model_TI_from_first_principles}
\begin{equation}\label{Helical_Dirac_Surface_States}
H_{surf} = {\bf \hat{z}}\cdot\left({\bf \sigma}\times {\bf k}\right) ,
\end{equation}
where $\bf \hat{z}$ is the unit vector perpendicular to the surface (located in the $xy$-plane), ${\bf \sigma}=(\sigma_x, \sigma_y, \sigma_z)$ are the  Pauli matrices and $\bf k$ is the 3D wave vector.

Topologically nontrivial materials can be classified via their symmetries and dimensions~\cite{Review_TI2,PhysRevB.78.195125,Ryu2010175,Periodic_Table_TI_and_SC} or by dimensional reduction~\cite{Top_FT_of_TR_invariant_insulators}. Three-dimensional systems are characterized by time reversal (TR) symmetry leading to each eigenstate of the above Hamiltonian being accompanied by its TR conjugate or Kramers partner \cite{PhysRevB.78.195125}. Experimentally, however, one observes an odd number of Dirac states. This is understood by realizing that the Dirac cones of the Kramer's pairs appear on each side of the surface of the material and the cone in empty space cannot be detected.  In addition, these surface states are  protected from backscattering by the TR symmetry, which makes them insensitive to spin-independent scattering - a useful feature for quantum computation applications \cite{Leek2007}. Chern Insulators (CIs) are essentially two-dimensional TIs and their low-energy  band structure, also  described by Eq.~\eqref{Helical_Dirac_Surface_States}, consists of even number of helical edge states. CIs have strong enough interband exchange energy, responsible for the hybridization of the surface states from the Dirac cone doublets.  CIs states are further described by a topological integer Chern number  $C\in \mathbb{Z}$ quantified as $C = \frac{1}{2}\sum_{i=1}^{N} \text{sign}(\Delta_{i})$, where  $N$ denotes the (even) number of Dirac cones and $\Delta_{i}$ is the mass gap of each Dirac cone. The mass gap can be tailored by an applied magnetic field or other means and it can be positive or negative.

The properties affecting the vdW/Casimir interactions in systems with topologically non-trivial phases are linked to the dimensionality and response characteristics of the involved Dirac materials. The optical conductivity components of TIs and CIs involving the low-energy Dirac carriers have been obtained within the standard Kubo approach or the quantum kinetics equation method \cite{Pablo_Casimir_CIs,Wang-Kong_Kerr_Faraday_effects_thin_TI}. Analytical representations for the longitudinal surface optical conductivity in the small temperature regime  $k_B T  \ll {\rm min}(|\mu_F|, |\Delta|)$ (here $\mu_F$ is the Fermi energy relative to the Dirac point) and small disorder have been found \cite{Wang-Kong_Kerr_Faraday_effects_thin_TI,Wang-Kong_Kerr_Faraday_effects_thin_TI2,Grushin_CI_model,TIs_finite_surface_gap1,TIs_finite_surface_gap2} with expressions similar to the ones for graphene (Eqs. \ref{gra-response-intra}, \ref{gra-response-inter}). In addition, the topologically protected surface states lead to a strong quantum Hall effect (QHE) without an external magnetic field whenever perturbations breaking the TR symmetry induce a gap in the band structure. For the low-energy carriers in Eq. \ref{Helical_Dirac_Surface_States} the associated surface  Hall conductivity has the following expression at imaginary frequency
\begin{equation}
\sigma_{xy}(i\omega) = - \frac{\alpha c \Delta}{2\pi\hbar\omega}\left[ \tan^{-1}\left(\frac{\hbar\omega}{2\epsilon_{c}}\right) - \tan^{-1}\left(\frac{\hbar\omega}{2|\Delta|}\right) \right].
\end{equation}
Here $\epsilon_c$ is the energy cutoff of the Dirac Hamiltonian, which we associate with the separation between the Dirac point and the closest bulk band. 

For the Casimir interaction involving CIs, it is important to note that these materials can exhibit a quantum anomalous Hall effect at zero frequency or in the absence of an external magnetic field. By tuning the mass gap (via doping or changing the magnetization of the involved material), one can eliminate the 2D optical conductivity $\sigma_{xx}(\omega = 0, |\mu_F| < |\Delta|) = 0$, while the Hall conductivity becomes $\sigma_{xy}(\omega = 0, |\mu_{F}| < |\Delta|) = \frac{\alpha c}{4\pi}\text{sign}(\Delta)$. After summing up the contributions from all Dirac cones, one obtains a {\it quantized} Hall conductivity in terms of the Chern number $\sigma_{xy}(\omega=0) = \frac{\alpha c}{2\pi}C$. 

Inducing a mass gap has important consequences for the surface Hall response in 3D TIs, as well. By applying an external magnetic field, it is possible to realize the fractional quantum Hall effect  with a quantized conductivity $\sigma_{xy}(\omega = 0, |\Delta| > |\mu_{F}|) = \frac{\alpha c}{2\pi}\left( \frac{1}{2} + n \right)$, where $n$ is an integer \cite{PhysRevB.65.245420, Review_TI2}. Nevertheless, one also needs to add the bulk dielectric response. Typically, a standard Drude-Lorentz model is sufficient, and authors have shown that specifically for the Casimir interaction a single oscillator for the dielectric function is enough to capture the characteristic behavior  \cite{Grushin_finite_freq_TI,TIs_finite_surface_gap1,Pablo_Casimir_TIs_FiniteT}. Therefore, the bulk response can be considered as $\epsilon(i\omega) = \epsilon_{0} + \frac{\omega_{e}^{2}}{\omega_{R}^{2} + \omega^{2} }$, where $\omega_e$ is the strength of the oscillator and $\omega_R$ is the location of the resonance.

The surface response properties dramatically affect the electrodynamics in topologically nontrivial materials in 3D. In fact, the electrodynamic interaction can be described via generalized Maxwell equations containing a magnetoelectric coupling due to the surface Hall conductivity. Equivalently, this generalized electrodynamics includes an $axion$ field $\theta({\bf r}, t)$ manifested in a Chern-Simmons term in the Lagrangian, $\mathcal{L}_{\theta} = \frac{\alpha\theta({\bf r}, t)}{2\pi^{2}}\textbf{E}\cdot \textbf{B}$, whose role is to preserve the TR symmetry in the Maxwell equations \cite{Magnetic_Monopole_from_axion_em}. While $\theta({\bf r}, t)$ depends on position and time in general, 
for topological insulators, this is a constant field, such that $\theta\neq 0$ in the bulk and $\theta = 0$ in the vacuum above the surface of the material. We further note that the quantization of the Hall effect in 3D TIs is inherited in the axion term according to  $\theta = (2n + 1)\pi$ ~\cite{Top_FT_of_TR_invariant_insulators, Essin2009}. 

The concept of axion electrodynamics was first proposed in high energy physics as a possible means to explain dark matter \cite{Magnetic_Monopole_from_axion_em,Peccei1977}, and  now axion-type of electromagnetic interactions appear in the description of condensed matter materials, such as TIs. The axion field originating from the topologically nontrivial surface states  leads to many new properties, including induced magnetic monopoles, quantized Faraday angle  in multiple integers of the fine structure constant, and large Kerr angle ~\cite{Magnetic_Monopole_from_axion_em, Wang-Kong_Kerr_Faraday_effects_thin_TI, Wang-Kong_Kerr_Faraday_effects_thin_TI2, Qi2009}. The modified electrodynamics due to the Chern-Simmons term with the associated boundary conditions is also of importance to the Casimir interaction, as shown  earlier  from a high energy physics perspective \cite{Bordag_Chern--Simons_surfaces}.

\begin{figure*}
\begin{minipage}{0.49\linewidth}
\includegraphics[angle=0,scale=0.5]{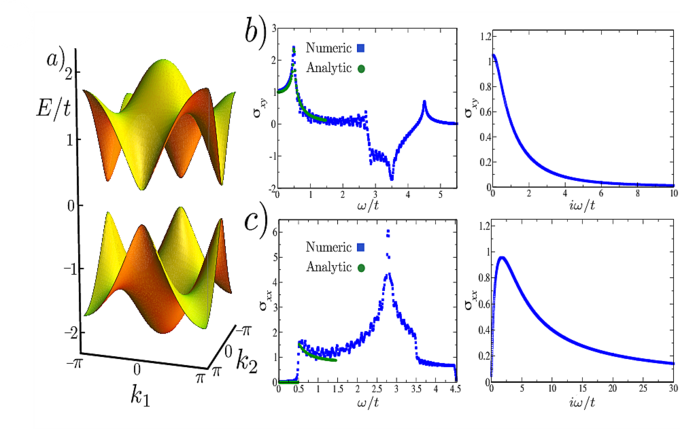}
\end{minipage}
\caption{(Color Online) (a) Energy band structure of a CI calculated via a generic two-band tight binding model \cite{Grushin_CI_model}. The low-momentum limit of the energy band structure is consistent with the Dirac Hamiltonian in Eq. \ref{Helical_Dirac_Surface_States}; (b) Real part of $\sigma_{xy}(\omega)$ (left panel) and $\sigma_{xy}(i\omega)$ (right panel); (c) Real part of $\sigma_{xx}(\omega)$ (left panel) and $\sigma_{xx}(i\omega)$ (right  panel) correspond to the energy band structure from (a). The calculations are performed with 
 $C=1$, $\Delta = 0.25 t$ and $\epsilon = 2.25 t$ ($t$ denotes the hopping integral for the employed lattice model; here it is taken to be equal to the frequency bandwidth). The conductivities are in units of $\alpha c/(2\pi)$. The analytically found $\sigma_{xx}(\omega)$ and $\sigma_{xy}(\omega)$ are in excellent agreement with numerically evaluated Kubo expressions. (Figure taken from \cite{Pablo_Casimir_CIs})}
\label{fig: condmain}
\end{figure*}

\subsection{Possibility of Casimir repulsion in topological materials}

The underlying electronic structure of the materials and their unconventional Hall response, however, open up opportunities to explore the Casimir effect in new directions. Fig. \ref{fig: condmain}(a) depicts the low-energy Dirac band structure for an appropriate lattice model for a CI with the associated longitudinal ($\sigma_{xx}$) and Hall ($\sigma_{xy}$) conductivities \cite{Pablo_Casimir_CIs,Grushin_CI_model}. The reflection matrices in the Lifshitz expression from Eq. \ref{Lifshitz_ret} have been determined for two semi-infinite TI substrates with isotropic surface conductivity $\sigma_{ij}$ and dielectric and magnetic bulk response properties taken as diagonal 3D matrices $\epsilon=\epsilon(\omega)\mathbb{I}$ and $\mu=\mu(\omega)\mathbb{I}$, respectively. Generalizations due to non-local effects and anisotropies in the response ~\cite{Pablo_Casimir_TIs_FiniteT}, as well as finite width substrates ~\cite{T_Matrix_Multilayers}, can also be included. The reflection coefficients for CIs  follow from the ones for the 3D TIs simply by setting $\varepsilon,\mu\to 1$  \cite{Quantized_Casimir_Force,Pablo_Casimir_CIs,Martinez13}. It turns out, however, that in all cases the surface Hall conductivity is a key component in understanding the asymptotic distance dependence, magnitude, and sign of the interaction. 

\begin{figure*}
\begin{minipage}{0.49\linewidth}
\includegraphics[angle=0,scale=0.25]{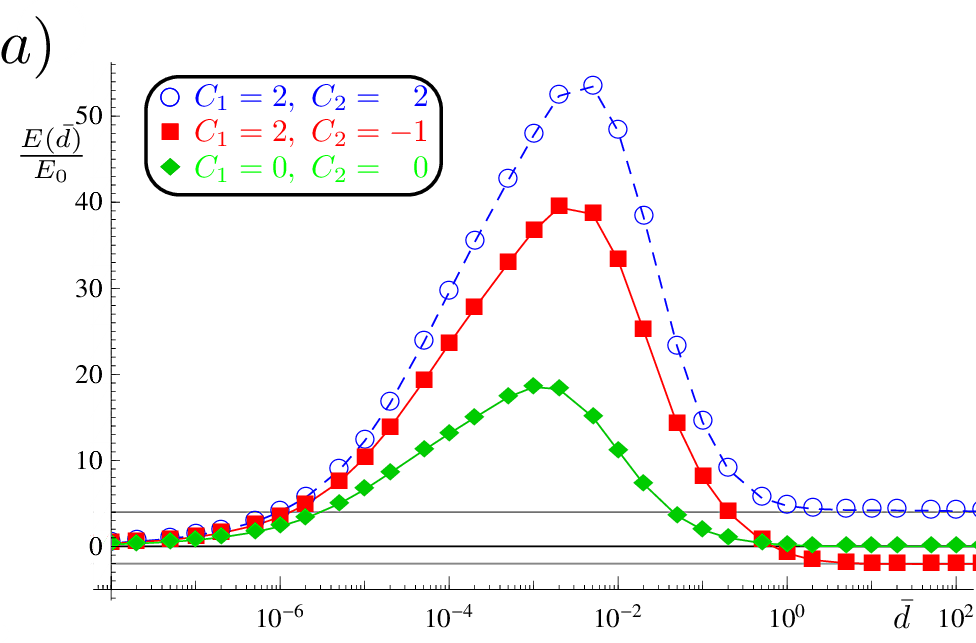}
\end{minipage}
\begin{minipage}{0.3\linewidth}
\includegraphics[angle=0,scale=0.25]{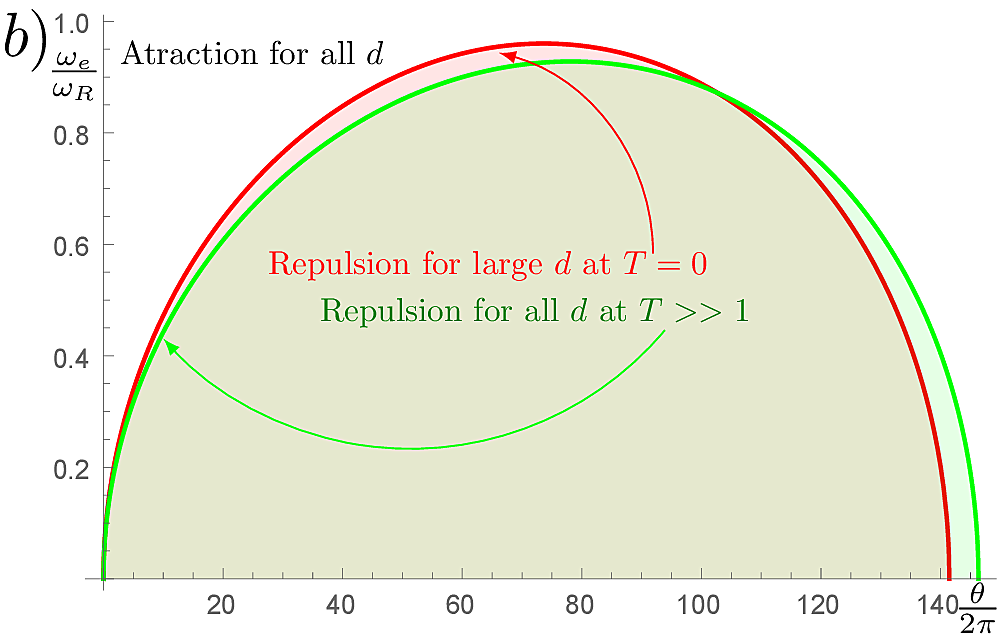}
\end{minipage}
\caption{\label{fig_Repulsion_Maps}
(Color Online) (a) Casimir interaction energy $E(d)$ between two CIs in units of $E_{0}(d)=-\hbar c \alpha^{2}/(8\pi^{2} d^{3})$ as a function of $\bar{d}=td/(\hbar c)$ for different values of $C_{1,2}$ and $|\Delta| = t$.  (Figure adapted from \cite{Pablo_Casimir_CIs}).
(b) A phase diagram $(\frac{\omega_{e}}{\omega_{R}},\frac{\theta}{2\pi})$ for the interaction energy between two semi-infinite TIs substrates for all separation scales. The parameters $\omega_{e}$ and $\omega_{R}$ correspond to the strength and location of the Drude-Lorentz oscillator, respectively. The repulsion for large separations at $T=0$ $K$ is given by the red region (outlined by the red line), while the repulsion for all separations at high $T$ is given by the green region (outlined by the green line). Here, $\frac{\theta}{2\pi}=(n+\frac{1}{2})$ and $\theta_1=-\theta_2=\theta$ for the substrates.  Repulsion is observed for a large range of Chern numbers.}
\end{figure*}

Fig. \ref{fig_Repulsion_Maps}(a) summarizes results for calculated Casimir energies at the quantum mechanical regime (low $T$ and/or large $d$). The graph indicates that there is a change of distance  dependence behavior when comparing the small and large $d$ asymptotics for interacting CIs. The analytical expressions for the conductivity components, which agree very well with the numerical Kubo formalism calculations according to Fig. \ref{fig: condmain}(b,c), are especially useful in better understanding of the underlying physics of the Casimir energy. It has been obtained that  the energy at small $d$ is determined by the longitudinal component of the conductivity and the interaction is always attractive \cite{Pablo_Casimir_CIs}.  For large $d$, however, it is possible to achieve repulsion if the two CIs have Chern numbers $C_1C_2<0$ and the Hall conductivity is much larger than the longitudinal one. The interaction energy in this case is a non-monotonic function of distance and it is quantized according to $E\sim C_1C_2$, which also indicates that  the strongest repulsion occurs for  materials with large Chern numbers. It is concluded that if repulsion is desired, one must search for CI materials with vanishing diagonal conductivity components and strong Hall conductivity capable of sustaining much enhanced $C$ numbers. 
Let us note that a quantized Casimir interaction may be typical for materials that can support a strong Hall effect. In fact, such phenomenon was predicted to exist in graphene/graphene systems at sufficiently large separations with an external magnetic field \cite{Quantized_Casimir_Force}. The associated Landau-level filling factors in the Hall conductivity lead to a quantization condition in the force, and similar findings have been reported for atom/graphene configurations \cite{Farina2014}.

The Casimir interaction between TIs has also been studied, in which case the bulk dielectric response is included  (Sec. III.D). Recent work \cite{Grushin_Casimir_TI} has shown that the energy has unique characteristics due to the balance between the bulk and surface states contributions mediated by the axion term $\theta$. Fig. \ref{fig_Repulsion_Maps}(b) summarizes numerical results for the interaction energy phase diagram showing repulsive and attractive regimes depending on the Drude-Lorentz parameters ($\frac{\omega_e}{\omega_R}$) and the surface contribution ($\theta$). 
Reported analytical calculations enable a better understanding of the important factors determining the interaction in various limits. It is found that if the bulk response is treated as a single Lorentz oscillator, it is possible to obtain Casimir repulsion at larger separations \cite{Pablo_PWS_TIs}, where the surface contributions through the Hall conductivity dominate the response provided $\theta_1\theta_2<0$. For shorter separations the bulk response becomes dominant, and the Casimir interaction is attractive. One also obtains attraction at all distance scales when $\theta_1\theta_2>0$  \cite{Pablo_PWS_TIs}. Other authors  have predicted that repulsion is also possible in the regime of short separations, however, this is considered to be  an artifact of a frequency independent surface conductivity taken in the calculations  \cite{Grushin_Casimir_TI, Pablo_Casimir_TIs_FiniteT, TIs_finite_surface_gap1, Casimir_TI_slabs}. In addition, several recent works have shown that the behavior of the Casimir interaction and the existence of repulsion, in particular, depend strongly on the magnitude of the finite mass gap, the applied external magnetic field, and the thickness of the TI slabs \cite{TIs_finite_surface_gap1,Casimir_TI_slabs}.  The thermal Casimir interaction between TIs has also been studied \cite{Pablo_PWS_TIs,Grushin_Casimir_TI}. The energy corresponding to the $n=0$ Matsubara term depends strongly on the axion fields. The interaction is found to be attractive when $\theta_1\theta_2>0$. However, thermal Casimir repulsion is obtained at all distances  for $\theta_1\theta_2<0$ as shown in Fig.  \ref{fig_Repulsion_Maps}(b). Similar considerations for repulsion may apply for CIs, since their thermal Casimir energy can be obtained analogously to be  
proportional to the surface Hall conductivities $\sigma_{xy}^{(1)}\sigma_{xy}^{(2)}$. These results indicate that topological Dirac materials, such as 2D CIs and 3D TIs, may be good candidates to search for a repulsive thermal Casimir interaction.  

The long-ranged dispersive interactions involving systems with nontrivial topological texture are complex phenomena. 
Materials with Dirac carriers lend themselves as templates where concepts, typically utilized in high energy physics, cross over to condensed matter physics with vdW/Casimir interactions as a connecting link. Topologically nontrivial features in the electronic structure and optical response properties result in unusual asymptotic distance dependences, an enhanced role of thermal fluctuations at all distance scales, and new possibilities of Casimir repulsion. Ongoing work in the area of Dirac materials will certainly continue stimulating further progress in the field of vdW/Casimir physics and further widening the scope of fluctuation-induced phenomena.

\section{Structured Materials}

The geometry of the interacting objects and the interplay with the properties of the materials is also of interest for tailoring the Casimir force. Structured materials, including metamaterials, photonic crystals, and plasmonic nanostructures, allow the engineering of the optical density of states by proper design of their individual components. As a result, one is able to manipulate the interaction utilizing complex, non-planar geometries. Recent experimental studies have begun the exploration of such geometry effects particularly with dielectric and metallic gratings \cite{Chan2008,IntravaiaNC13}. The theoretical description has been challenging due to the complex dependence of dispersive interactions upon non-planar boundary conditions. One approach relies on effective medium approximations, where the emphasis is on models of the dielectric response of the composite medium as a whole. The second approach deals with particular boundary conditions via computational techniques. While the interaction of electromagnetic waves with metallic and dielectric structures of complex shapes is well established in classical photonics, the main challenge stems from the inherently broad band nature of Casimir interactions, where fluctuations at all frequencies and wave-vectors have to be taken into account simultaneously.

\subsection{Metamaterials}

Electromagnetic metamaterials are composites consisting of conductors, semiconductors, and insulators, that resonantly interact with light at designed 
frequencies. The individual components make up an ordered array  with unit cell size much smaller than the wavelength of radiation. As a result, an electromagnetic wave impinging on the material responds to the overall combination of these individual scatterers as if it were an effectively homogeneous system. Metamaterials were speculated almost 50 years ago by Victor Veselago  \cite{Veselago1968}, who was the first to explore materials with negative magnetic permeability in optical ranges. However, it was over twenty years ago that John Pendry proposed the workhorse metamaterials' structure, the split-ring resonator (SRR), that allowed an artificial magnetic response and was a key theoretical step in creating a negative index of refraction \cite{Pendry1999}. David Smith and colleagues were the first to experimentally demonstrate composite metamaterials, using a combination of plasmonic-type metal wires and an SRR array to create a negative effective permittivity 
$\epsilon_{\rm eff}(\omega)$ and a negative effective permeability  $\mu_{\rm eff}(\omega)$ in the microwave regime \cite{Shelby2001}. Many exotic phenomena have been discovered afterwards, including negative index of refraction, reversal of Snell's law, perfect focusing with a flat lens, reversal of the Doppler effect and Cherenkov radiation, electromagnetic cloaking, and transformation optics. 

Such materials are of great interest for Casimir force modifications. Casimir repulsion was predicted by Boyer \cite{Boyer1974} between a perfectly conducting plate and a perfectly permeable one, but it may also occur between real plates as long as one is mainly (or purely) nonmagnetic and the other mainly (or purely) magnetic 
\cite{Kenneth02}. The latter possibility has been considered unphysical \cite{Iannuzzi2003}, since naturally occurring materials do not show strong
magnetic response at near-infrared or optical frequencies, corresponding to gaps $d=0.1-10 \mu$m.  However, recent progress in nanofabrication has resulted in metamaterials with magnetic response in the visible range
of the spectrum \cite{Shalaev2007}, fueling the hope for ``quantum levitation". 

\begin{figure}[h]
\centerline{\includegraphics[width=7.8cm]{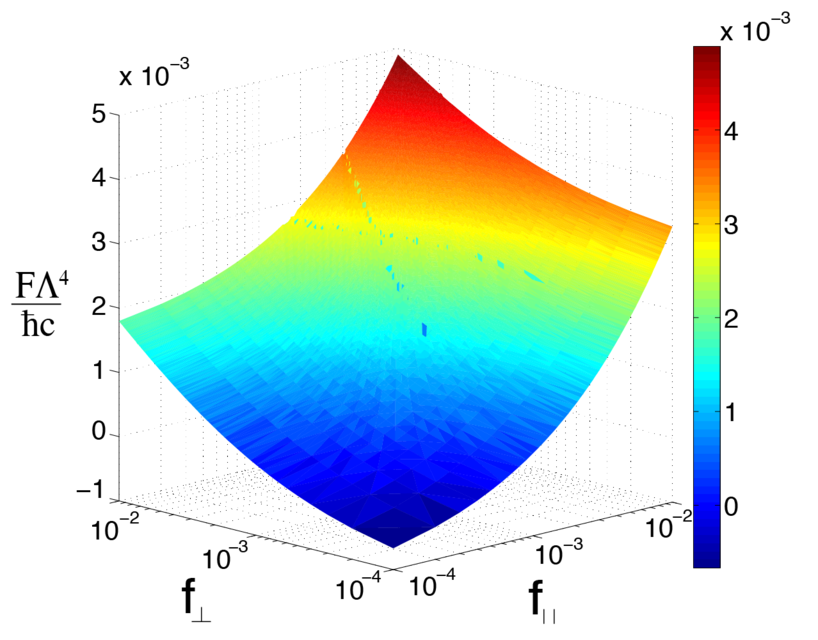}}
\caption{(Color Online) Casimir force per unit area $A$ between a metallic semi-space and an anisotropic metallic-based planar magnetic metamaterial with a weak Drude background. The filling factors $f_{\parallel}$ and $f_{\perp}$, parallel and orthogonal to the vacuum-metamaterial interface, account for the fraction of metallic structure contained in the metamaterial. The SRR Drude parameters are $\Omega_D=2 \pi c/\Lambda=1.37 \times 10^{16}$ rad/sec and $\gamma_D=0.006 \;  \Omega_D$ (corresponding to silver) and
its Drude-Lorentz parameters are 
$\Omega_e/\Omega_D=0.04$, $\Omega_m/\Omega_D=0.1$, $\omega_e/\Omega_D=\omega_m/\Omega_D=0.1$, and $\gamma_e/\Omega_D=\gamma_m/\Omega_D=0.005$. The Drude parameters for the metallic semi-space are $\Omega=0.96 \; \Omega_D$ and $\gamma=0.004 \; \Omega_D$. Temperature is set to zero, and the distance between the bodies is fixed at $d=\Lambda$. Figure taken from \cite{Rosa2008}.
}
\label{fig:anisotropy}
\end{figure}

The Casimir force for structured materials with unit cells much smaller than the wavelength of light can be calculated via Eq. \ref{Lifshitz_ret} for magnetodielectric media with reflection coefficients for two identical substrates ($\mathbb{R}_1=\mathbb{R}_2=\mathbb R$) found as:
\begin{equation}
R^{(ss)}=\frac{\mu_{\rm eff}(i\omega)\bar{q}-\bar{k}}{\mu_{\rm eff}(i\omega)\bar{q}-\bar{k}}, R^{(pp)}=\frac{\epsilon_{\rm eff}(i\omega)\bar{q}-\bar{k}}{\epsilon_{\rm eff}(i\omega_n)\bar{q}-\bar{k}} ,
\end{equation}
where $\bar{q}=\sqrt{k^2_{\parallel}+(\omega/c)^2}$ and $\bar{k}=\sqrt{k^2+\mu_{\rm eff}\epsilon_{\rm eff}\omega^2/c^2}$. Calculations based on this approach have suggested that left-handed metamaterials might lead to repulsion \cite{Henkel2005,Leonhardt2007}. 
Metamaterials, however,
typically have narrow-band magnetic response and are anisotropic. Thus questions naturally arise concerning the validity of such predictions for real systems.
Given that the Lifshitz formula is dominated by low-frequency modes $\omega< c/d$, a repulsive force is in principle possible for a passive left-handed
medium as long $\mu_{\rm eff}(i \omega)$  is sufficiently larger than $\epsilon_{\rm eff}(i \omega)$ in that regime. Then the repulsion is a consequence of the low-frequency response behavior and not of the
fact that the medium happens to be left-handed in a narrow band about some real resonant frequency.
Application of the Lifshitz formalism requires the knowledge of 
$\epsilon_{\rm eff}(i\omega)$ and $\mu_{\rm eff}(i \omega)$ for a large range, up to the order of $\omega=c/d$. Such functions can be evaluated via the Kramers-Kronig relations in terms of 
$\epsilon_{\rm eff}(\omega)$ and $\mu_{\rm eff}(\omega)$ at real frequencies.  The point about the broad band nature of the response properties is very important,
as it shows that knowledge of a metallic-based metamaterial near a resonance is not sufficient
for the computation of Casimir forces: the main contribution to 
$\epsilon_{\rm eff}(i \omega)$ and $\mu_{\rm eff}(i \omega)$ typically comes from
frequencies lower than the resonance frequency. This also implies that repulsive forces, if any,  are in
principle possible without the requirement that the metamaterial resonance should be near the frequency scale
defined by the inverse of the gap of the Casimir cavity.

\begin{figure}[h]
\centerline{\includegraphics[width=7.2cm]{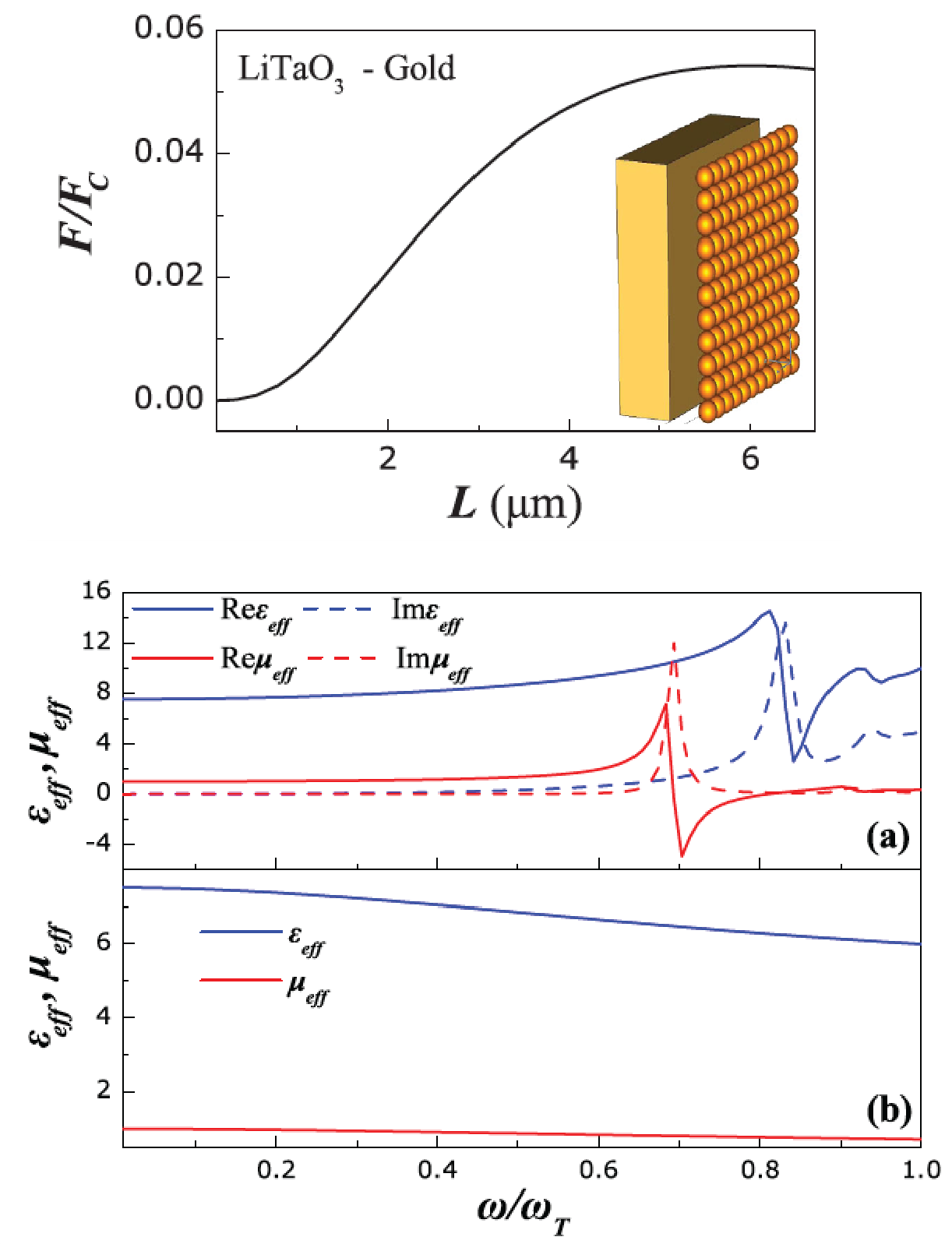}}
\caption{(Color Online) Top: Normalized Casimir force as a function of separation between a gold semi-space and a 2D metasurface made of close-packed (lattice constant $a=11.24 \mu$m) LiTaO$_3$ spheres of radius $r=5.62 \mu$m, calculated via scattering theory. 
The optical response of LiTaO$_3$ is described by a single-resonance Drude-Lorentz model, 
$\epsilon(\omega) = \epsilon_{\infty} (1+\frac{\omega_L^2-\omega_T^2}{\omega_T^2-\omega^2-i \omega \gamma})$, where $\epsilon_{\infty}=13.4$, the transverse and longitudinal optical phonon frequencies are $\omega_T=26.7 \times 10^{12}$ rad/sec and $\omega_L=46.9 \times 10^{12}$ rad/sec, and the dissipation parameter 
is $\gamma=0.94 \times 10^{12}$ rad/sec. The gold Drude parameters are taken as $\hbar \Omega_p = 3.71$ eV and $\Omega_p \gamma_p^{-1}=20$.
Bottom: effective permittivity and permeability of the close-packed LiTaO$_3$ spheres as a function of (a) real and  (b) imaginary frequencies as calculated by the Maxwell-Garnett effective medium theory. Figures taken from \cite{Yannopapas2009}.
}
\label{fig:casimir-magnetic}
\end{figure}

In typical metamaterial structures, the effective electric permittivity $\epsilon_{\rm eff}(\omega)$ and magnetic permeability $\mu_{\rm eff}(\omega)$ close to the metamaterial resonance are well described in terms of a Drude-Lorentz model,
\begin{equation}
\epsilon_{\rm eff}(\omega), \mu_{\rm eff}(\omega) = 1 - \frac{\Omega^2_{e,m}}{\omega^2-\omega^2_{e,m} + i \gamma_{e,m} \omega}
\label{EMA-MM}
\end{equation}
in which $\Omega_e (\Omega_m)$ is the electric (magnetic) oscillator strength, $\omega_e (\omega_m)$ is the metamaterial electric (magnetic) resonance frequency, and
$\gamma_e (\gamma_m)$ is a dissipation parameter. These parameters depend mainly on the sub-wavelength geometry of the unit cell, which can be modeled as a LRC circuit. 
For metamaterials that are partially metallic, such as SRRs (operating
in the GHz-THz range) and fishnet arrays (operating in the near-infrared or optical) away from resonance, it is reasonable
to assume that the dielectric function also has a Drude background $\epsilon_D(\omega)=1-\frac{\Omega^2_D}{\omega (\omega + i \gamma_D)}$ (here $\Omega_D$ is the plasma frequency and $\gamma_D$ is the Drude dissipation rate).  As the Drude background clearly overwhelms 
the resonant contribution at low frequencies, it contributes substantially to the Casimir force between metallic metamaterial structures. Effects of anisotropy, typical in the optical response of 3D metamaterials and of 2D metasurfaces, can also be incorporated 
\cite{Rosa2008}. Fig. \ref{fig:anisotropy}, which depicts the Casimir force between two identical planar 3D uni-axial metamaterials that have only electric anisotropy, shows that the interaction is always attractive.

A key issue here is the realization that it is incorrect to use the above Drude-Lorentz expressions when computing dispersion interactions \cite{Rosa2008a}. Indeed, although Eqs. (\ref{EMA-MM}) are valid close to a metamaterial resonance, they do not hold in a broad-band frequency range. In particular, calculations based on Maxwell's equations in a long wavelength approximation for SRRs result in a slightly different form for the effective magnetic permeability \cite{Pendry1999}
\begin{equation}
\mu_{\rm SRR}(\omega) = 1 - \frac{f \omega^2}{\omega^2 - \omega_m^2 + i \gamma_m \omega} ,
\label{EMA-Correct}
\end{equation}
where the filling factor $f<1$ is a geometry dependent parameter. The crucial difference between Eqs. (\ref{EMA-MM}) and (\ref{EMA-Correct}) is the $\omega^2$ factor in the numerator of the latter, a consequence of Faraday's law \cite{Rosa2009}. Although close to the resonance
both expressions give almost identical behaviors, they differ in the low-frequency limit: $\mu_{\rm eff}(i \omega)>1$ while $\mu_{\rm SRR}(i \omega)<1$. The fact that all passive materials have $\epsilon(i \omega)>1$ implies that Casimir repulsion is impossible for any magnetic metamaterial made of metals and dielectrics  \cite{Rosa2008}. This conclusion is confirmed by scattering theory calculations, that do not rely on any effective medium or homogenization approximations. For example, in
\cite{Yannopapas2009} the Casimir force was computed exactly for 2D metasurfaces  made of a square close-packed array of non-magnetic microspheres of LiTaO$_3$ (an ionic material) or of CuCl (a semiconductor).
Although the systems are magnetically active in the infrared and optical regimes, the
force between finite slabs of these materials and metallic slabs is attractive since the effective electric permittivity at imaginary frequencies is larger than the magnetic permeability. In Fig. \ref{fig:casimir-magnetic} we show the Casimir force (normalized with respect to the ideal zero-temperature Casimir force $F_C$) between a gold plate and a 2D  LiTaO$_3$ metasurface together with the effective permittivity and permeabilities of a close-packed LiTaO$_3$ crystal. The results confirm that the Casimir interaction is attractive in magnetic metamaterials made of non-magnetic meta-atoms. In contrast, intrinsically magnetic meta-atoms could potentially lead to Casimir repulsion. Naturally occurring ferromagnets do not show magnetic response in the infrared and optical regimes, as needed by the Casimir effect, but small magnetic nanoparticles (e.g. few nanometer-sized Ni spheres) become super-paramagnetic in the infrared. A realization of the original idea for Casimir repulsion by Boyer
was then proposed based on a metasurface made of such intrinsically magnetic nano particles \cite{Yannopapas2009}.

Chiral metamaterials made of metallic and dielectric meta-atoms were also proposed as candidates for Casimir repulsion 
\cite{Zhao2009}. When described by an effective medium theory, such systems posses an effective magneto-electric response that modifies the standard constitutive relations in Maxwell's equations as ${\bf D} = \epsilon_0 \epsilon {\bf E} + i \kappa_m  {\bf H}/c$ and ${\bf B} = \mu_0 \mu \epsilon {\bf H} - i \kappa_m  {\bf E}/c$. 
Close to a resonance, the magneto-dielectric coefficient $\kappa_m$ can be modeled as $\kappa_m(\omega)=\frac{\omega_{\kappa_m} \omega}{\omega^2 - \omega^2_{\kappa_m r} + i \gamma_{\kappa_m} \omega}$. 
For such materials the reflection matrix is no longer diagonal and there is polarization mixing. 
Repulsive Casimir forces and stable nanolevitation was predicted for strong chirality (large values of $\omega_{\kappa_m}/\omega_{\kappa_m r}$) \cite{Zhao2009}. However,
these results were shown to be incompatible with the passivity and causal response of the materials \cite{Silveirinha2010b}, which implies that the condition
${\rm Im} [\epsilon(\omega)] {\rm Im} [\mu(\omega)] - ( {\rm Im} [\kappa_m(\omega)] )^2 > 0$ must be satisfied.  
This relation imposes a limit of the strength of the imaginary part of $\kappa_m$, and 
results in an attractive Casimir force between chiral metamaterials made of metallic/dielectric meta-atoms for any physical values of the magneto-electric coupling
\cite{Silveirinha2010}. These theoretical arguments were also confirmed by full-wave simulations of chiral metamaterial structures \cite{McCauley2010a}, and it was shown that microstructure effects (i.e. proximity forces and anisotropy) dominate the Casimir force for separations where chirality was predicted to have a strong influence. Still, chiral metamaterials may offer a way to strongly reduce the Casimir force \cite{Zhao2010}.

\subsection{Photonic crystals}

Photonic crystal are man-made electromagnetic structures that, unlike metamaterials, have unit cell sizes on the order of the wavelength of light. The most important property of photonic crystals made of low-loss dielectric periodic structures occurs when the wavelength is about twice their period \cite{Joannopoulos2008}. Many exotic phenomena are found, including the appearance of photonic band gaps preventing light from propagating in certain directions with specified frequencies,
the localization of electromagnetic modes at defects, and  the existence of surface states that bound light to the surface for modes below the light line. Photonic crystals were co-discovered in 1987 by Eli Yablonovitch, who proved that spontaneous emission is forbidden when a three-dimensional periodic structure has an electromagnetic bandgap which overlaps with an electronic band edge \cite{Yablonovitch1987}, and by Sajeev John, who showed that strong localization of photons could take place in disordered dielectric superlattices \cite{John1987}. The simplest possible photonic crystal, a 1D multilayered stack made of materials of alternating dielectric constants,  had been already investigated more than a century ago by Lord Rayleigh. Today photonic crystals come in different fashions, including complex 3D structures (e.g., the Yablonovite \cite{Yablonovitch1991}), periodic dielectric waveguides, and photonic-crystal slabs and fibers. 

Photonic crystals offer great flexibility in designing atomic traps close to surfaces at sub-micrometer distances allowing the integration of nanophotonics and atomic physics with a host of exciting quantum technologies. Trapping atoms near surfaces is determined by the Casimir-Polder force. However, in analogy to Earnshaw's theorem of electrostatics,  there are no stable Casimir-Polder (or Casimir) equilibria positions for any arrangements of non-magnetic systems,  provided the electric permittivities of all objects are higher or lower than that of the medium in-between them \cite{Rahi10:PRL}. For example, there is no stable equilibrium position for a ground-state atom above a metallic/dielectric structure.\footnote{A corollary of this theorem is that there is no Casimir repulsion for any metallic/dielectric-based metamaterial treated in the effective-medium approximation. Hence, when applied to dispersive interactions, effective medium is a good approach only at separations larger than the unit cell dimensions of the metamaterial. At short distances, displacements of structured Casimir plates might lead to repulsion that, however, must be compatible with the absence of stable equilibria.} Fortunately, no such constraints exist for excited state atoms, or when the trapping potential energy is the superposition of the Casimir-Polder interaction and an external optical trapping field. 

\begin{figure}[h]
\centerline{\includegraphics[width=7.0cm]{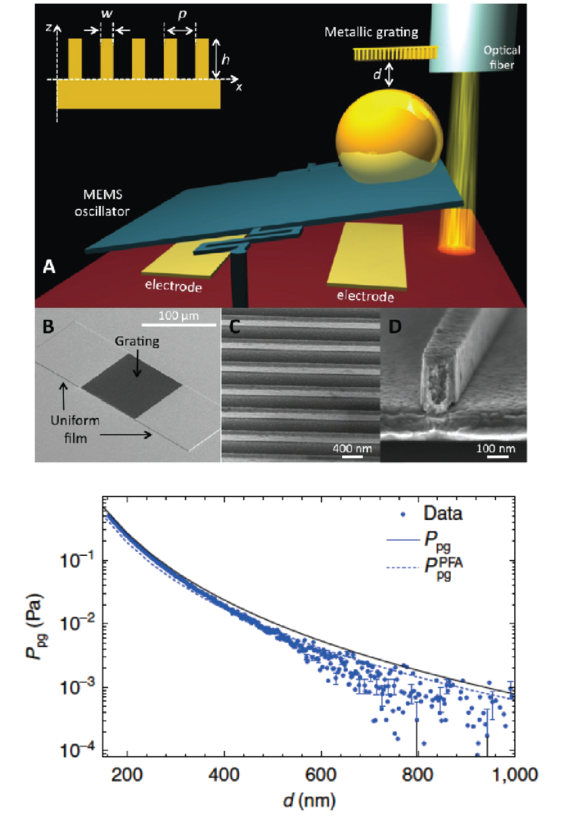}}
\caption{(Color Online) Top: A) Schematic of the experimental configuration used to measure the Casimir force between a gold-coated sphere and a gold nanostructured grating. One of the many nanostructures used in the experiment is shown (SEM images in B-D). The radius of the Au sphere is $R=151.7$ $\mu$m.
Bottom: Plane-grating pressure as a function of sphere-grating separation for a lamellar grating
with period $p=250$ nm, tooth-width $w=90$ nm, and height $h=216$ nm. Experimental measurements of sphere-grating force gradient divided by $2 \pi R$ (dots with error bars), and plane-grating pressure computed with the proximity force approximation (dashed lines) and exactly using scattering theory (solid line). Figure taken from \cite{IntravaiaNC13}.
}
\label{fig:Casimir-nanograting}
\end{figure}

The Casimir-Polder interaction can be calculated for an atom in state $l$ with polarizability $\alpha_l(\omega)$ considering Eq. \ref{Lifshitz_ret} for a rarefield dielectric, as shown in  \cite{Lifshitz1980,Dzyalosh,milonni}. Typically, the arising Green's function is solved via computational FDTD techniques (to be reviewed in next section).  It has been shown that the Casimir-Polder force 
between a ground-state atom and a 1D dielectric grating can trap atoms along the lateral directions of the dielectric surface \cite{Contreras-Reyes2010}. However, there is no trapping along the directions parallel to the grating's grooves. Fully stable traps in 3D can be obtained utilizing photonic crystals, such as 1D periodic dielectric waveguides  \cite{Hung2013}. These proposed structures support a guided mode suitable for atom trapping within a unit cell, as well as a second probe mode with strong atom-photon interactions. The combination of the light-shifts from a laser beam together with the Casimir-Polder force from the dielectric nanostructure results in a fully stable, 3D atomic trap. Aligning the photonic band gap edges with selected atomic transitions substantially enhances the atom-photon interactions, since the electromagnetic density of state diverges due to a van Hove singularity. These ideas have been recently implemented experimentally with a Cs atom trapped within a 1D photonic crystal waveguide consisting of two parallel SiN nanobeams with sinusoidal corrugation \cite{Goban2014}. The measured rate of emission into the guided mode along the 1D waveguide was $\Gamma_{\rm 1D}=0.32 \Gamma'$, where $\Gamma'$ is the decay rate into all other channels. Such a high coupling rate is unprecedented in all current atom-photon interfaces, and paves the way for studying novel quantum transport and many-body phenomena in optics. 
Other works involving atom-surface dispersive interactions in close proximity to photonic crystals include resonant dipole-dipole energy transfer \cite{Bay1997} and enhanced resonant forces \cite{Incardone2014}
between atoms with transition frequencies near the edge of the photonic bandgap, and strong localization of matter waves  mediated by quantum vacuum fluctuations in disordered dielectric media \cite{Moreno2010a}.

\subsection{Plasmonic nanostructures}

Metallic nanostructures can support collective electromagnetic modes, such as  surface plasmons (also known as surface plasmon polaritons), which can propagate along the surface, decay exponentially away from it, and have a characteristic frequency of the order of the plasma frequency \cite{Maier2007}. Surface plasmons affect the Casimir interaction in a non-trivial manner \cite{Intravaia2005}, and this point was also discussed for Dirac materials in Sec. III. When written in terms of real frequencies, the Lifshitz formula, Eq. \ref{Lifshitz_ret}, for planar systems has a term arising from the propagative modes, which gives an attractive force at all distances. There is a second term associated with the evanescent hybrid plasmonic modes, which results in an attractive force at short distances (shorter than the plasma wavelength) and a repulsive one at longer distances. There is a subtle cancellation between the attractive and repulsive terms at large separations, resulting in an always attractive force between planar metallic surfaces for all separations. This observation suggests that metallic nanostructures at scales below the plasma wavelength can potentially enhance the repulsive contribution due to plasmons  and lead to a suppression of the Casimir force. Nanostructured metallic surfaces with tailored plasmonic dispersions have already impacted classical nanophotonics, with applications ranging from extraordinary light transmission \cite{Ebbesen1998} to surface-enhanced Raman scattering \cite{Nie1997}. Metallic structures with strong deviations from the planar geometry and possessing geometrical features on very small scales are also likely to give significant new insights into potential Casimir devices.

In addition to plasmons associated with the metallic nature of the plates, there is another type of plasmonic excitations, the so-called spoof plasmons, that arise from geometry and exist even for perfectly reflecting surfaces. Pendry and co-workers \cite{Pendry2004,Garcia-Vidal2005} proposed engineered dispersion by periodically nanostructuring surfaces by perforating perfect electrical conductors. The resulting surfaces support surface modes that have dispersion similar to real surface plasmons in metals, but with the effective plasma frequency determined by the geometric parameters of the perforation. Spoof plasmons are also present in nanostructures made of real metals,    enhance the modal density of states, and  modify the Casimir interaction in nanostructured metallic cavities \cite{Davids2014}.

Besides computations of the interaction in complex systems, including nanostructured surfaces \cite{emig04_2,Lambrecht09,Davids2010,Intravaia2012a,Guerout2013,Antezza2014}, advances in Casimir force  measurements have also been reported. However, the experimental progress has been limited due to difficulties associated with the reliable fabrication and the measurement of the force. Using an {\it in situ} imprinting technique, whereby the corrugation of a diffraction grating was imprinted onto a metallic sphere by mechanical pressure, the lateral Casimir force between two axis-aligned corrugated surfaces was measured as a function of their phase shift \cite{Chen2002}, and the normal Casimir force between them was also measured as a function of the angle between their corrugation axes \cite{Banishev2013b}. Nanostructured lamellar gratings made of highly-doped Si have been used to measure the Casimir interaction with a metallic sphere
\cite{Chan2008}, with conclusive evidence of the strong geometry dependency and non-additivity of the force. More recently, a strong Casimir force reduction through metallic surface nanostructuring has been reported \cite{IntravaiaNC13}. In Fig. \ref{fig:Casimir-nanograting} the experimental setup is shown, consisting of a plasmonic nanostructure in front of a metallic sphere attached to a MEMS oscillator. A deep metallic lamellar grating with sub-100 nm features strongly suppressed the Casimir force, and for large inter-surface separations reduced it beyond what would be expected by any existing theoretical prediction. Existing state-of-the-art theoretical modeling, based on the proximity force approximation for treating the curvature of the large-radius sphere ($R=151.7 \mu$m, much larger than any geometrical length scale in the system), and an exact {\it ab initio} scattering analysis of the resulting effective plane-grating geometry, did not reproduce the experimental findings. The development of a full numerical analysis of the sphere-grating problem, capable of dealing with the disparate length scales present in the experiment \cite{IntravaiaNC13} with plasmonic nanostructures, remains an open problem.

\begin{figure}[h]
\centerline{\includegraphics[width=6cm]{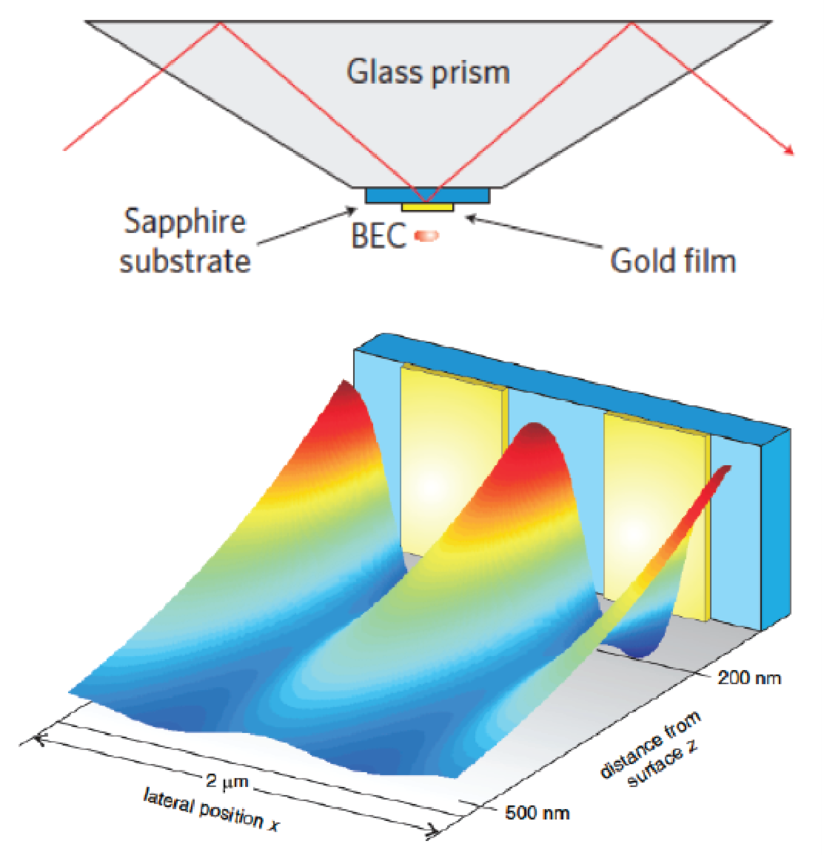}}
\caption{(Color Online) Top: Schematic configuration to measure the Casimir-Polder potential probed by a Bose-Einstein condensate diffracted from a plasmonic nanostructured being excited by an external laser field in the Kretschmann configuration. Bottom: Atom-grating potential landscape arising from the combination of a repulsive evanescent-wave potential and the 
Casimir-Polder attraction. The external laser field has a power of $P=211$ mW. The evanescent field from the grating modulates the repulsion, and the attractive Casimir-Polder potential is the strongest on top of the gold stripes. At a distance of 200 nm, the potential is laterally modulated with an amplitude of $\Delta E/k_B = 14 \mu$K. Figures taken from  \cite{Stehle2011,Bender2014}.
}
\label{fig:Casimir-BEC}
\end{figure}

Nanostructured surfaces have also been used in studies of atom-surface dispersion interactions. Casimir-Polder forces between a single atom or a Bose-Einstein condensate above a grating have been measured using different methods  \cite{Grisenti1999,Oberst2005,Perreault2005,Pasquini2006,Zhao2008}, and the near- and field-field scaling laws of the Casimir-Polder potential were verified. Theoretical proposals have also been put forward to measure the Casimir-Polder potential at corrugated surfaces with Bose-Einstein condensates \cite{Messina2009,Moreno2010}. In a series of recent experiments, ultracold atoms were utilized to survey the potential landscape of plasmonically tailored nanostructures. In \cite{Stehle2011} a Rb Bose-Einstein condensate was accelerated towards Au plasmonic microstructures whose plasmons were excited by external laser fields in a Kretschmann configuration  (Fig. \ref{fig:Casimir-BEC}). A blue-detuned laser beam generates an evanescent optical field that repels the atoms from the surface, 
while the atom-grating Casimir-Polder interaction produces an attractive potential. This combination results in a potential barrier that can be mapped by classical or quantum reflection measurements. Diffraction measurements of Bose-Einstein condensates from metallic nanogratings have allowed to locally probe the Casimir-Polder potential
\cite{Bender2014}, revealing information about its landscape  (Fig. \ref{fig:Casimir-BEC}) in agreement with theoretical calculations based on the scattering approach to atom-grating Casimir interactions.

\section{Non-trivial Boundary Conditions}

Casimir interactions are fundamentally changed in the presence of
macroscopic objects due to the shapes of boundaries and interfaces,
which lead to complex and highly non-additive wave
effects~\cite{Rodriguez11:review,Dalvit:book,BuhmannI:book,ReidRo12:review,Rodriguez14:review}. Understanding
the ways in which non-trivial shapes and boundary
conditions affect the force has not only shed light on various
ways to design forces
 used to combat unwanted Casimir effects in NEMS and MEMS,
but continues to reveal regimes and situations where the often-employed
proximity force approximation (PFA) and pairwise summation (PWS)
approximation fail dramatically. Such structures can also lead to
forces that differ significantly from the attractive, monotonically
decaying force laws associated with planar bodies and/or dilute,
atomic media.

Early studies of Casimir forces focused on simple
geometries, e.g. planar bodies and generalizations thereof, by
employing sum-over-mode formulations where the zero-point energy of
electromagnetic fields (field fluctuations) rather than dipolar
interactions (charge fluctuations) were
summed~\cite{Casimir1948,milonni}. The equivalence of these two
perspectives comes from the
fluctuation--dissipation theorem, relating the properties (amplitude
and correlations) of current fluctuations in bodies to the
thermodynamic and dissipative properties of the underlying
media~\cite{Lifshitz1956,Eckhardt84,Lifshitz1980}. Ultimately, the
connection between current and field fluctuations arises from the
well-known dyadic electromagnetic Green's function~\cite{Jackson98}:
\begin{equation}
G_{ij}({\bf r},{\bf r}';\omega) = \left\{ \left[\nabla \times \nabla \times - \varepsilon({\bf r},\omega) \omega^2\right]^{-1} {\bf \hat{e}}_j \delta({\bf r}-{\bf r}') \right\}_i
\end{equation}
where $\hat{\bf e}_j$ is the unit vector. 
The connection to sum-over-mode formulas arises from the
trace of the Green's function being related to the electromagnetic
density of states $\rho(\omega) = \frac{1}{\pi}
\frac{d(\omega^2\varepsilon)}{d\omega} \mathrm{Tr}\,\mathrm{Im}\,
G_{ij}(\bf{r},\bf{r},\omega)$, which when integrated $\sum_\omega \rho(\omega) =
\int d\omega \rho(\omega)$ leads to the famous $\mathcal{E} =
\sum_\omega \frac{\hbar \omega}{2}$
formula~\cite{vanKampen68,Gerlach71,Rodriguez07:PRA}.  Although this
formulation was originally developed in special geometries involving
perfectly metallic conductors, where Hermiticity leads to well-defined
modes, it has also been extended to handle other situations of
interest such as open structures and lossy
dielectrics~\cite{Genet03:scat,Enk95,Graham09,Mochan06,Davids2010,Milton10,Intravaia12}. Despite
these generalizations, the sum-over-mode approach poses practical
challenges for computations in general structures due to the
cumbersome task of having to compute all of the modes of the
system~\cite{Ford93,Enk95:torque,Rodriguez07:PRA}.

Instead, more powerful applications of the fluctuation--dissipation
theorem exist in which Green's functions are directly employed to
compute energy densities and stress tensors (momentum transport)
rather than modal contributions to the energy, reducing the problem to
a series of classical scattering calculations: scattered fields due to
known incident fields/sources. This latter viewpoint was originally
employed by Lifshitz and others to calculate forces between planar
dielectrics bodies~\cite{Dzyaloshinskii61,Lifshitz1980}, and it turns out
to be much more useful when dealing with complex geometries. The advantage comes from the fact that the
Green's function does not need to be obtained analytically, as was done for
planar bodies, but it can be routinely and efficiently computed
numerically via classical electromagnetism. These ideas lie at the
center of recently developed general-purpose techniques, schematically shown in Fig. \ref{fig:fig1}, for computing
forces in complex structures that boil down to a series of classical
scattering calculations of Green's
functions~\cite{Rodriguez07:PRA,Pasquali08,Pasquali09,RodriguezMc09:PRA,McCauleyRo10:PRA,Xiong09,Xiong10}
or related quantities such as scattering
matrices~\cite{gies03,emig03_2,emig04_2,Gies06:worldline,Rahi09:PRD,Lambrecht06,Emig07,Kenneth08,MiltonPa10,
ReidRo09,Reid11,Reid12:FSC,Atkins13}.

Despite their relative infancy, these methods have already led to a
number of interesting predictions of unusual Casimir forces in a wide
range of structures, including
spheres~\cite{gies06:PFA,Neto08,Emig08:sphpl},
cylinders~\cite{emig06}, cones~\cite{Maghrebi10},
waveguides~\cite{Rodriguez07:PRL,RahiRo07,Zaheer07,Rodriguez-Lopez09,Pernice10},
and patterned
surfaces~\cite{emig03_1,emig04_2,Rodrigues06,RodriguezJo08:PRA,Lambrecht09,Davids2010,Guerout2013},
among
others~\cite{Rodriguez08:PRL,RodriguezMc10:PRL,RodriguezWo10,Broer12}. Here
we provide a concise but inclusive exposition of the main techniques
employed in state-of-the-art calculations along with discussions of
their suitability to different kinds of problems, all the while
focusing on representative results that reveal the highly non-additive
character of Casimir forces.

\subsection{Scattering methods}

\begin{figure*}[tb!]
  \centering \includegraphics[width=1\textwidth]{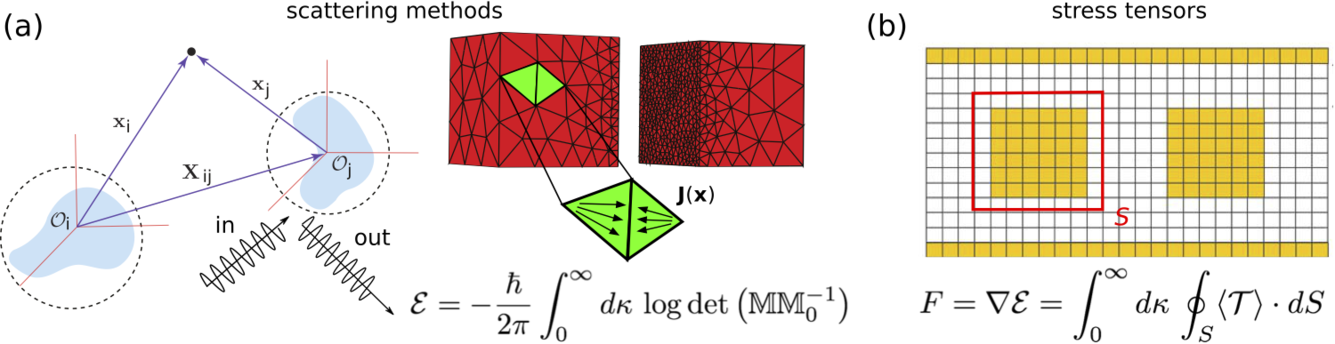}
 \caption{(Color Online) Schematic illustration of numerical methods
    recently employed to compute Casimir interactions in complex
    geometries: (a) scattering methods where the field unknowns are
    either incident/outgoing propagating waves (left) or vector
    currents ${\bf J}({\bf x})$ defined on the surfaces of the bodies
    (right), and the resulting energies are given by Eq. \ref{eq:Escat}; (b)
    stress--tensor methods in which the force is obtained by
    integrating the thermodynamic Maxwell stress tensor over a
    surface surrounding one of the bodies utilizing the fluctuation--dissipation 
    theorem. }
   \label{fig:fig1}
\end{figure*}

A sophisticated and powerful set of techniques for calculating Casimir
forces are scattering methods (Fig. \ref{fig:fig1}(a)). While these approaches come in a
variety of flavors, they often rely on formulations that exploit
connections between the electromagnetic density of
states~\cite{gies03,emig03_2,emig06,Kenneth08} or path-integral
representations of the electromagnetic
energy~\cite{Kardar99,Lambrecht06,Emig07,Kenneth08,Rahi09:PRD,MiltonPa10,ReidRo09,Reid11},
and classical scattering matrices ~\cite{Dalvit:book}. Regardless
of the chosen starting point, the Casimir energy turns out to be given
in the form:
\begin{equation}
\mathcal{E} = -\frac{\hbar}{2\pi} \int_0^\infty d\omega \, \log \mathrm{det}
(\mathbb{M} \mathbb{M}^{-1}_0),
\label{eq:Escat}
\end{equation}
where  the matrix $\mathbb{M}$ is
\begin{equation*}
  \mathbb{M} = \left(\begin{array}{lll}
    \mathbb{M}^{(11)} & \mathbb{M}^{(12)} & \ldots \\
    \mathbb{M}^{21} & \mathbb{M}^{(22)} & \ldots \\
    \vdots & \vdots & \ddots
    \end{array}\right)
\end{equation*}
whose diagonal blocks $\mathbb{M}^{(\alpha\alpha)}$ are precisely the
scattering matrices of isolated bodies and whose off-diagonal blocks
$\mathbb{M}^{(\alpha\beta)}$ encapsulate interactions and scattering
among the bodies~\cite{Lambrecht06,Rahi09:PRD}. Multiplication
by the inverse matrix $\mathbb{M}^{-1}_0$ ensures that
the divergent, self-interaction energy of the bodies in isolation
(separations $d \to \infty$) is subtracted, leaving behind a finite
quantity~\cite{Rahi09:PRD}.  

Although Eq.~\ref{eq:Escat} may appear largely unrelated to the
Lifshitz formula (Eq. \ref{Lifshitz_ret}), the connection between the two
becomes apparent when considering extended structures.  Specifically,
given two semi-infinite, periodic bodies the formula can be written in the more
familiar form, $\mathcal{E} = \int_0^\infty d\omega \, \log \mathrm{Det}
\left(1 - \mathbb{R}^{(1)} \mathbb{M}^{(12)} \mathbb{R}^{(2)}
\mathbb{M}^{(21)} \right)$, where $\mathbb{R}^{(\alpha)}$ are the
reflection matrices of each individual half-space and
$\mathbb{M}^{(\alpha\beta)}$ are translation matrices that describe
wave propagation between
them~\cite{Rahi09:PRD,Lambrecht06,Lambrecht09}. For planar bodies, as discussed in Sec. III.A, the
scattering matrices can be expressed in a Fourier basis and the above
expression reduces to the Lifshitz formula~\cite{Lifshitz1956},
originally obtained via direct evaluation of the Maxwell stress
tensor.

It is also possible to derive a slightly different scattering formula,
known as the TGTG formula~\cite{Kenneth08,Klich09}, in which the
energy between two arbitrary bodies is expressed as
\begin{equation}
\mathcal{E} =
\int_0^\infty d\kappa \, \log \det \big(1 - \mathbb{T}^{(1)}
\mathbb{G}^{(12)}_0 \mathbb{T}^{(2)} \mathbb{G}^{(21)}_0\big),
\end{equation}
where $\mathbb{T}^{(\alpha)}$ are the $T$ operators appearing in the
Lippmann-Schwinger equation (related to the scattering matrices of
individual bodies), and $\mathbb{G}_0^{(\alpha\beta)}$ are the
homogeneous Green's functions of the intervening medium, describing
the wave propagation.  Note that even though these formulations may
appear to be completely divorced from the original picture of dipole
fluctuations, the fact that the energy is described by the scattering
properties of the bodies is not surprising. In particular, as
discussed further below, at equilibrium it is possible to describe the
statistics of field fluctuations independently of the corresponding
current sources of the
fluctuations~\cite{Eckhardt84,Lifshitz1980}. Intuitively, one can
consider Casimir interactions as arising from the scattering and
momentum-exchange of vacuum electromagnetic fields originating from
radiating sources infinitely far away (rather than within the bodies)
and that ultimately end up equilibrating as they get scattered,
absorbed, and re-emitted by the bodies.

Equation~\ref{eq:Escat} was originally exploited to study forces
between highly symmetric structures, e.g. spheres and cylinders, where
the corresponding propagators, scattering, and translation matrices
can be expanded in terms of convenient, de-localized free-wave
solutions of the Helmholtz
equation~\cite{Kenneth08,Rahi09:PRD,Lambrecht09,MiltonPa10,Maghrebi10},
as illustrated in Fig.~\ref{fig:fig1}(a). The resulting spectral
methods~\cite{Balian78,Mazzitelli06,Dalvit06,Lambrecht06,Emig07,Milton08,Kenneth08,Lambrecht09,Rahi09:PRD}
are advantageous in a number of ways: First, they yield analytical
results that offer insight into the properties of the Casimir force at
asymptotically large separations~\cite{Rahi09:PRD} or under
assumptions of dilute
media~\cite{Milton08,Milton08:scat,Golestanian09,Bitbol13}. Second,
the trace operations for smooth and high-symmetry structures can be
efficiently implemented due to the very high-order and possibly even
exponential convergence of the basis
expansions~\cite{boyd01:book,Dalvit:book}. Finally, since
the energy expressions involve simple products of scattering matrices
having well-studied properties, this formulation is 
well-suited for establishing general constraints on the signs and
magnitudes of forces under various circumstances.  Of particular
importance is the recent demonstration that the force between any two
mirror-symmetric bodies must always be
attractive~\cite{Kenneth02,KennethKl06}, resolving a long-standing
question about the sign of the internal pressure or self-force on
a perfectly metallic, isolated sphere (the limit of two opposing
hemispheres)~\cite{Boyer68,Milton78,Brevik88,Bordag2001}. Similarly,
recent works have shown that stable suspensions (local equilibria)
between vacuum-separated, non-magnetic bodies are generally
impossible~\cite{Lambrecht97,Rahi10:PRL}.

For more complicated bodies lacking special symmetries, involving
sharp corners, or where non-uniform spatial resolution is desired, it
is advantageous to employ localized basis functions. More commonly,
the unknowns are defined on a generic mesh or grid and the resulting
equations are solved numerically, examples of which are the finite
difference~\cite{Taflove00}, finite element~\cite{Jin02}, and
boundary-element~\cite{bonnet99,chew01} methods. The latter category
are closely related to scattering
methods~\cite{Rodriguez07:PRA,Xiong10,ReidRo09,Reid11}; in the
surface-integral equation formulation of electromagnetic scattering,
the scattering unknowns are fictitious electric and magnetic currents
defined on the surfaces of the bodies, illustrated in
Fig.~\ref{fig:fig1}(a), and expanded in terms of an arbitrary basis of
surface vector fields~\cite{chew01}.  The connection to scattering
problems comes from the fact that incident and scattered fields are
related to the current unknowns via homogeneous Green's functions
(analytically known); not surprisingly, this
formulation leads to a similar trace expression for the Casimir energy
given in Eq.~\ref{eq:Escat}, except that the elements of
$\mathbb{M}$ consist of overlap integrals among the various surface
basis functions. A powerful
implementation of this approach is the boundary-element method (BEM),
where the current unknowns are expanded in terms of localized basis
functions (typically, low-degree polynomials) defined on the elements
of some discretized surface, as illustrated in Fig.~\ref{fig:fig1}(a). As
a result the $\mathbb{M}$ matrices turn out to be none other than the
well-studied BEM matrices that arise in classical scattering
calculations~\cite{chew01}. Such a formulation allows straightforward
adaptations of sophisticated BEM codes, including recently published,
free and widely available software packages~\cite{Reid:scuffem}. Like
most numerical methods, the BEM method can handle a wide range of
structures, including interleaved bodies with corners, and enables
non-uniform resolutions to be employed as needed.

\subsection{Stress tensor methods}

Although originally conceived as a semi-analytical method for
computing forces in planar
bodies~\cite{Jaekel91,Zhou95,Klimchitskaya00,Tomas02}, leading to the
famous Lifshitz formula (Sec. III.A), the stress tensor approach can also be
straightforwardly adapted for numerical
computations~\cite{Rodriguez07:PRA,RodriguezJo08:PRA,Rodriguez08:PRL,RodriguezMc09:PRA,McCauleyRo10:PRA,Xiong09}
since it relies on repeated calculations of Green's functions. In this
formulation (schematically shown in Fig. \ref{fig:fig1}(b)), the Casimir force on an object is expressed as an
integral of the thermodynamic, Maxwell stress tensor $\langle T_{ij}
\rangle = \varepsilon \left(\langle E_i E_j \rangle - \frac{1}{2} \sum_k \langle E_k
E_k \rangle\right) + \left(\langle H_i H_j \rangle - \frac{1}{2}
\sum_k \langle H_k H_k \rangle\right)$ over an arbitrary
surface $S$ surrounding the object,
\begin{equation}
  \bf{F} = \int_0^\infty d\kappa \iint_S \langle \bf {T} \rangle \cdot \bf {dS}.
\end{equation}
Similar to scattering methods, here the picture of fluctuating dipoles
is masked by an equivalent scattering problem involving fields rather
than fluctuating volume currents, whereby the correlation functions
$\langle E_i E_j \rangle, \langle H_i H_j \rangle \sim G_{ij}$, a
consequence of the fact that at equilibrium currents and field
fluctuations become thermodynamically equivalent~\cite{Eckhardt84}.

Beyond special--symmetry structures where the Green's functions can be
expanded in a convenient spectral
basis~\cite{Jaekel91,Zhou95,Klimchitskaya00,Tomas02}, recent
implementations of the stress-tensor method for arbitrary geometries
exploit general-purpose techniques, such as the finite-difference
method illustrated in Fig.~\ref{fig:fig1}(b), where space is divided into
a uniform grid of finite resolution, and the resulting matrix
equations for the Green's functions are solved
numerically~\cite{strikwerda89,anderson99,Taflove00}.  Early
implementations include both finite-difference
frequency-domain~\cite{Rodriguez07:PRA,Xiong09} and
time-domain~\cite{RodriguezMc09:PRA,McCauleyRo10:PRA} methods.

Because Casimir forces involve broad bandwidth fluctuations time
domain methods are advantageous in that $G_{ij}(\bf{r},\bf{r}',\omega)$ at
all frequencies can be computed at once via Fourier
transforms~\cite{Taflove00}. While the finite difference stress-tensor
method does not offer the efficiency and sophistication of other
formulations and discretization schemes, such as the BEM
fluctuating--surface current method~\cite{Reid11}, they are
compensated by their flexibility and generality. For instance, they
are extremely simple to implement (leading to many free and
easy-to-use numerical packages~\cite{Oskooi10:Meep}, can handle 
many different kinds of boundary conditions and
materials (including anisotropic and even nonlinear dielectrics), and
are well understood. A BEM implementation of the stress-tensor method
was also first suggested in Ref.~\cite{Rodriguez07:PRA} and
subsequently implemented by Ref.~\cite{Xiong09}, although for small
problems the trace formulas above provide a simpler and more efficient
alternative since they do not require repeated integration over
surfaces and involve only products of BEM matrices. On the other hand,
the stress tensor method offers computational advantages for large
problems since it involves repeated evaluation of Green's functions,
or matrix-vector products, making it an ideal candidate for
applications of fast-solver (iterative) techniques~\cite{chew97}.

\begin{figure*}[tb!]
  \centering \includegraphics[width=0.58\textwidth]{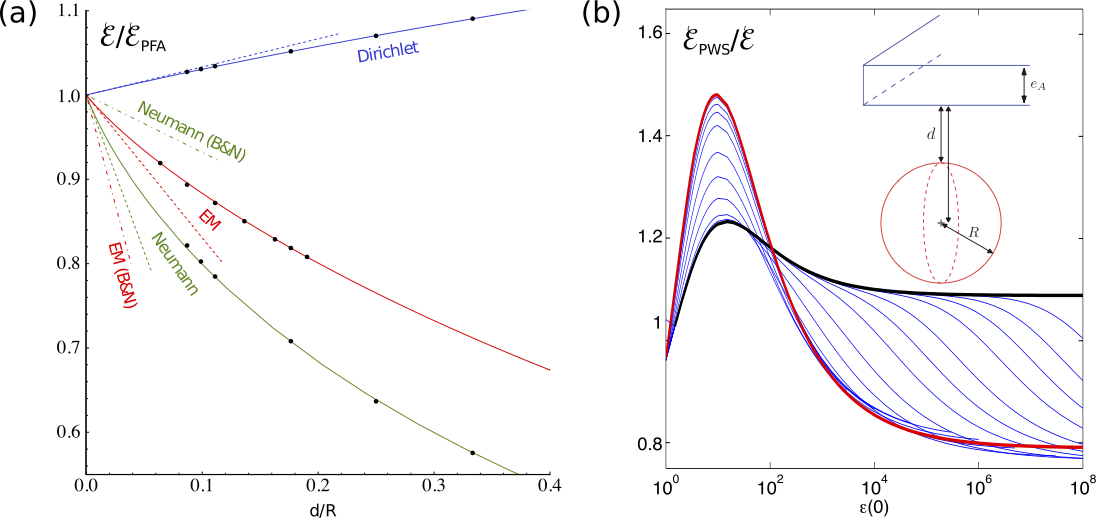}
  \caption{(Color Online) Results illustrating the range of validity of PFA and PWS
    approximations in the sphere--plate geometry, involving a sphere
    of radius $R$ separated from a semi-infinite plate by a
    surface--surface distance $d$. The exact Casimir energy is
    computed by application of the scattering method. (a) Ratio of the
    exact and PFA energies as a function of $d/R$ for 
    perfectly metallic conductors~\cite{Bimonte11}. Numerical data
    (dots) are compared to the first correction of the PFA (dashed
    lines), described in the text, and also to a fit performed via
    Pad\'{e} approximants (solid curves), in which the energy ratio is
    given by $\frac{\mathcal{E}}{\mathcal{E}_\mathrm{PFA}} = 1 +
    \frac{1}{3} (1 - \frac{60}{\pi^2}) \frac{d}{R} + \frac{8}{100}
    \big(\frac{d}{R}\big)^2 \log \frac{d}{R}$. (b) Ratio of the PWS
    and exact Casimir energies as a function of the static
    permittivities of the sphere and plate, for various ratios
    $d/R$ (blue curves)~\cite{Bitbol13}. The
    red and black curves represent equivalent results for the
    plate--plate and sphere--plate geometries in the limit of infinite
    separations.}
   \label{fig:fig2}
\end{figure*}

\subsection{Casimir interactions in complex geometries}
While PFA and PWS approximations provide simple, quickly solvable, and
intuitive expressions for forces in arbitrary geometries,
they are uncontrolled when pushed beyond their limits of validity and
have been shown to fail (even qualitatively) in the simplest of
structures~\cite{Bordag06,gies06:PFA,Rodriguez11:review,Dalvit:book,Bitbol13}. Increased
demand for experimental guidance has stimulated recent efforts in
quantifying the validity and accuracy of PFA.

Although PFA is technically only applicable in geometries with 
smooth, large-curvature objects and small separations, it has
nevertheless been heuristically applied in the past to study a wide
range of other situations~\cite{Lambrecht08,Rodriguez14:review}. For instance, a number
of recent works have employed scattering methods to investigate
extensions of PFA in the sphere--plate geometry at large separations
in the idealized limit of perfect conductors~\cite{Bimonte11,Fosco12},
where the PFA energy takes on the closed-form expression
$\mathcal{E}_\mathrm{PFA} = \frac{\pi^3 \hbar c R}{1440 d^2}$, where
$R$ denotes the sphere radius. The plotted ratio of energies in
Fig.~\ref{fig:fig2}(a) shows that PFA increasingly overestimates the
energy as $d \to \infty$, which is to be expected since at $d/R \gg
1$, the interaction approaches that of a dipole above a plate
exhibiting a significantly faster decay $\sim
\frac{1}{d^6}$~\cite{Buhmann07}. Higher-order perturbative PFA
corrections for large curvatures $d/R \ll 1$ have also been
obtained~\cite{Bordag06,Lambrecht08,Bimonte11,Fosco12}
(Fig.~\ref{fig:fig2}(a)). Techniques based on Pad\'{e} approximants,
which constrain the force at short and large separations using
gradient and multipole expansions (Fig.~\ref{fig:fig2}(a)), have also
culminated in analytical expressions~\cite{Bimonte11}, leading to the hope that similar methods
can be applied to more complex geometries. The situation is more
complicated in cases involving realistic metals and finite
temperatures, as illustrated by recent predictions of  {\it geothermal 
effects} involving nontrivial interplay between geometry, materials,
and temperature in the sphere--plate
geometry~\cite{Neto08,CanaguierDurand10,Weber10,Weber10:geothermal}.

The PWS approximation, applicable in the limit of large separations
and dilute media, relies on dividing the object into small elements
("atoms") and summing the corresponding vdW and
Casimir-Polder
interactions~\cite{Bergstrom97,Milton08,Golestanian09, Veble07}. The presence
of multiple scattering, otherwise absent in the limit of weak coupling
or dilute media~\cite{Milton08}, has long been known to significantly
modify the underlying two-body force laws~\cite{Axilrod43}. Despite
these shortcomings, PWS approximations have been recently applied to
numerically approximate interactions in complex
geometries~\cite{Tajmar04,Sedmik06}, especially in the field of
microfluidics~\cite{Stone01,Parsegian}. While PWS approximations are
strictly applicable in the limit of dilute media, recently they were shown to lead to larger errors in the
experimentally relevant case of dielectric
materials~\cite{Bitbol13}. This situation is illustrated in
Fig.~\ref{fig:fig2}(b), which shows that PWS underestimates the
energy in the perfect-metal limit ($\varepsilon \to -\infty$) by
$\approx 20\%$, is exact in the dilute limit of $\varepsilon \to 1$,
and is (surprisingly) most inaccurate at intermediate $\varepsilon
\sim 10$ where it overestimates the energy by roughly $60\%$. Such
counter-intuitive results shed light on the complexities associated
with dilute approximations, since a heuristic argument based on the
screening of fields in materials with large dielectric contrasts would
predict strictly monotonically increasing deviations. By examining
interactions between compact objects at asymptotically large
separations, it is also possible to obtain perturbative corrections to
Casimir--Polder
forces~\cite{Balian78:multscat,Golestanian00,emig06,Emig08:sphpl,Milton08:scat,Rahi09:PRD,Golestanian09,Stedman2014}. Formal derivations of PWS approximations in the limit of dilute media as well
as perturbative corrections applicable in systems with larger index
contrasts have also been
developed~\cite{Milton08,Golestanian09,RodriguezLopez09}.

\begin{figure*}[tb!]
  \centering \includegraphics[width=1.0\textwidth]{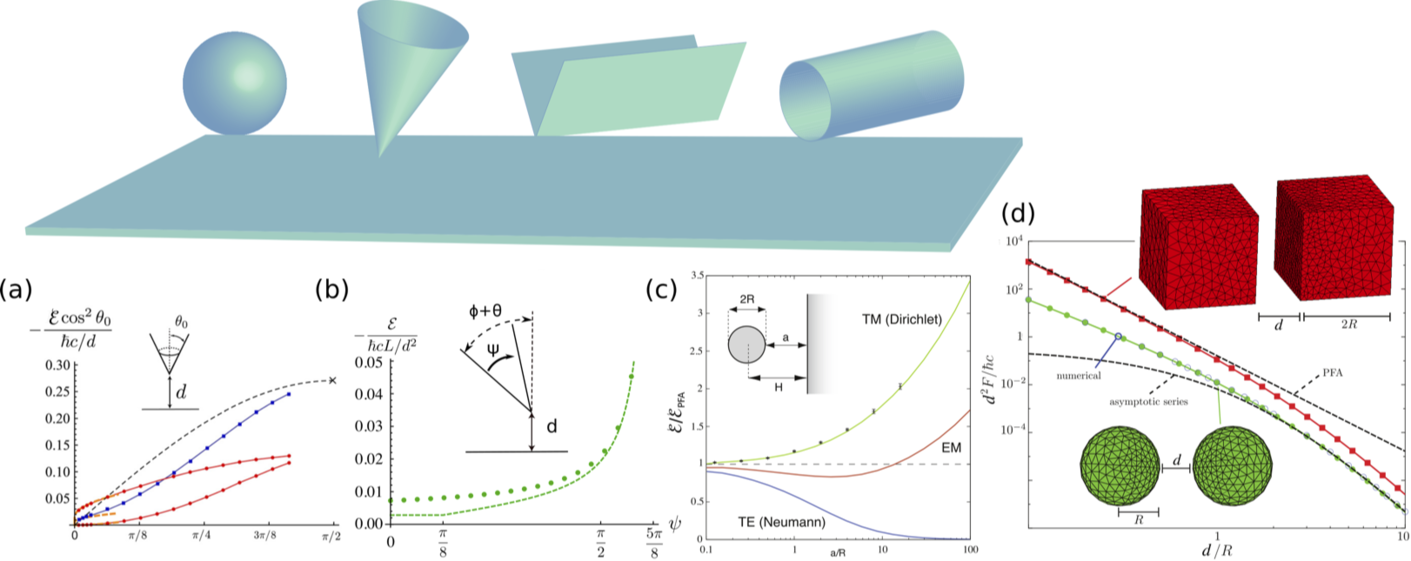}
  \caption{(Color Online) Selected results of Casimir interactions involving compact
    bodies suspended above planar objects or interacting with other
    compact bodies, calculated via scattering matrix techniques in
    combination with spectral methods.  (a) Casimir energy of a
    perfectly conducting, vertically oriented cone of semi-opening
    angle $\theta_0$ suspended above a perfectly conducting plate by a
    fixed distance $d$~\cite{Maghrebi10}.  The PFA approximation is
    shown as the dashed line. (b) Casimir energy of a perfectly
    conducting wedge and plate, with one face at a fixed angle
    $\varphi_0 + \theta_0$, as a function of the full opening angle
    $\psi = 2\theta_0$~\cite{Maghrebi10}. The PFA prediction, in
    dashed lines, is compared to the exact calculation, indicated by
    solid circles. (c) Ratio of the exact Casimir and PFA energies of
    the perfectly conducting cylinder--plate structure shown in the
    inset, decomposed into both TE and
    TM contributions~\cite{emig06}. (d) Casimir force
    between perfectly conducting metallic cubes (red squares) or
    spheres (green circles), as computed by the BEM method of
    Ref.~\cite{Reid13:cubes}, divided by the corresponding PFA
    forces. Results for spheres are compared to computations performed
    using scattering matrix methods (blue circles).}
   \label{fig:fig3}
\end{figure*}

At intermediate separations that are on the order of the sizes of the objects and for realistic
materials, neither PFA nor PWS, nor perturbative corrections thereof
can accurately predict the behavior of the Casimir force. However, it
is precisely this regime that is most easily tackled by numerical
methods. Application of scattering methods to the study of compact
bodies interacting with planar objects have led to a number of
interesting predictions, a select number of which are illustrated in
Fig.~\ref{fig:fig3}. In geometries involving perfect-conductor bodies
with special symmetries such as spheres, cones, wedges, or cylinders,
scattering-matrix methods have been employed to obtain both numerical
and semi-analytical
results~\cite{emig06,Mazzitelli06,Neto08,Rahi09:PRD,Emig09:ellipsoids,Dalvit:book}. Other,
more complicated shapes such as waveguides, disks, cubes, tetrahedral
particles, and capsules are less amenable to spectral methods, but
have nevertheless been studied using brute-force
techniques~\cite{Rodriguez07:PRL,ReidRo09,Reid13:cubes,Atkins13}. Figure~\ref{fig:fig3}
shows that the energy of a cone with a semi-opening angle $\theta_0$
and a substrate vanishes logarithmically $\mathcal{E} \sim
-\frac{\hbar c}{d} \frac{1}{\log \theta_0}$ as $\theta_0 \to 0$, a
type of divergence that is characteristic of lines and other
scale-invariant objects~\cite{Maghrebi10}. In contrast, the PFA energy
is predicted to vanish linearly as $\theta_0 \to 0$. For a tilted
wedge, the PFA energy remains constant until the back surface of the
wedge becomes visible to the plate, while exact results indicate
smoothly varying angle dependence despite the screening
effects~\cite{Maghrebi10}. Earlier calculations of forces between
cylinders, spheres and ellipsoidal bodies and plates have also
demonstrated unexpectedly weak decay rates and other interesting
non-additive modifications \cite{emig06,Emig07,Emig09:ellipsoids,Mazzitelli06,Dalvit06,Neto08}. For
more complicated structures, such as the pair of cubes shown in
Fig.~\ref{fig:fig3}, it is more convenient to employ brute-force
techniques like the BEM method~\cite{Reid13:cubes}.

\begin{figure*}[tb!]
  \centering \includegraphics[width=0.65\textwidth]{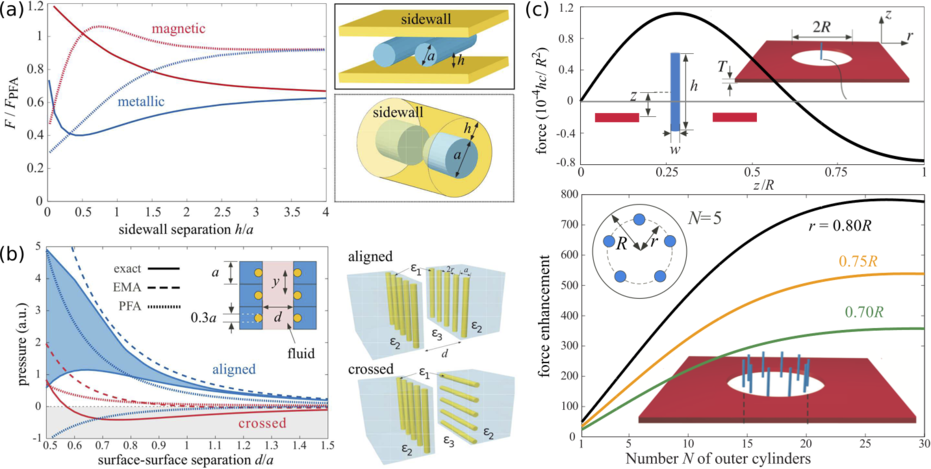}
  \caption{(Color Online) Selected results illustrating unusual Casimir effects from
    non-additive interactions in complex structures. (a) Casimir force
    between either translationally invariant waveguides (solid lines)
    or cylindrically symmetric rods (dotted lines), normalized by the
    corresponding PFA force, as a function of their separation from
    adjacent sidewalls a distance $h$
    apart~\cite{Rodriguez07:PRL,McCauleyRo10:PRA}. Configurations of
    either perfect-electric (blue) or perfect-magnetic (red)
    conductors are considered. (b) Casimir pressure between two
    patterned structures involving periodic arrays of cylinders
    embedded in a semi-infinite substrate, as a function of their
    surface--surface separation $d$, computed by application of both
    scattering and FDTD
    methods~\cite{RodriguezMc09:PRA,McCauleyRo10:PRA,McCauleyRo11:fluid}. Exact
    calculations (solid lines), PFA (dotted lines), and
    effective-medium theory (dashed lines) are also shown. (c) Casimir
    force between a small, anisotropic particle or an array of
    particles and a plate with a hole~\cite{LevinMc10,McCauleyRo11},
    demonstrating repulsion for vacuum-separated metallic bodies.}
  \label{fig:fig4}
\end{figure*}

Unusual Casimir interactions in multi-body geometries have also been
recently studied~\cite{Rodriguez11:review,Dalvit:book}. For
instance, application of numerical methods (first employing stress
tensors~\cite{Rodriguez07:PRL} and subsequently scattering
matrices~\cite{RahiRo07}) in a structure composed of two metallic
co-planar waveguides suspended above adjacent metal sidewalls
(Fig.~\ref{fig:fig4}(a)) reveal that the attractive Casimir force per
unit length between the waveguides varies non-monotonically as a
function of their separation from the sidewalls $h$. Large deviations
from PFA can be explained from the fact that PFA is unable to
accurately capture the competing effects of TE and
TM fields at small and large
$h$~\cite{Hertzberg07,Zaheer07,RahiRo07}. Extensions of this geometry
to situations involving finite rods, such as the cylindrically
symmetric geometry of Fig.~\ref{fig:fig4}(a) where the sidewalls are
joined to form a cylindrical tube and described by either
perfect-electric or perfect-magnetic boundary
conditions~\cite{McCauleyRo10:PRA}, demonstrate the importance of
dimensionality and boundary conditions on the behavior of the force.
Along similar lines, structures involving periodic arrays of finite
cylinders on slabs (Fig.~\ref{fig:fig4}(b)) reveal strong variations in
the force depending on whether the arrays are aligned or crossed, even
leading to changes in its sign at close separations when the system is
immersed in a
fluid~\cite{RodriguezMc09:PRA,McCauleyRo10:PRA,McCauleyRo11:fluid}. Exact
calculations are also compared to predictions based on PFA, showing
significant, qualitative deviations. An effective medium theory
description of the problem, in which the slabs are treated as
homogeneous, anisotropic dielectrics, gives surprisingly accurate
predictions down to separations of the order of the period.

Objects with nontrivial geometry can also be utilized to obtain
repulsive Casimir interactions in vacuum~\cite{Rodriguez14:review}. A
proof-of-principle of the feasibility of repulsion in vacuum was
recently demonstrated using BEM and FDTD numerics in a structure
involving a small, elongated particle above a plate with a
hole~\cite{LevinMc10}, shown schematically in
Fig.~\ref{fig:fig4}(c). Due to constraints on the size of the
particles and hole as well as on the lengthscales needed to observe
these effects, the force in that geometry turns out to be too small
(atto-Newtons) for current experimental
detection~\cite{LevinMc10}. However, extensions to multiple particles
(attached to a substrate) have demonstrated thousand-fold force
enhancements without the need to change hole radii or
lengthscales~\cite{McCauleyRo11}, as illustrated in
Fig.~\ref{fig:fig4}(c). Interestingly, the enhancement can be
understood as arising not only from the presence of additional bodies,
but from increased repulsion due to the larger polarizability of the
particles as they interact with fringing fields near the edge of the
plate~\cite{Eberlein11,Milton11,Milton12}. One can also show that the
interaction between a polarizable particle and a perfect-metal wedge
or half-plate is repulsive~\cite{Milton12}, provided that the wedge is
sufficiently sharp and that the particle is sufficiently anisotropic.
Similar results should extend to vdW interactions on
molecules and atoms near structured surfaces, but the main challenge
in these systems is the need to attain a large degree of particle
anisotropy. Recent calculations show that Rydberg atoms cannot
achieve a high enough anisotropy~\cite{Ellingsen10}. Regardless of
their current experimental observability and practical considerations,
these recent theoretical predictions demonstrate that geometry can
prove to be a powerful resource for shaping Casimir forces.

\section{Soft and Biological Materials \label{VII}}

Besides the prominent role of fluctuation-induced interactions in inorganic materials systems, vdW forces have many other interesting manifestations. The adhesion of the gecko, with no help from glues, suction or interlocking, is perhaps the most popular example for vdW interactions in biological and bio-related matter. Researchers have shown experimentally that vdW interactions between the gecko spatular toes and hydrophobic surfaces in air are responsible for the gecko clinging to substrates \cite{Autumn-2002,Autumn-2002a,MLee-book}. Other experiments suggest that while geckos indeed use no glue, they do leave "footprints" of residue identified as phospholipids with phosphocholine head groups \cite{Stark1}. On the other hand, the contact surface between the gecko's toes and the substrate is saturated with methylene moieties of the phospholipids and contains no water. This is furthermore consistent with predominantly hydrophobic surface of gecko setae \cite{AStark2}, landing additional support to the mostly vdW origin of the adhesion force. Currently there is much gecko-inspired interest in constructing materials with similar adhesion properties \cite{Suh}. Single strand vertical arrays of cylindrical pillars produced by electron-beam lithography and etched into an array of vertical round shaped pillars are expected to show behavior similar to gecko toes. Dry adhesives for robotic applications based upon the characteristics of vertical and angled flaps from polydimethylsiloxane (PDMS) are also being considered for applications \cite{Yu-2011}.

\subsection{The importance of aqueous solvent}

The most important defining characteristics of vdW interactions in soft- and bio-matter comes from the presence of a solvent, i.e., water \cite{isra}. The interaction between two substrates with dielectric function $\varepsilon(\omega)$ separated by a water layer with $\varepsilon_w(\omega)$ can be described by the standard Lifshitz formula (Eq.\ref{Lifshitz_ret}) \cite{Bordag:book}. The fact that typically $\varepsilon(0) \ll \varepsilon_w(0)$ makes the  $n= 0$ Matsubara term quite important, and can account for about 50 \% or more of the total value of the Hamaker coefficient \cite{Ninham-1970}. For  lipid-water systems \cite{Pabst-2015} retardation effects are not important even at very large distances \cite{Ninham-1970b, Parsegian}. The significant thermal effects from the $n=0$ term show that results for vdW interactions in standard condensed media cannot be simply transcribed into the soft-matter context. Also, apart from its large static dielectric constant, the full  dielectric spectrum of water leads to non-monotonic features in the vdW interaction between ice and water vapor \cite{elbaum_application_1991} or hydrocarbon films \cite{Safran-ice} across a liquid water layer in the retardation regime \cite{elbaum2}. 

Solvent effects are also important when their dispersion properties are in a certain relation with those of interacting materials. Recent work by Capasso $et$ $al.$ \cite{Munday07,Munday09}, as well as previous work by various authors \cite{Milling96,Meurk97, Lee01:casimir,Lee02:casimir, Feiler08}, made it clear that for specific asymmetric interaction geometries, a solvent whose dielectric permittivity $ \varepsilon_m(\omega)$ is between those of the interacting bodies 1 and 2, $ \varepsilon_1(\omega) > \varepsilon_m(\omega) >  \varepsilon_2(\omega)$, can create repulsive vdW interactions. Though in principle this solvent-mediated Casimir-Lifshitz levitation has been known since the appearance of the Lifshitz theory \cite{Dzyaloshinskii61}, it has not been used to effectively control the sign of the vdW interaction. Solvent mixtures with low molecular weight solutes such as glucose and sucrose also affect the dielectric properties of the solution and can thus modify the vdW interactions  \cite{Glucose}. Solvent-like effects could be important also for two graphene sheets separated by atomic hydrogen gas, with one sheet adsorbed on a $\rm SiO_2$ substrate, while the other is freestanding \cite{Bostrom-H}, as discussed in Sec. IIIC.

Electrolyte screening is also a defining feature for bio-matter in aqueous environments. The presence of salt ions screens the $n=0$ Matsubara frequency in the Hamaker coefficient. The existence of this screening is connected with the fact that the $n=0$ Matsubara term actually corresponds to the classical partition function of the system, and for confined Coulomb fluids, such as inhomogeneous electrolytes, can lead to a thoroughly different form of the $n=0$ term in the full Matusbara sum of the Lifshitz theory \cite{Podgornik-Zeks,Perspective}, as discussed below. The screening of vdW interactions in electrolyte solutions can be derived in a variety of ways, most simply by replacing the Laplace equation with the linearized Debye-Huckel equation \cite{Parsegian,isra}. In this approach, the free ions present in the aqueous solution are taken into account just like in the case of bad conductors \cite{pitaevskii_thermal_2008}, where the number of charge carriers is small and obeys the Boltzmann statistics. The zero frequency Matsubara term in Eq. \ref{Lifshitz_ret} then leads to  an approximate $n=0$ Hamaker coefficient ${\cal H}_0(d) = (3/4) k_BT (1+2\kappa_0 d) e^{-2 \kappa_0 d}$ \cite{Parsegian}. This expression is obviously screened with twice the Debye screening length $\kappa_0^{-1} = 8\pi \ell_B n_0$, where $\ell_B \simeq 0.7$ nm is the Bjerrum thermal length and $n_0$ is the bulk salt concentration.

\begin{figure}[ht]
\includegraphics[scale=0.45]{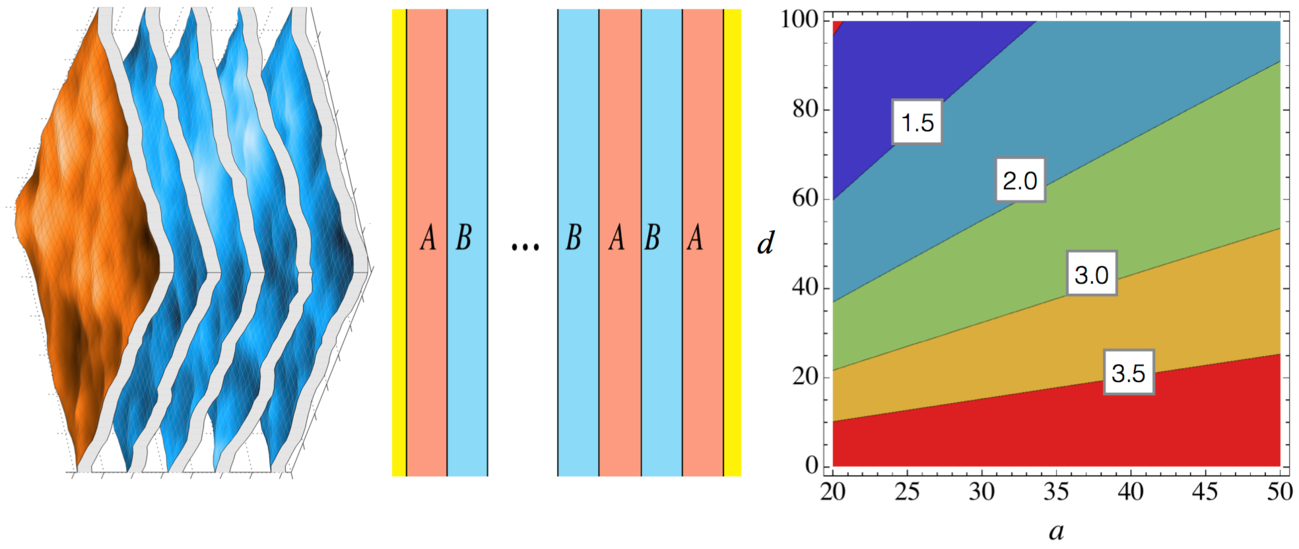}
\caption{(Color Online) Multilamellar array of lipid membranes, each composed of a hydrocarbon core with two surface hydrophillic headgroup layers. Left: a realistic presentation with fluctuating positions of the membranes (for details see \cite{Pabst-2015}). Middle: a model system with a rigid array of alternating solvent (B) - membrane (A) - solvent (B) regions. Right: The Hamaker coefficient, ${\cal H}(a, d)$ in [zJ], as a function of the separation between membranes, $d$, and the thickness of the lipid bilayers, $a$. The Hamaker coefficient is calculated based on the full dispersion spectra of water and lipids (hydrocarbons) \cite{Podgornik-2006}. }
\label{Fig1_RP}
\end{figure}

In many bio-systems, the vdW free energy is typically cast into the form of the Hamaker-type approximation ${\cal F}(d, T)=- \frac{{\cal H}(d)}{12 \pi d^2}$, with the separation-dependent Hamaker coefficient that can be calculated exactly via the Lifshitz formalism, when accurate experimental data for the dielectric properties are available.  For example, the Hamaker coefficients for lignin and glucomannan interacting with cellulose, titania and calcium carbonate in vacuum, water and hexane, are found within a relatively narrow range of $\sim 35 - 58$ zJ for intervening vacuum and $\simeq 8-17$ zJ for an intervening aqueous medium \cite{Bergstrom}, with the dielectric response properties extracted via spectroscopic ellipsometry \cite{Bergstrom2}. The Hamaker coefficients for the interactions of the wood components with common additives in paper such as $\rm TiO_2$, and $\rm CaCO_3$ in water were obtained as $\simeq 3-19$ zJ \cite{Bergstrom} and can explain important adhesion, swelling and wetting phenomena ubiquitous in paper processing.

The long-range interaction between proteins is also of vdW nature \cite{Leckband1,Leckband2}. Estimates for protein-protein interactions across water or dilute salt solutions report Hamaker coefficients  mostly within the range $\simeq 10-20$  zJ \cite{Farnum}. $\cal H$ for interacting proteins, such as bovine serum album (BSA), has been found to be  $\simeq 12$  zJ by considering a Drude-Lorentz model for the dielectric function and the zero Matsubara frequency term included \cite{Lenhoff1,Lenhoff2}. Using the anisotropic coarse-grained model of a protein on the level of amino acid residues, one can calculate effective polarizabilities of bovine pancreatic trypsin inhibitor (BPTI), ribonuclease inhibitor, and lysozyme in an aqueous solution \cite{Zhao-2}. These results have to be approached with caution, however. Accurate frequency-dependent polarizabilities are rarely  available either from theoretical or experimental studies \cite{Bagchi}, thus one has to rely on plausible but probably unrealistic model approximations \cite{Zhao-1}. The same is true for the static dielectric constant that shows pronounced variation from the inside to the periphery of the protein \cite{Alexov}. The anisotropic optical spectrum of collagen, a fibrous protein, has been calculated by $ab$ $initio$ methods \cite{poudel_partial_2014} and used to estimate the corresponding non-isotropic Hamaker coefficients \cite{Dryden-GeckoNatMat}. The vdW interactions between collagen fibers show a substantial angle-independent component of the Hamaker coefficient $\simeq 9.3$ zJ. The origin of the angular dependence of vdW interactions is in fact twofold: the morphological anisotropy, given by the shape, and the material anisotropy, given by the dielectric response tensor. Both contribute to the general angular dependence and consequently torques between biological macromolecules \cite{Hopkins-aniso}, see below.

\subsection{ Lipid membranes}

In general,  the vdW interaction is of fundamental importance for the stability of biological matter \cite{Thompson-2009} and for membrane arrays in particular \cite{Petrache-2003,Maggs3}. The essential component of a membrane is the lipid bilayer, a planar layer of finite thickness composed of a hydrocarbon core with hydrophyllic boundaries facing the aqueous solution \cite{Nagle-2004}. vdW interactions between lipid membranes were in fact the first example of using Lifshitz theory in condensed media \cite{Adrian-nature}. Calculated non-retarded Hamaker coefficients were found to be in the range $1- 10$ zJ. The importance of the ionic screening of the zero frequency Hamaker term in electrolyte solutions for membranes  has also been carefully quantified  \cite{Ninham-1970, Petrache-2003}, see above. As already stated, the very high static dielectric constant of water \cite{Parsegian} leads to an anomalously large contribution to the entropy of vdW interactions, which remains unretarded for all separations as it corresponds mostly to the $n=0$ Matsubara term. However, taking into account electrolyte screening at sufficiently large salt concentrations reverses the anomalous effect of the water dielectric constant, so that retardation effects emerge from a combination of electrolyte screening and standard retardation screening \cite{Ninham-1970}.

Most experiments yielding the strength of the non-retarded Hamaker coefficients are actually based on multilamellar interaction geometries that allow for detailed osmotic stress small-angle X-ray scattering (SAXS) studies \cite{Nagle-2004, Pabst-2015} (see Fig. \ref{Fig1_RP}). One also needs to consider the non-pairwise additive vdW effects in multilamellar geometries that can be significant \cite{Narayanaswamy-2013}. The interaction of a pair of two lipid membranes with thickness $a$ at a separation $d$ in a multilamellar stack yields for the interaction surface free energy density ${\cal F}(a, d)$:

\begin{eqnarray}
{\cal F}(a, d)  & \simeq & - \frac{k_BT}{ 4\pi ~(a+d)^{2}}
\left[ \frac{1}{2} \left(\zeta(2,\frac{d}{a+d}) - 2 \zeta(2, 1) +
\zeta(2,\frac{d + 2a}{a+d})\right) \overline{\Delta}^{2}(0) + \right. \nonumber\\
& & + \left. \sum_{n=1}^{\infty}\left({\cal Z}(2 + y,\frac{d}{a+d}) - 2
{\cal Z}(2 + y, 1) +
{\cal Z}(2 + y,\frac{d + 2a}{a+d})\right) \overline{\Delta}^{2}(\imath \omega_n) \right] , 
\label{fin-1}
\end{eqnarray}
where $\zeta(m,n)$ is the zeta-function and $y =  2 \frac{\omega_n}{c} \sqrt{\epsilon_{B}(\imath \omega_n)}(a+d)$.  The function ${\cal Z}(2 + y,x)$ is exponentially screened with $y$  according to \cite{Podgornik-2006} and $\overline\Delta(\omega) = \left(  \frac{\rho_{A}\epsilon_{B} -    \rho_{B}\epsilon_{A} }  {\rho_{A}\epsilon_{B} +    \rho_{B}\epsilon_{A} } \right)$, where $\epsilon_{A}(\omega)$ and  $\epsilon_{B}(\omega)$ are the permitivities the lipid and water, respectively. Also, $\rho_{A,B} ^2 = Q^{2} - \frac{\epsilon_{A,B}\omega ^{2}}{c^{2}}$, where $Q$ is the magnitude of the transverse wave vector. (for details see Ref. \cite{Podgornik-2006}).

The non-additive effects in ${\cal F}(a, d)$ vanish at $d \ll a$, where the interaction is obviously reduced to that of two semiinfinite lipid regions across water. Approximating the water response function by one Debye and twelve Lorentz oscillators \cite{dagastine_dielectric_2000,Roth1}, and the lipid response function by four Lorentz oscillators in the ultraviolet regime \cite{Parsegian} yields ${\cal H}(a = 4$ nm, $d \sim a) = 4.3$ zJ. The usually quoted theoretical result with no retardation effects \cite{Parsegian} is ${\cal H}(a, d \sim a) = 3.6$ zJ, while experimentally determined ${\cal H}$ is typically in the range $2.87 - 9.19$ zJ for dimyristoyl phosphatidylcholine (DMPC) and dipalmitoyl phosphatidylcholine (DPPC) lipid multilayers \cite{hamaker}. The lipid bilayer thickness dependence is clearly seen in recent experiments with dioleoyl phosphocholine/distearoyl-phosphocholine/cholestrol (DOPC/DSPC/Chol) mixtures that phase separate into liquid-ordered (Lo) and liquid-disordered (Ld) domains with ${\cal H}=4.08$ zJ for Ld and ${\cal H}=4.15$ zJ for Lo domains \cite{Pabst-2015}.

Other fluctuation-induced Casimir-like interactions \cite{Kardar99} are also of relevance in the context of lipid membranes. Among these, the Helfrich interactions due to steric repulsion between fluctuating membranes have received very detailed attention, see \cite{Freund,Bing-Sui}. However, more directly related to the Casimir effect are the thermal height fluctuations of membranes, constrained on average to be planar, that can couple in various manners to the local membrane composition. For example, embedded macromolecules, such as proteins \cite{Phillips}, cause modifications of the height fluctuations due to the spatial variation of the effective membrane rigidity at the position of the inclusion. Thus, there is an elastic Hamiltonian \cite{Deserno1} in terms of the membrane height function $h({\bf x})$ \cite{Lipowsky}    
\begin{equation}
H[h({\bf x})] = \int d^2{\bf x} \left[ {\textstyle\frac12} \kappa_r({\bf x}) \left(\nabla^2 h({\bf x})\right)^2 + \overline{\kappa}_r({\bf x})\left({\partial^2h({\bf x})\over \partial x^2} {\partial^2 h({\bf x})\over \partial y^2}-\left({\partial^2h({\bf x})\over \partial x\partial y}\right)^2\right)\right], 
\label{cgwnuiel}
\end{equation}
where  the local bending rigidity $\kappa_r({\bf x})$ and the local Gaussian rigidity $\overline{\kappa}_r({\bf x})$ are position dependent upon ${\bf x}=(x,y)$ denoting the in-plane coordinates for the membrane projected area \cite{Dean-misc}.  In single component membranes, where $\kappa_r$ and $\overline{\kappa}_r$ are constant, the $\overline{\kappa}_r$ term is zero when the membrane is a free-floating sheet, since by virtue of the Gauss-Bonnet theorem it only depends on the boundary and topology of the membrane \cite{David-book}. The height correlator can be found analytically in this case \cite{Deserno1}. In the presence of elastic inclusions, both moduli contain an unperturbed constant part, ${\kappa}_{0,r}$ and $\overline{\kappa}_{0,r}$, as well as the position dependent parts $ \Delta\overline\kappa_r({\bf x}), \Delta\kappa_r({\bf x}')$ that vanish outside of the inclusions. Other parametrizations of the effect of inclusions are also possible and have been considered \cite{Netz-incl}. Alternatively, membrane inclusions can be also considered as curvature sources \cite{Domm-Fournier}. To the above energy one can also add surface energy and an external potential energy when appropriate \cite{Zandi-2013}. Apart from the order of derivatives in the fluctuating field, second for membranes and first for electrostatic field, and the local bending rigidity taking the role of the local dielectric permittivity, $H[h({\bf x})]$  is completely analogous to the electrostatic field Hamiltonian \cite{Perspective} or indeed to the Hamiltonian of critical mixtures \cite{Trondle10}. Thus one can expect that thermal fluctuations effects will also be present.

Indeed, inclusions in the membrane, which modify its local mechanical properties, can experience fluctuation-induced forces between them \cite{Goulian-1993, Golestanian-1996, Golestanian-1996a, Bartoloa, Park-1996, Zandi-2011,Yolcu-2011}. The interactions are energetically of the order of $k_BT$, and can in certain circumstances, in particular for tensionless membranes, be long-ranged and therefore potentially exhibit an important effect on the organization of the membrane \cite{Machta-2012}.  Several studies have considered coupling of membrane inclusions to the membrane curvature via the elastic stress \cite{Bitbol,Zandi-2011,Yolcu-2011} or to topological defects in orientational disorder \cite{Korolev-2011, Golestanian-1996}. Assuming that $\kappa_r({\bf x})$ and $\overline{\kappa}_r({\bf x})$ are small, one can calculate the cumulant expansion of the partition function \cite{Goulian-1993}. The effective two-body interaction between regions deviating from the background rigidity $\kappa_{0,r}$ is then obtained in the simple form
\begin{equation}
H_2 = {k_B T \over 4\pi^2 \kappa_{0,r}^2}\int d^2{\bf x} d^2{\bf x'}{ \Delta\overline\kappa_r({\bf x})\Delta\kappa_r({\bf x}')
\over |{\bf x}-{\bf x}'|^4} + \dots.\label{H2}
\end{equation}
after expanding to the lowest order in the deviation from the constant values of the rigidities, $\Delta\overline \kappa_r(\bf r)$ and  $\Delta \kappa_r(\bf r')$, that vanish outside of the inclusion. When the separation between local regions (inclusions) characterized by change in rigidities is much larger than the size of the regions, the first order term in a multipole expansion of the energy between the two regions therefore decays as the fourth power of separation.  Notably, both $\kappa_r({\bf x})$ and $\overline{\kappa}_r({\bf x})$, have to be present in order to have a fluctuation interaction. On the other hand, considering membrane inclusions as curvature sources one can bypass these constraints, at the same time also strongly enhancing and increasing the range of the interactions \cite{Domm-Fournier}. These Casimir-like forces are dominated by fluctuations and their variance also shows a characteristic dependence on the separation  \cite{Bitbol} as well as pronounced many-body aspects \cite{Fournier-aggreg}, as expected for Casimir-like interactions.

Scattering methods developed for the electromagnetic Casimir effect \cite{Emig-2} and discussed in Sec. V have been employed for the interaction between two membrane embedded disks discs of radius $R$ to all orders, leading to asymptotic result for large separations \cite{Zandi-2011}. The effective field theory formalism also affords an efficient framework for the computation of membrane fluctuation-mediated interactions \cite{Deserno-1,Yolcu-2011}. In the case of broken cylindrical symmetry of the inclusions, the fluctuation interaction retains the same separation dependence, but its strength depends on the two orientation angles as $\cos{2 \theta_1} \cos{2 \theta_2}$ and $\cos{4(\theta_1 + \theta_2)}$ \cite{Golestanian-1996a,Park-1996}. This of course implies the existence of fluctuation or vdW torques, see below.

Experimentally, fluctuation mediated interactions between membrane inclusions might be difficult to measure directly, if at all feasible. More promising seems to be the detection of their consequences, like fluctuation-induced aggregation of rigid membrane inclusions \cite{Fournier-aggreg,Weikel-2001} or through their effect on the miscibility of lipid mixtures in multicomponent membranes \cite{Dean-misc,Machta-2012}. 

\subsection{van der Waals torques}

Biological materials are typically anisotropic in terms of shapes as well as response properties \cite{Hopkins-aniso}. Such anisotropy leads to vdW torques, which have been first studied by Kats \cite{Kats} for the special case of isotropic boundaries with anisotropic intervening material, and independently by Parsegian and Weiss \cite{Weiss-torque} who studied the inverse case of bodies with anisotropic dielectric response interacting across an isotropic medium in the non-retarded limit \cite{Kornilovitch}. The full retarded Lifshitz result was obtained only much later in a veritable {\sl tour de force} by Barash \cite{Barash1}, following previous partial attempts \cite{Barash2}, leading to a series of recent developments \cite{Enk95:torque,Shao,Capasso1,Capasso-erratum}. The general Lifshitz formulae for the interaction between two anisotropic half spaces or even an array of finite size slabs \cite{Veble2} are algebraically very complicated and untransparent, with little hope of a fundamental simplification \cite{Philbin}. Morphological anisotropy effects are seen either between anisotropic bodies \cite{Emig09:ellipsoids,Emig-2} or between surfaces with anisotropic decorations such as corrugations \cite{Banishev2013b,Roya2}. vdW-like torques have been predicted also between anisotropic inclusions within fluctuating membranes \cite{Golestanian-1996a,Park-1996}. 

\begin{figure}[ht]
\includegraphics[scale=0.6]{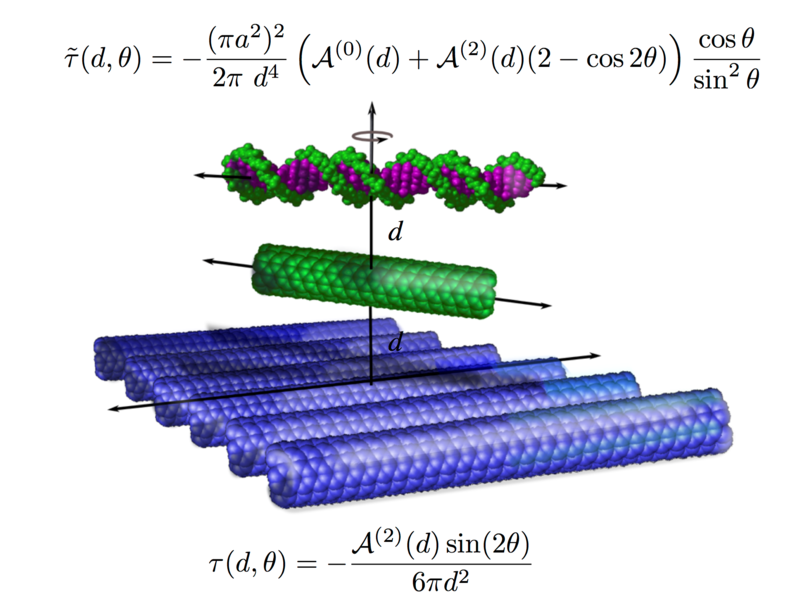}
\caption{(Color Online) ) Forces and torques between anisotropic molecules/molecular aggregates. Top: $\tilde\tau(d, \theta)$, torque between two inclined long cylindrical molecules  (above DNA below SWCNT) as a function of the closest separation $d$ and angle of inclination of their cylindrical axes $\theta$, calculated within the Lifshitz theory. The two Hamaker coefficients ${\cal A}^{(0)}(d)$ and ${\cal A}^{(2)}(d)$ depend on the dielectric anisotropy of the interacting materials as well as on the separation due to retardation effects, but they do not depend on mutual orientation. Bottom: $\tau(d, \theta)$, torque per unit surface area between two planar semi-infinite slabs at separation $d$ with angle of inclination $\theta$ between their anisotropic uniaxial tensors. Only the surface layer of the bottom slab and one molecule of the upper slab are shown. Again, the Hamaker coefficient ${\cal A}^{(2)}(d)$ depends on the dielectric anisotropy of the interacting materials as well as on the separation due to retardation effects, but not on the orientation of the dielectric tensors. Adapted from \cite{Dryden-GeckoNatMat}. (Graphics courtesy J.C. Hopkins).}
\label{Fig2-RMP-Casimir}
\end{figure}

The first attempt to evaluate the vdW interaction between two cylinders comes from Barash and Kyasov \cite{Barash-1989}. Results for two infinitely long anisotropic cylinders can be obtained in a dilution process \cite{Parsegian}, such that the presence of dielectric cylinders can be considered as a small change of the dielectric permittivity of two semi-infinite regions \cite{pitaevskii_thermal_2008}. This approach leads to interactions between infinite cylinders of radius $R$ at minimum separation $d$, at any angle of inclination $\theta$ in non-retarded \cite{Rajter1} as well as retarded limits \cite{Siber1}, but also between cylinders and anisotropic semiinfinite layers \cite{Hopkins-aniso,Saville}. The vdW interaction free energy for inclined cylinders is
\begin{equation}
 {\cal F}(d, \theta) = - \frac{ (\pi R^{2})^2}{2\pi~d^{4}\sin{\theta}}\left( {\cal A}^{(0)}(d)  + {\cal A}^{(2)}(d) \cos 2\theta \right),
\label{Glt}
\end{equation}
where $1/\sin{\theta}$ stems from the shape anisotropy and the $\cos 2\theta$ dependence associated with the material anisotropy, Fig. \ref{Fig2-RMP-Casimir}.  The Hamaker coefficients, ${\cal A}^{(0)}(d)$ and ${\cal A}^{(2)}(d) $, are functions of  separation and the relative anisotropy measures, but do not depend explicitly on the angle of inclination $\theta$. They also depend on the material types and Matsubara sampling frequencies \cite{Siber1,DrydenStark-gecko}. 
In the symmetric interaction case \cite{Hopkins-aniso} both ${\cal A}^{(0)}(d)$ and ${\cal A}^{(2)}(d) $ decompose into a square and thus cannot be negative or change sign. In the asymmetric case, however, the sign of the interaction as well as the sign of the torques are more complicated, as they depend on the perpendicular and parallel dielectric response of the interacting bodies. They do not follow the general rule for interacting planar bodies 1 and 2 across a medium $m$, with a change of sign implied by the sequence, $ \varepsilon_1(\omega) > \varepsilon_m(\omega) >  \varepsilon_2(\omega)$, see above.  

vdW torques between semi-infinite anisotropic materials also imply torques between anisotropic cylindrical molecules, such as filamentous graphitic systems of metallic and semiconducting single-walled CNTs \cite{Rajter1,Siber1,Rajter2}, multiple composites of DNA \cite{zheng_molecular_2009, young_using_2014, sontz_dna_2012}, type I collagen \cite{cheng_electrochemical_2008}, and polystyrene \cite{jin_superhydrophobic_2005}. DNA optical properties \cite{Pinchuk1} have been converted into the corresponding separation dependence vdW free energy in the case of pairs of nucleotides \cite{Pinchuk3,Pinchuk2}. A static dielectric constant of $8$ for DNA, that enters the $n=0$ Matsubara term,  was recently measured inside single T7 bacteriophage particles by electrostatic force spectroscopy \cite{Cuervo}. Optical dispersion data are also available for single nucleotides, nucleosides and derivatives, synthetic polynucleotides (polyuridylic acid, polyadenylic acid, poly-AU), various nucleic acids, such as RNA and native bacterial DNAs in aqueous solutions \cite{Voet}, or wet and dry polymerized oligonucleotides and mononucleotides \cite{Zalar,Silaghi} as well as dry DNA thin films \cite{Sonmezoglu}.  Optical properties of DNA oligonucleotides (AT)10, (AT)5(GC)5, and (AT-GC)5 using {\sl ab initio} methods  and UV-Vis decadic molar absorbance measurements show a strong dependence of the position and intensity of UV absorbance features on oligonucleotide composition and stacking sequence \cite{Schimelman-PCCP}. The calculated Hamaker coefficients for various types of DNA molecules are overall small but depend on the base-pair sequence details and could control the finer details of the equilibrium assembly structure \cite{Grzybowski-2009}. 
In fact, the stacking sequence dependence of the optical properties has important repercussions for the molecular recognition between two approaching DNA molecules that depends on vdW interactions \cite{Bing}. The angle-independent part of the Hamaker coefficient is $\simeq 5 zJ$, while the angular part is effectively zero when the zero-frequency Matsubara component is fully screened by the electrolyte solution. At least for DNA molecules, it then appears that the anisotropy effects stem purely from the shape anisotropy, whereas this is not the case for CNTs. Among the fibrous proteins collagen also shows strong vdW interactions with silica resulting in a Hamaker coefficient that is 39\% larger than that of the silica-(GC)10 DNA interaction at 5 nm separation \cite{Dryden-GeckoNatMat}. 

Though the vdW torque, defined as $\tau(d, \theta) = -\frac{\partial {\cal F}(d, \theta)}{\partial \theta}$, is eminently measurable, it has however not been measured directly yet \cite{Capasso2,Capasso3,Chen}. The anisotropy that engenders the vdW torque can be of different origins: it can result either from anisotropy of the dielectric response of the interacting bodies or from their asymmetric shape \cite{Hopkins-aniso}. Both give  effective Hamaker coefficients that depend on the mutual orientation of the dielectric or shape axes \cite{Dryden-GeckoNatMat}. The material anisotropy can also be either intrinsic, or a consequence of arrays of nanoparticles embedded in an isotropic background \cite{Raul2}, and/or a consequence of the action of external fields \cite{Raul1}. It remains unclear which anisotropic effects would be best suited for accurate experiments.

\subsection{Electrostatic fluctuations}

The $n=0$ ("static") Matsubara term corresponds to the classical partition function of the Coulomb system and can have a form very different from the one derived from the Lifshitz theory. For an interacting Coulomb fluid such as a confined electrolyte or plasma, a counterion only system, or a system of dipoles or polarizable particles in an inhomogeneous dielectric background,  the $n=0$ Matsubara term corresponds to the free energy of fluctuations around the mean-field in the range of parameter space where the mean-field (Poisson-Boltzmann or weak coupling) approximation holds (for details,  see \cite{Perspective}). This leads to effects such as screening of the Hamaker coefficient, universal value for the Hamaker coefficient or its anomalous separation dependence. In effect, the $n=0$ Matsubara term actually corresponds to Gaussian or one-loop electrostatic potential fluctuations around the mean-field for a fully coupled system \cite{Podgornik-Zeks, Netz-vdW} and thus presents a first order correction to the description of Coulomb fluids on the mean-field level \cite{Holm}. Formally this follows from an effective non-Gaussian ``field-action" $S[\phi]$ that one can derive from an exact field-theoretic representation of the confined Coulomb fluid partition function in terms of the fluctuating local electrostatic potential \cite{Edwards-Lenard, Podgornik89b}. The mean-field theory is then defined as the saddle-point of this field action \cite{Perspective}. 

Thermal fluctuations around the saddle-point at the first-order loop expansion representing the contribution from correlated Gaussian fluctuations around the mean-field or saddle-point solution \cite{Book-Coulomb}, lead to a thermal fluctuation-induced vdW-like attraction in the form of the trace-log of the ``field-action" Hessian 
\begin{equation}
F = - k_BT ~{\rm Tr Log}  \left( \frac{\delta^2 S[\phi]}{\delta \phi({\bf r}) \delta \phi({\bf r}')}\bigg\vert_{\phi({\bf r})=-\rm{i} \psi_{\mathrm{PB}}({\bf r})}\right) + {\cal O}(\phi^3),
\end{equation}
where $\phi({\bf{r}})=-\rm{i} \psi_{\mathrm{PB}}({\bf r})$ is the saddle-point (Poisson-Boltzmann) electrostatic potential configuration. This second-order correction universally lowers the interaction pressure between surfaces and thus leads to an attractive contribution to the total interaction pressure. It also includes a term that exactly cancels the zero-frequency contribution in the Lifshitz theory \cite{Podgornik89b} so it should be viewed as a substitute for the $n=0$ Lifshitz term. As a rule, this fluctuation attraction is weaker than the repulsive leading-order saddle-point contribution, thus the total interaction remains repulsive \cite{Netz1,Netz2}. In certain models that either assume charge asymmetry \cite{Kanduc1}, surface condensation or adsorption of counterions on (fixed) charged boundaries, the repulsive mean-field effects are strongly suppressed and the total interaction is then mostly  due to thermal fluctuations \cite{Lau1,Ha,Lau2}. This can happen in the case of oppositely charged surfaces where the thermal fluctuation interactions can become dominant \cite{Kanduc1,Lau3,Burak}. 

A related problem of thermal electrostatic fluctuations is presented by the Kirkwood-Shumaker (KS) interactions that exist between macroions with dissociable charges, such as proteins \cite{Lund}. Originally this interaction was obtained from a perturbation theory around an uncharged state \cite{Shumaker,Lund}. The KS interaction is similar to the thermal vdW interaction but it corresponds to monopolar charge fluctuations \cite{Natasa1} and is thus in principle much longer ranged. Monopolar fluctuation cannot arise for fixed charges on interacting bodies and some surface charging mechanism or charge regulation, where the macroion surfaces respond to the local electrostatic potential with a variable effective charge, is needed \cite{Borkovec-review}.  Formally charge regulation can be described by another, nonlinear source term in the field-action, $f_S(\phi({\mathbf{r}}))$, that involves the fluctuating potential at the surface (S) \cite{Natasa2}. Nonlinearity of this field-action is essential as a linear dependence on the fluctuating potential, in fact, corresponds to a fixed charge that cannot exhibit monopolar charge fluctuations. These are given by the surface capacitance determined by the second derivative of $f_S(\phi )$ with respect to the surface potential, and the KS interactions depend quadratically on this capacitance. The KS interaction between two macroions in the asymptotic regime between particles $1$ and $2$ then assumes the form ${\cal F} \sim -\frac{{\cal C}_1 {\cal C}_2}{{R}^2}$.  The exact form of $f_S$ and thus the capacitance $\cal C$ is not universal as it depends on the surface-ion interaction \cite{Tomer,FleckNetz,Natasa2}.  Although calculating the KS interaction is challenging, exact solutions are available  in 1D, demonstrating a rich variety of behaviors due to charge regulation and the ensuing correlated fluctuations \cite{Maggs1}. Some of these 1D properties transfer also to the more realistic 3D models  between globular proteins with dissociable surface charge groups \cite{Natasa2}.
Similar anomalously long-range monopolar fluctuations and concurrent Casimir/vdW-like interactions can also result from a different mechanism where monopolar charge fluctuations result not from charge regulation but rather from nano-circuits with capacitor components, where fluctuating charges are transferred through the wire connection in a capacitor system \cite{Drosdoff-2015}.

The link between Coulomb interactions and thermal Casimir/vdW interactions has been implicated also in some theoretical approaches to the Hofmeister or specific ion effects \cite{Ninham-rev}. These works motivated numerous investigations of non-electrostatic ion-specific interactions between ions and surfaces and their role in modifying surface tension of electrolyte solutions or indeed the solution behavior of proteins. In fact, the standard Onsager-Samaras result was only recently realized to be fluctuational in nature \cite{Tomer}. Ninham and coworkers as well as others \cite{Edwards} made  attempts to include vdW interactions into a complete theory of ion interactions in confined aqueous solutions \cite{Ninham1,Ninham2,Ninham3}. The major problem in including  vdW interactions into the description of inhomogeneous electrolytes is that the contributions of fixed charges and ion polarizability are in general not additive \cite{Demery-non}. However, they sometimes can be  approximated by an additive contribution on the strong coupling level, provided that the polarizability of the ions is large enough. A popular Ansatz that simply adds a vdW ion-polarizability dependent contribution to the electrostatic potential of mean force has thus a very limited range of validity.

\section{Experiments probing materials aspects of vdW/Casimir interactions}

The theory and computation of vdW/Casimir and related fluctuation-induced interactions  are experiencing an expansion paralleled by numerous recent materials and structured systems discoveries. Experimental efforts have also been reported probing how different materials can be utilized to modulate this subtle force. Most recent experiments have concentrated primarily on structured systems and some biological matter, as discussed in this review. Additionally, novel measurements have served as a propeller to the field not only to validate certain theoretical predictions, but also to identify new problems.

Atomic force microscopy (AFM) techniques have been employed in vdW measurements giving unprecendented insights into smaller and hetergoeneous systems. Pulling single molecules with an AFM tip from metallic surfaces \cite{Tautz-NatComm} confirms the asymptotic $d^{-3}$ force law and quantifies the non-additive part of the vdW interaction. Non-additive effects have also been demonstrated in adhesion measurements in various tribological environments, as well \cite{Jacobs1,Jacobs2}. Another recent report gives a clear evidence of the vdW screening capabilities in graphene/$\mathrm{MoS_2}$ heterostructures giving insight into adhesion properties of graphene and other layered materials \cite{Tsoi-ACSNano}. These non-additive and screening effects are particularly challenging for theory, which has motivated developments in first-principles calculations methods, as discussed in Sec. II. 

The first experimental measurement of the Casimir force involving graphene has been reported in \cite{Banishev2013}. Good agreement with theory taking into account the Dirac spectrum has been achieved for the studied graphene/$\mathrm{SiO_2}$ setting \cite{Banishev2013, Mostepanenko2015}. However, many issues need to be investigated further. For example, more precise comparison with theory is needed, which on the other hand requires measurements of free standing graphene interactions. Schemes to determine the asymptotic distance dependence and temperature effects are absent. Experiments in this direction will be extremely desirable as they can serve as a validator for the numerous theoretical predictions. Reliable experiments for stacks of graphene are also much desired in order to determine directly the binding energy of graphite and settle the wide range of values reported through indirect measurements \cite{Crespi1998, Hertel2004}. It would also be very interesting to seek experimental knowledge of vdW/Casimir interactions involving other systems with a Dirac spectrum. Probing novel topological phases in such dispersive interactions using topological and Chern insulators can be very beneficial.   

Experimental measurements in structured materials continue to shed
light on many aspects of the Casimir force, from non-additive effects
in grating and related
geometries~\cite{IntravaiaNC13,Banishev2013b,Chan2008}, to repulsive
interactions in interleaved structures~\cite{RodriguezJo08:PRA,Tang-2015},
to strong temperature corrections arising at large
separations~\cite{Sushkov10} or in situations involving structured
magnetic media~\cite{Bimonte15}. Experiments at nano-metric scales,
involving objects with non-trivial surface topology due to roughness
or patch charges, are also beginning to push the boundaries of
theoretical techniques~\cite{Lamoreaux10:review,Kim10}. For instance,
recent AFM characterizations of the surface morphology of planar
objects~\cite{Zwol08:rough} coupled with predictions based on the
above-mentioned state-of-the-art simulation techniques~\cite{Broer12}
reveal that the presence of roughness on the scale of their
separations manifests as strong deviations in the power-law scaling of
the force~\cite{Broer13}. Challenges that continue to be addressed in
current-generation experiments include the need to control and
calibrate materials properties. Specifically, accurate comparison
between theory and experiments require accurate knowledge of the dielectric response  
over a broad spectral region, spurring recent efforts to
characterize numerous material properties from DC to ultraviolet
wavelengths~\cite{Zwol09,Sedighi14}. Finally, a number of experiments
employing novel fabrication techniques have begun to explore Casimir
forces in on-chip, integrated systems~\cite{Zou13,Yamarthy12} where
parallelism is no longer a key impediment, removing the need for
external instrumentation needed to bring objects closer together and
paving the way for applications where the force is exploited in
conjunction with other effects (e.g. mechanical or optical actuation)
to enable new
functionalities~\cite{Pernice10,Rodriguez11-APL,Yamarthy12}.

On the other hand, small-angle X-ray scattering techniques are useful for measuring vdW potentials in soft- and bio-matter systems. Their contribution to the elucidation of the details of long-range interactions in multilamellar membrane array context is crucial and continuing \cite{Pabst-2015}. It remains to be seen whether SAXS coupled to osmotic stress could provide also a vdW component to the interaction potential between filamentous bio-molecules \cite{Yasar}, allowing also a determination of the Hamaker coefficient that could be compared with calculations \cite{Schimelman-PCCP,poudel_partial_2014}. While the Hamaker coefficients for general proteins could be approaching solid predictions \cite{Eifler_collagen_2014}, their experimental determination is marred by the limitations inherent in the second virial coefficient determination of the global characteristics of the interaction potential \cite{Prausnitz-BJ}. Novel methodologies such as colloid-probe AFM could be a new potential source of valuable data on intermolecular potential \cite{Borkovec-AFM}, including protein-protein interactions at all values of separation \cite{AFM-protein}.

\section{Future Outlook}

The Casimir/vdW force has manifestations in many parts of physics as discussed at length in this review. The quest for a fundamental understanding of this ubiquitous and subtle force bridges concepts from condensed matter and high energy physics, which has become much more apparent with recent discoveries of novel materials. This field has also stimulated the development of computational methods at the atomistic levels as well as larger scale with the goal of taking into account  the collective and non-additive nature of the same dispersive interaction. This particular direction has also been stimulated due to materials science expansion and better design of devices. Nevertheless, the field can become even broader with several eminent problems awaiting solutions.  

Materials with Dirac spectrum hold much promise to discover new science about the vdW/Casimir interaction. Unusual behavior of the Casimir force in terms of sign, magnitude, distance dependence, and other factors has been found in graphene, TIs and CIs, as discussed in this review. But there are many new entering players with emergent properties. 2D TIs, such as $\mathrm{HgTe/Cd/Te}$, Bi bi-layers, and InAs/GaSb; 3D TIs, such as $\mathrm{Bi_{1-x}Sb_{x}}$, $\mathrm{Bi_2Te_3}$, Heusler alloys, and topological crystalline insulators, such as $\mathrm{SnTe}$, $\mathrm{Pb_{1-x}Sn_xSe}$, have prominent spin-orbit interaction, which coupled with the Dirac spectrum can lead to diverse optical response \cite{Balatsky-2014}. 3D Weyl and Dirac semimetals, such as $\mathrm{Cd_3Al_2}$, $\mathrm{Na_3Bi}$ can also be put in this category. Recent reports show that the quantum electrodynamics in Weyl semimetals results in a non-trivial response due to the associated axion field, which can lead to a repulsive Casimir interaction \cite{Grushin-2012,Galitski-2015}. However, external electric and magnetic fields together with temperature and doping can modulate the electronic structure and optical properties by creating new topological phases, such as valley polarized materials for example. These are yet to be studied in the context of the vdW/Casimir interaction. 

Another interesting direction to explore originates from nonlocality, especially at very small separations. For
instance, at nano-metric scales, the combination of structured
materials, thermal as well as dielectric inhomogeneities, and
non-local effects associated with atomic-scale physics, can
potentially conspire to affect fluctuation phenomena. Preliminary
works studying non-local material effects have begun to shed light on
these issues~\cite{Esquivel-Sirvent06,Despoja11,Luo14}. One promising
set of scattering techniques that could be used to tackle these
emerging regimes are volume-integral equation methods, related to
surface-integral equations but involving volume rather than surface
currents inside the bodies~\cite{Polimeridis14}. Another set of
techniques that are beginning to pave the way for fundamentally new
designs in nanophotonics but which have yet to be exploited in Casimir
computations are large-scale optimization
methods~\cite{Bendsoe03}. While such brute-force explorations require
careful and efficient formulations due to the large number of required
calculations, the above-mentioned numerical developments offer hope
that such an approach to design is within reach. Finally, although
many of the interesting, non-additive Casimir effects predicted thus
far still remain out of reach of current experiments, perhaps related
physical principles can be employed to discover other structures where
non-monotonicity and/or repulsion is larger and more experimentally
accessible.

Transformation optics, a powerful method for solving Maxwell's equations in curvilinear coordinates \cite{Leonhardt2006,Pendry-2006}, may offer a different perspective to the vdW/Casimir effect, especially for systems that have sizes  comparable to their separation. It can be an efficient numerical approach for vdW calculations by taking into account nonlocal effects for absorption and scattering spectra, electromagnetic modes and field enhancement. Transformation optics has been a powerful tool for optics design with applications, such as perfect lensing and cloaking \cite{Pendry-2006}. Recent reports have shown interesting physical insight for vdW interactions in 3D objects with nonlocal dielectric properties \cite{Luo-2014,Pendry-2013}. An exciting future direction can be to examine many of such predictions and applications in the context of vdW/Casimir interactions for new directions of control and manipulations.  

We further note that fluctuation-induced phenomena go beyond the dipolar fluctuations that give rise to the vdW/Casimir force. Charge and potential fluctuations, beyond the situations discussed in bio-materials, may be very interesting in solid state devices. Dispersive forces of charged objects are much less studied. Systems with reduced dimensionality may be used to investigate much longer ranged monopolar fluctuation forces which can exist on their own or be entangled with the "traditional" dipolar fluctuations \cite{Bimonte-2007,Drosdoff-2015}. This practically unexplored direction holds promise to expand fluctuation induced interactions beyond dipolar excitations and further broaden the perspective of Casimir-like phenomena.    

We also would like to mention the puzzle about the relaxation properties of conduction carriers and their role in the Lifshitz theory. Depending on the dielectric model used, Drude-like or plasma-like, different magnitudes of the thermal Casimir force between metallic or magnetic systems are predicted. The crux of the issue is the correct description of the low-frequency optical response of the materials. In \cite{Sushkov10} the Casimir force between metallic samples was measured, and the authors interpreted their results in agreement with the Drude model, after subtracting a force systematics due to electrostatic patches that was modeled and fitted to the total observed force. Recent independent measurements of patch distributions on metallic samples used in Casimir force experiments report different strengths and scaling laws for the patch contribution to the total force \cite{Behunin2014,Garrett2015}. In contrast, several other Casimir experiments \cite{Decca2005,Banishev-2013} seem to be in agreement with the plasma model description, which is surprising given that this model neglects dissipation effects in metals. More recently,  a proposal was put forward \cite{Bimonte-2014,Bimonte-2014a} based on the isoelectronic technique (which eliminates the need for electrostatic corrections due to patches) that makes a significant step forward towards a resolution of this controversial issue. With this set up, it becomes possible to strongly enhance the discrepancy between the predictions for the Casimir force based on either model for the dielectric response. Preliminary measurements  \cite{Decca-2015} are in favor of theoretical extrapolations to low frequency based on the plasma model. To date, a basic understanding of this fundamental problem in Casimir physics is still missing. Perhaps one way to resolve this issue is to consider materials with reduced dimensions and novel phases. For example, the thermal fluctuations effects are much more prominent for graphene, which can be a possible direction to explore in this context. A different pathway could be that more sophisticated models for the response properties are needed. Nevertheless, a possible resolution to this open problem may be found by improving our understanding of materials properties.

\section{Conclusions}
A broad perspective in materials and their properties was given to the field of van der Waals and Casimir interactions. This comprehensive review shows that this a broad area where materials have played an important role in motivating the development of new theoretical models and computational approaches as well as advances in experimental techniques. Materials may hold the answers of several open  problems in fluctuations-induced phenomena, which are of fundamental and applications relevance.      

\section{Acknowledgements}
L. M. W. acknowledges financial support from the US Department of Energy under contract DE-FG02-06ER46297. D. A. R. D. was supported by the LANL LDRD program. A.T. thanks the European Research Council (ERC StG VDW-CMAT) for funding.  P. R.-L. acknowledges financial support from People Programme (Marie Curie Actions) of the European Union's Seventh Framework Programme (FP7/2007-2013) under REA grant agreement No.302005. A. W. R. acknowledges financial support from the US National Science Foundation under Grant No. DMR-1454836. R. P. would like to acknowledge the support of the U.S. Department of Energy, Office of Basic Energy Sciences, Division of Materials Sciences and Engineering under Award DE-SC0008176. R. P. would like to thank V. Adrian Parsegian, Roger H. French, Wai-Yim Ching, Nicole F. Steinmetz, and Jaime C. Hopkins for their input in preparing this review.

\bibliography{References}

\end{document}